\pdfoutput=1

\documentclass[11pt,twoside,a4paper,cmspaper,final,collab]{cms-tdr}
\def\svnVersion{291017}\def\cmsCernNo{2013-037}\def\cmsCernDate{\today}\def\cmsMessage{Published in Physical Review D as \href{http://dx.doi.org/10.1103/PhyRevD.91.092005}{\doi{10.1103/PhyRevD.91.092005.}}}

\begin{document}\cmsNoteHeader{EXO-12-060}

\hyphenation{had-ron-i-za-tion}
\hyphenation{cal-or-i-me-ter}
\hyphenation{de-vices}
\RCS$HeadURL: svn+ssh://svn.cern.ch/reps/tdr2/papers/EXO-12-060/trunk/EXO-12-060.tex $
\RCS$Id: EXO-12-060.tex 285229 2015-04-19 23:51:48Z alverson $
\newlength\cmsFigWidth
\ifthenelse{\boolean{cms@external}}{\setlength\cmsFigWidth{0.49\textwidth}}{\setlength\cmsFigWidth{0.49\textwidth}}
\newlength\cmsFigWidthTwo
\ifthenelse{\boolean{cms@external}}{\setlength\cmsFigWidthTwo{0.4\textwidth}}{\setlength\cmsFigWidthTwo{0.49\textwidth}}
\ifthenelse{\boolean{cms@external}}{\providecommand{\cmsLeft}{top}}{\providecommand{\cmsLeft}{left}}
\ifthenelse{\boolean{cms@external}}{\providecommand{\cmsRight}{bottom}}{\providecommand{\cmsRight}{right}}
\ifthenelse{\boolean{cms@external}}{\providecommand{\CL}{C.L.\xspace}}{\providecommand{\CL}{CL\xspace}}
\ifthenelse{\boolean{cms@external}}{\providecommand{\CLend}{C.L.\xspace}}{\providecommand{\CLend}{CL.\xspace}}
\providecommand{\CLs}{\ensuremath{\mathrm{CL}_\mathrm{s}}\xspace}
\ifthenelse{\boolean{cms@external}}{\providecommand{\NA}{\ensuremath{\cdots}}}{\providecommand{\NA}{\text{---}}}

\providecommand{\Wprime}{\PWpr}
\providecommand{\Zprime}{\cPZpr}
\providecommand{\sm}{standard model\xspace}
\providecommand{\MT}{\ensuremath{M_\mathrm{T}}\xspace}
\providecommand{\MTlower}{\ensuremath{M_\mathrm{T}^\mathrm{min}}\xspace}
\providecommand{\Mchi}{\ensuremath{M_\chi}\xspace}
\providecommand{\ET}{\ensuremath{E_\mathrm{T}}\xspace}
\providecommand{\pT}{\ensuremath{p_\mathrm{T}}\xspace}
\providecommand{\MPT}{\ensuremath{\vec{p}\kern.1em_{\text{T}}^{\text{miss}}}\xspace}
\providecommand{\invpb}{\pbinv}
\providecommand{\invfb}{\fbinv}
\providecommand{\WprimeMuNu}{$\Wprime \to \mu\nu $\xspace}
\providecommand{\WprimeENu}{$\Wprime \to \Pe\nu $\xspace}
\providecommand{\WprimeKK}{\ensuremath{\cmsSymbolFace{W}_\mathrm{KK}}\xspace}
\providecommand{\WprimeKKn}{\ensuremath{\cmsSymbolFace{W}^{\mathrm{(n)}}_\mathrm{KK}}\xspace}
\providecommand{\WprimeKKtwo}{\ensuremath{\cmsSymbolFace{W}^\mathrm{(2)}_\mathrm{KK}}\xspace}
\providecommand{\WMuNu}{$\PW \to \mu \nu $\xspace}
\providecommand{\pb}{\ensuremath{\unit{pb}}\xspace}
\providecommand{\fb}{\ensuremath{\unit{fb}}\xspace}
\providecommand{\POWHEG}{{\textsc{powheg}}\xspace}
\providecommand{\LMET}{$\ell$ + \MET}
\providecommand{\PZ}{\cPZ\xspace}
\providecommand{\tPZ}{\cPZ\xspace}
\providecommand{\tPW}{\PW\xspace}
\providecommand{\TeVmOne}{TeV\ensuremath{^{\mathrm{-1}}}\xspace}

\cmsNoteHeader{EXO-12-060}
\title{Search for physics beyond the standard model in final states with a lepton and missing transverse energy in proton-proton collisions at
\texorpdfstring{$\sqrt{s}=8\TeV$}{ sqrt(s) = 8 TeV}}

\date{\today}

\abstract{
A search for new physics in proton-proton collisions having final states with an electron or muon and missing transverse energy is presented.
The analysis uses data collected in 2012 with the CMS detector, at an LHC center-of-mass energy of 8\TeV, and corresponding to an integrated luminosity of
19.7\fbinv.
No significant deviation of the transverse mass distribution of the charged lepton-neutrino system from the standard model prediction is found.
Mass exclusion limits of up to 3.28\TeV at 95\% confidence level for a $\PWpr$ boson with the same couplings as that of the standard model $\PW$ boson are
determined.
Results are also derived in the framework of split universal extra dimensions, and exclusion limits on Kaluza--Klein \WprimeKKtwo states are found.
The final state with large missing transverse energy also enables a search for dark matter production with a recoiling $\PW$ boson, with limits set on the mass and the
production cross section of potential candidates.
Finally, limits are established for a model including interference between a left-handed $\PWpr$ boson and the standard model $\PW$ boson, and for a compositeness
model.
}

\hypersetup{%
pdfauthor={CMS Collaboration},%
pdftitle={Search for physics beyond the standard model in final states with a lepton and missing transverse energy in proton-proton collisions at sqrt(s) = 8 TeV},%
pdfsubject={CMS},%
pdfkeywords={CMS, physics, W' decay}}

\maketitle

\section{Introduction}
\label{sec:intro}

The standard model (SM) of particle physics is a theory of the structure of matter, describing the properties of
all known elementary particles and the forces between them.
Being studied experimentally for five decades, its predictions have been verified with very high precision.
Despite the great success of the SM, beyond the SM (BSM) physics addresses a variety of open issues. To name a few examples:
the relationship of the electroweak and gravitational energy scales must be understood and incorporated in the theory;
an underlying concept is needed to explain the origin of the observed three fermion families;
astrophysical observations indicate the presence of dark matter (DM) not described in the standard model.
Many SM extensions predict additional heavy gauge bosons, including models with extended gauge sectors, designed to achieve gauge coupling unification, and theories with extra spatial dimensions.
BSM physics can be detected through observation of significant deviations from SM predictions.

The search presented in this paper is sensitive to deviations from the SM prediction for the transverse mass spectrum of events with a charged lepton (electron or muon) and
one or more particles that cannot be directly detected (neutrino, dark matter particle) in the final state.
Additionally, events are allowed to include an arbitrary number
of jets, as they may originate from initial state radiation.   	
Interpretations of the observations are made in the context of various theoretical models:
the sequential standard model (SSM) with a \Wprime boson~\cite{reference-model},
a helicity-non-conserving contact interaction model (HNC-CI)~\cite{HNCM-CI},
a dark matter (DM) model with a DM particle recoiling against a $\PW$ boson~\cite{DMModelSummery,DMModelSummery2,DMSummery},
the sequential standard model with a \Wprime boson interfering with the \tPW boson (SSMS, SSMO)~\cite{criticalPaper,WprimeHelicity,dudkov},
split universal extra dimensions (split-UED)~\cite{PhysRevD.79.091702,JHEP04(2010)081}, and
a \TeVmOne model~\cite{TeVref1, TeVref2, TeVref3, TeVref4}, the latter two predicting
an additional spatial dimension.

Since the discovery of the $\PW$ boson, experiments have scrutinized the lepton and missing transverse energy channel for evidence of physics beyond the standard model.
Neither searches by the Tevatron experiments D0~\cite{d0-limit} and CDF~\cite{cdf-limit},  nor searches carried out previously at the LHC experiments
ATLAS~\cite{atlas-2010, atlas-1fb-limit, atlas-2011limit, ATLAS:2014wra} and CMS~\cite{2010cms-electron-limit,2010cms-muon-limit,2011limit,2012ichep-limit}, have found any indication for such a deviation.
The present analysis improves upon the discovery potential of its predecessors.
It is based on data from an integrated luminosity of $19.7\pm0.5\fbinv$~\cite{LUM-13-001} of proton-proton collisions at a center-of mass energy of 8\TeV, recorded in 2012 with the CMS detector~\cite{Chatrchyan:2008zzk} at the CERN LHC.

The search for new physics is carried out in the transverse invariant mass distribution.
The shape of the distribution is taken into account by using a binned likelihood
method. This approach is especially important as the examined theories predict very different signal event distributions. While the SSM \Wprime boson can be discovered at
very high transverse mass, the DM and CI models manifest themselves as event excesses at lower values of the transverse mass. A \Wprime boson interfering with the standard
model $\PW$ boson can even lead to a deficit of events in some regions compared to the SM prediction.

This paper is structured as follows. Section \ref{sec:detector} describes the experimental setup. The theoretical models are explained in Section \ref{sec:models}. A
discussion of the event reconstruction and selection criteria in Section \ref{sec:selection} is followed by a presentation of the transverse mass distribution of the
selected events in Section \ref{sec:results}. In Section \ref{sec:background} detailed information about the relevant background processes and their prediction is given. A
thorough determination of the uncertainties (Section \ref{sec:uncertainties}) is essential in order to interpret the results. Limit-setting procedures are explained in
Section \ref{sec:limittools}. In Section \ref{sec:limits} the limits in terms of the different signal models are derived. A summary of the results is given in Section
\ref{sec:summary}.

\section{CMS detector}
\label{sec:detector}

The central feature of the CMS apparatus is a superconducting solenoid of 6\unit{m} internal diameter, providing a magnetic field of 3.8\unit{T}. Within the superconducting solenoid volume are a silicon
pixel and strip tracker, a lead tungstate crystal electromagnetic calorimeter (ECAL), and a brass/scintillator hadron calorimeter (HCAL), each composed of a barrel and two endcap sections. Muons are
measured in gas-ionization detectors embedded in the steel flux-return yoke outside the solenoid. Extensive forward calorimetry complements the coverage provided by the barrel and endcap detectors.

The electromagnetic calorimeter consists of 75\,848 lead tungstate crystals which provide coverage in pseudorapidity $\abs{ \eta }< 1.479 $ in a barrel region (EB) and $1.479 <\abs{ \eta } < 3.0$ in two
endcap regions (EE).

The ECAL energy resolution for electrons with a transverse energy $\ET \approx 45$\GeV from $\Z \to \Pe \Pe$ decays is better than 2\% in the central region of the ECAL barrel $(\abs{\eta} < 0.8)$, and
is between 2\% and 5\% elsewhere~\cite{Chatrchyan:2013dga}. For high energies, which are relevant for this analysis, the electron energy resolution slightly improves~\cite{EXO-12-061}.

Muons are measured in the pseudorapidity range $\abs{\eta}< 2.4$, with detection planes made using three technologies: drift tubes, cathode strip chambers, and resistive plate chambers. Matching muons
to tracks measured in the silicon tracker results in a relative transverse momentum resolution in the barrel of
about 1\% for muons with a transverse momentum \pt of up to about 200\GeV and
better than 10\% for high momentum muons of \pt$\sim$1\TeV~\cite{Chatrchyan:2012xi}.

A more detailed description of the CMS detector, together with a definition of the coordinate system used and the relevant kinematic variables, can be found in Ref.~\cite{Chatrchyan:2008zzk}.

\section{Physics models and signal simulation}
\label{sec:models}

Many models of new physics predict \tPW-boson-like particles decaying with an experimental signature of a charged lepton $\ell$ and missing transverse energy \MET, which may flag the presence of a non-interacting particle. \MPT is defined as ${-}\sum \ptvec$ of all reconstructed particles with \MET being the modulus of \MPT.
These additional heavy vector bosons may arise in models with more symmetry groups, extra dimensions, compositeness, or other scenarios. Their presence may be detected as a feature in the observed spectrum of transverse mass, defined as
\begin{equation}
\MT = \sqrt{2  \pt^\ell  \MET  \bigl(1 - \cos [\Delta \phi(\ell,\MPT)]\bigr) },
\label{eqn:mt}
\end{equation}
where $\Delta \phi(\ell,\MPT)$ is the azimuthal opening angle between the directions of the missing transverse energy and that of the charged lepton. The spectrum is expected to be dominated by the \tPW boson background, which has a very small cross section at high \MT. Most new physics models predict high-\pt leptons, which should be identifiable in the low-background region.

\begin{figure*}[hbtp]
\centering
\includegraphics[width=.8\textwidth]{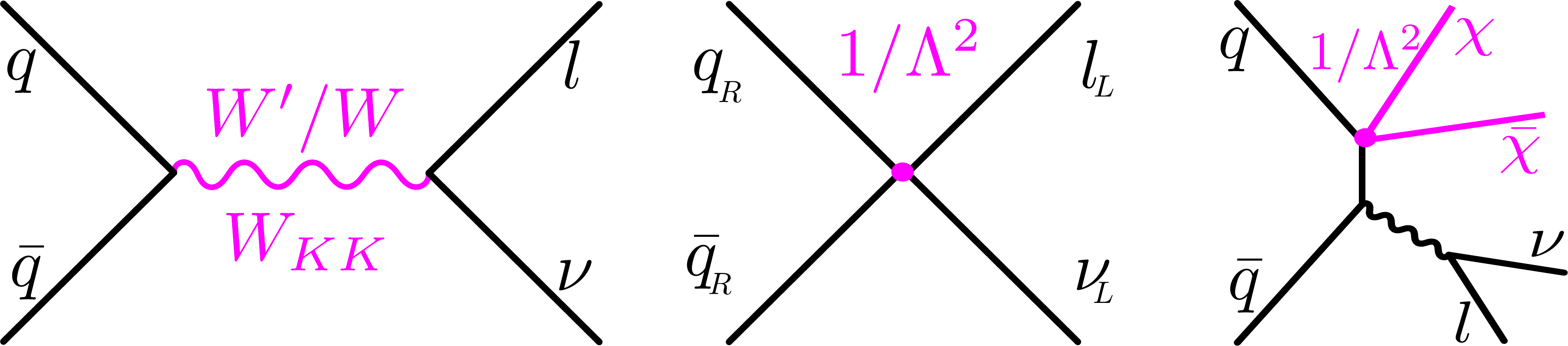}
\caption{Production and decay of an SSM \Wprime or \WprimeKK boson (left); HNC-CI (center); DM single \tPW boson production (right).}
\label{fig:feynman}
\end{figure*}

This section summarizes the new physics models used for interpretation of the observations, along with model-specific assumptions and details of the generator programs used
for production of simulated signal event samples. All generated signal events are processed through a full simulation of the CMS detector based on
\GEANTfour~\cite{Agostinelli:2002hh,Allison:2006ve}, a trigger emulation, and the event reconstruction chain. An overview of the models considered is given in
Table~\ref{tab:models}. Three representative signal examples (sequential standard model, contact interactions, and dark matter) are used as examples in the distributions
throughout the paper. Diagrams of these three signals are shown in Fig.~\ref{fig:feynman}.
In addition, specific model variations are discussed: the sequential standard model with \PW-\Wprime interference (SSMO and SSMS) and two models with one extra dimension and selective particle
couplings to this dimension.
All limits are given at 95\% confidence level (\CL) unless stated otherwise.

The analysis is performed in two channels: the $\Pe+\MET$ and the $\mu+\MET$ channel, where the charged lepton is required to be prompt.
Final states where the electron or muon originates from, \eg, a $\tau$ decay, are not considered as a signal. Therefore, the results can be interpreted for each coupling individually. Only the dark matter model is exempt from that rule. In this model the lepton is produced via a standard model interaction as shown in  Fig.~\ref{fig:feynman}, and thus events with non-prompt leptons originating from $\tau$ decays are also considered.

\begin{table*}[hbtp]
\renewcommand{\arraystretch}{1.1}
\topcaption{Overview of models considered, with the relevant model parameters. The models are explained in the text with more detail.}
\begin{tabular}{p{0.3\textwidth} c p{0.5\textwidth}}
\hline
\hline
Model name & Parameters & Description\\
\hline
Sequential standard model without interference (SSM)~\cite{reference-model}          & $M_{\Wprime}$ & The SSM \Wprime boson does not interfere with the \tPW boson. It has the same coupling strength to fermions as the \tPW boson and its decay width is determined by its mass.\\
\hline
Helicity-non-conserving contact interaction model (HNC-CI)~\cite{HNCM-CI}    & $\Lambda$   & A four-fermion contact interaction model. Quarks and leptons are composite objects of fundamental constituents. No interference effects occur in this model.\\
\hline
Dark matter effective theory (DM)~\cite{DMModelSummery,DMModelSummery2,DMSummery}	                    & $\Mchi,\Lambda, \xi$   & A dark matter model with \tPW-boson radiation. Fermionic dark matter particles have a effective coupling to quarks. An SM \tPW boson recoils against the pair of dark matter particles.\\
\hline
Sequential standard model with same sign couplings (SSMS)~\cite{criticalPaper,WprimeHelicity,dudkov}     & $M_{\Wprime}$ & The SSMS \Wprime boson interferes with the SM \tPW boson and couples in the same way to fermions. This leads to a destructive interference for $M_\PW < \MT < M_{\Wprime}$ and to a constructive interference for $\MT > M_{\Wprime}$. The coupling strength can vary, resulting in different widths.\\
\hline
Sequential standard model with opposite sign couplings (SSMO)  & $M_{\Wprime}$ & Similar to SSMS, with the \tPW and \Wprime boson couplings to quarks having the opposite sign. This leads to a constructive interference for $M_\PW < \MT < M_{\Wprime}$. The coupling strength can vary, resulting in different widths.\\
\hline
Split universal extra {dimensions} model (split-UED)~\cite{PhysRevD.79.091702,JHEP04(2010)081}            & $\mu$, $R$  & The tower of \WprimeKKn Kaluza--Klein excitations has the same couplings as the \tPW boson. Only if the degree of excitation $n$ is even \WprimeKKn boson couples to SM fermions.
The LHC is expected to be sensitive only to the second excitation in the tower ($n = 2$). The size of the extra \mbox{dimension} $R$ determines the mass of the \WprimeKKn boson. Interference with the SM \tPW boson is not considered.\\
\hline
\TeVmOne model with a \mbox{single} additional spatial dimension (\TeVmOne)~\cite{TeVref1, TeVref2, TeVref3, TeVref4}             & $M_\mathrm{C}$     & SM \tPW bosons propagate into the additional dimension as Kaluza--Klein states. Their coupling constant to fermions is $\sqrt{2}$ times  larger than that of the SM \tPW boson. The compactification scale is denoted $M_\mathrm{C}$ in this model.\\
\hline
\label{tab:models}
\end{tabular}
\end{table*}

\subsection{The Sequential Standard Model \texorpdfstring{\Wprime}{W'} boson}
\label{sec:models-ssm}
In the SSM, the \Wprime boson, as shown in Fig.~\ref{fig:feynman} (left), is considered to be a heavy analogue of the SM \tPW boson, with similar decay modes and branching fractions. These are modified by the presence of the
$\cPqt\cPaqb$ decay channel, which opens up for \Wprime boson masses above 180\GeV. Dedicated searches in this channel are described in Refs.~\cite{tbanalysis-d0,WprimeTBCMS,WprimeTBAtlas,WprimeTBCMSCharge,WprimeTBCMS7TeV,Aad:2014xra}. This analysis considers \Wprime boson masses of ${\geq}300\GeV$, yielding a predicted branching fraction ($\mathcal{B}$) of
about 8.5\% for each of the leptonic channels studied. Under these assumptions, the width of a 1\TeV \Wprime boson would be about 33\GeV. Decays of the \Wprime boson via \PW\tPZ are assumed to be
suppressed; dedicated searches for these decays can be found in Refs.~\cite{WprimeWZCMS,Khachatryan:2014xja}.

The SSM~\cite{reference-model} is a benchmark model used as a reference point for experimental \Wprime boson searches for more than two decades. The Tevatron experiments D0 and CDF
established mass exclusion limits of around 1.00\TeV~\cite{d0-limit} and 1.12\TeV~\cite{cdf-limit}, respectively.  The \Wprime boson searches were among the first analyses to be performed at the LHC,
exploiting the large center-of-mass energy. The LHC experiments ATLAS and CMS have more recently raised the \Wprime boson mass exclusion to values around
2.5\TeV~\cite{2010cms-electron-limit,2010cms-muon-limit,2011limit,2012ichep-limit,atlas-2010, atlas-1fb-limit, atlas-2011limit}.

In accordance with previous analyses~\cite{d0-limit,cdf-limit,2010cms-electron-limit,2010cms-muon-limit,2011limit,2012ichep-limit,atlas-2010, atlas-1fb-limit, atlas-2011limit},
no interference with the SM \tPW boson is considered.
The absence of interference can be interpreted as the result of a V+A coupling of the SSM \Wprime boson.
The signature of a charged high-momentum lepton and missing transverse energy would be observed in the decays of such a \Wprime boson predicted
by left-right symmetric models~\cite{FourthColor,LeftRightNatur,LeftRight,Senjanovic:1978ev}. This particle is
typically assumed to have a heavy right-handed neutrino among its decay products~\cite{Min,MS,Mohapatra:1980yp}. However, the mass of the right-handed neutrino is not
constrained, and it could be light as long as it does not couple to SM weak bosons.
The transverse-mass signature is a Jacobian peak, similar to that of the SM \tPW boson but at much higher masses, as shown by the blue line in Fig.~\ref{fig:signal} (top left). With increasing
\Wprime-boson masses the phase space for production in $\Pp\Pp$ collisions at 8\TeV decreases, because of constraints from the PDFs,
leading to a growing fraction produced off-shell at lower masses.

The simulation of data samples in the SSM is performed at leading order (LO) with \PYTHIA6.4.26~\cite{Sjostrand:2006za}, using the CTEQ6L1 parton distribution functions
(PDF)~\cite{CTEQ6L1}. A \Wprime boson mass dependent K-factor is used to correct for next-to-next-to-leading order (NNLO) QCD cross sections, calculated using
$\FEWZ$~\cite{fewz,Gavin:2012kw}. The K-factors vary from 1.363 to 1.140. Table~\ref{tab:xsec} shows the LO and NNLO cross sections for this model.
The NNLO corrections decrease with \Wprime-boson masses up to around 2.5\TeV.
For higher masses, the K-factor increases and becomes similar to the low-mass values, because of the increased fraction of off-shell production (see Table~\ref{tab:xsec}).

\begin{table*}[hbtp]
\topcaption{Signal production cross sections.}
\label{tab:xsec}
\centering
\begin{tabular*}{\textwidth}{@{\extracolsep{\fill}}l c c c }
\hline
\hline
\multicolumn{4}{c}{Sequential SM \Wprime boson} \\
Particle mass & $\sigma_{\mathrm{LO}} \,\mathcal{B}$ (pb)& K-factor & $\sigma_{\text{NNLO}} \,\mathcal{B}$ (pb)  \\
$m_{\Wprime} = 300\GeV $ & 110       & 1.4    & 150       \\
$m_{\Wprime} = 900\GeV $ & 1.5       & 1.3    & 2.0       \\
$m_{\Wprime} = 2000\GeV$ & 0.021     & 1.2    & 0.026    \\
$m_{\Wprime} = 3000\GeV$ & 0.0013    & 1.2    & 0.0015   \\
$m_{\Wprime} = 4000\GeV$ & 0.00025    & 1.3    & 0.00033   \\
\hline
\end{tabular*}
\begin{tabular*}{\textwidth}{@{\extracolsep{\fill}}l c c c c}
\multicolumn{5}{c}{Contact interactions in the helicity-non-conserving model} \\
$\Lambda (\TeV)$ & 3 & 4 & 7 & 9\\
 $\sigma_{\mathrm{LO}} \times B$ (pb)& 0.54 &0.17 &0.018 &0.0067 \\
\hline
\end{tabular*}
\begin{tabular*}{\textwidth}{@{\extracolsep{\fill}}l c c c | c c c}
\multicolumn{7}{c}{Dark matter}\\
interference parameter $\xi$   & 1 & 0 & -1 & 1 & 0 & -1\\
Particle mass &\multicolumn{3}{c|}{ $\sigma_{\mathrm{LO}} \,\mathcal{B}$ (pb) } & \multicolumn{3}{c}{ $\chi$-proton cross section (pb)}\\
\multicolumn{7}{c}{Spin-independent $\Lambda = 200\GeV$} \\
$\Mchi = 3\GeV$ & $3.1$ & $7.4$ & $26.5$  & $3.6$  & $1.6$  & $0.4$  \\
$\Mchi = 100\GeV$ & $2.9$ & $7.1$ & $25.2$  & $6.0$  & $2.7$  & $0.7$  \\
$\Mchi = 300\GeV$ & $1.9$ & $4.8$ & $17.2$  & $6.1$  & $2.7$  & $0.7$  \\
$\Mchi = 500\GeV$ & $1.0$ & $2.5$ & $9.1$  & $6.1$  & $2.7$  & $0.7$  \\
$\Mchi = 1000\GeV$ & $0.1$ & $0.3$ & $0.9$  & $6.1$  & $2.7$  & $0.7$  \\
\multicolumn{7}{c}{Spin-dependent $\Lambda = 200\GeV$} \\
$\Mchi = 3\GeV$ & 3.1 & 7.4 & 26.5  & $0.2$  & $0.8$  & $1.9$  \\
$\Mchi = 100\GeV$ & 2.5 & 6.4 & 22.8  & $0.3$  & $1.4$  & $3.2$  \\
$\Mchi = 300\GeV$ & 1.2 & 3.1 & 11.1  & $0.4$  & $1.4$  & $3.3$  \\
$\Mchi = 500\GeV$ & 0.5 & 1.2 & 4.3  & $0.4$  & $1.4$  & $3.3$  \\
$\Mchi = 1000\GeV$ & 0.03 & 0.1 & 0.2  & $0.4$  & $1.4$  & $3.3$  \\
\hline
\end{tabular*}
\begin{tabular*}{\textwidth}{@{\extracolsep{\fill}}l c c}
\multicolumn{3}{c}{Models with interference of \tPW and \Wprime bosons (\tPW boson LO cross section subtracted)} \\
Particle mass &SSMS  $\sigma_{\mathrm{LO}} \,\mathcal{B}$ (pb) &SSMO  $\sigma_{\mathrm{LO}} \,\mathcal{B}$ (pb)  \\
$m_{\Wprime} = 300\GeV$  & $33\pm59$         & $90\pm81$           \\
$m_{\Wprime} = 500\GeV$  & $11\pm58$         & $21\pm57$           \\
$m_{\Wprime} = 1000\GeV$ & $0.12\pm0.85$     & $1.562\pm0.099$     \\
$m_{\Wprime} = 2000\GeV$ & $-0.030\pm 0.040$ & $0.0460\pm0.0079$   \\
$m_{\Wprime} = 3000\GeV$ & $-0.0064\pm0.0013$& $0.0125\pm0.0018$   \\
\hline
\end{tabular*}
\begin{tabular*}{\textwidth}{@{\extracolsep{\fill}}l c c c c}
\multicolumn{5}{c}{split-UED \WprimeKKtwo boson} \\
& \multicolumn{2}{c}{$\mu = 0.05\TeV$} & \multicolumn{2}{c}{$\mu = 10\TeV$}   \\
Particle mass & $\sigma_{\mathrm{LO}} \,\mathcal{B}$ (pb) & $\sigma_{\text{NNLO}} \,\mathcal{B}$ (pb)& $\sigma_{\mathrm{LO}} \,\mathcal{B}$ (pb)   & $\sigma_{\text{NNLO}}\,\mathcal{B}$ (pb) \\
$m_{\WprimeKKtwo} = 300\GeV $ & 42        & 56         &   250     & 340     \\
$m_{\WprimeKKtwo} = 500\GeV $ & 2.3       & 3.1        &    37     &  51    \\
$m_{\WprimeKKtwo} = 1000\GeV$ & 0.030     & 0.040      &   2.0     & 2.7    \\
$m_{\WprimeKKtwo} = 2000\GeV$ & 0.00013   & 0.00016    &   0.050   & 0.061  \\
$m_{\WprimeKKtwo} = 3000\GeV$ & 0.0000014 & 0.0000016  &   0.0042  & 0.0048 \\
\hline
\end{tabular*}
\begin{tabular*}{\textwidth}{@{\extracolsep{\fill}}l c}
\multicolumn{2}{c}{\TeVmOne model (\tPW boson LO cross section subtracted)} \\
Particle mass & $\sigma_{\mathrm{LO}} \,\mathcal{B}$ (pb) \\
$M_C = 2000\GeV $ & $-0.966\pm0.025$     \\
$M_C = 2600\GeV $ & $-0.079\pm0.014$    \\
\hline
\end{tabular*}
\end{table*}

\subsection{Contact interactions in the helicity-non-conserving model}
\label{sec:models-ci}
Another interpretation of an enhancement in the \LMET final state can be made in terms of a specific four-fermion contact interaction (CI), shown schematically in Fig.~\ref{fig:feynman} (middle). The considered model assumes that quarks
and leptons are composite objects with fundamental constituents~\cite{preons}, motivated by the observation of mass hierarchies in the fermion sector.

At energies much lower than the binding energy, denoted $\Lambda$, the quark and lepton compositeness manifests itself as a four-fermion CI. The CI
between two quarks, a neutrino, and a charged lepton is described by the helicity-non-conserving (HNC) model~\cite{HNCM-CI}. The corresponding cross section is
$\sigma_{\mathrm{CI}\to \mu\nu}= (\pi\hat{s})/(12\Lambda^4)$, where $\hat s$ is the center-of-mass energy of the partons. Typical cross sections for different values of $\Lambda$ are shown in Table~\ref{tab:xsec}. In the HNC model there is no interference of the final state with the SM \tPW boson because of the difference in the chiral structures. The \MT spectrum for CI would
yield a non-resonant excess, increasing with \MT relative to the SM expectation, shown in Fig.~\ref{fig:signal} (top right). Until recently, no limit on the compositeness energy scale had been set in the muon channel
in the HNC-CI model. A previous version of this analysis~\cite{2012ichep-limit} set a
limit of 10.5\TeV and updated the previous CDF limit in the electron channel from $\Lambda = 2.81\TeV$~\cite{cdf-CIlimit}.

Signal samples for this model were produced with \PYTHIA at LO. There are no existing higher-order calculations for this model, and LO cross sections are used.

\subsection{Dark matter}
\label{sec:models-dm}

One commonly used method to describe direct dark matter production at colliders is the use of an effective field theory (EFT)~\cite{DMModelSummery,DMModelSummery2,DMSummery}. The
matrix element is a four-fermion contact interaction with two quarks in the initial state and two fermionic dark matter particles ($\chi \overline{\chi}$) in the final state. This process would not result in any directly detectable final state objects.
The process may be triggered and analyzed through observation of a SM \tPW boson
recoiling
against the dark matter, as shown in Fig.~\ref{fig:feynman} (right). A search in this leptonic decay mode complements established monojet and
monophoton searches~\cite{CMS-PAS-EXO-12-048,CMS-monophoton-2012} and has the
advantage of lower SM background, as well as the presence of a lepton to trigger the event. This study of the single-lepton channel follows the strategy
outlined in
Ref.~\cite{monoLepton}.
Because the \pT sum of non-interacting particles must be balanced by the charged one, we use two-body decay kinematics for the reconstruction of these events.

Under the assumption of a weakly-interacting particle, different couplings are possible. In analogy with the SM weak interaction, the following two couplings are assumed:
\ifthenelse{\boolean{cms@external}}{
\begin{description}
\item[\mdseries{Spin-independent vector coupling}]
\begin{equation*}\frac{1}{\Lambda^2} \overline{ \chi} \gamma^{\mu} \chi ~\cdot~  \lambda_i \bar q_i \gamma_{\mu} q_i;\end{equation*}
\item[\mdseries{Spin-dependent axial-vector coupling}]
\begin{equation*}\frac{1}{\Lambda^2} \overline{ \chi} \gamma^{\mu}\gamma^5 \chi ~\cdot~ \lambda_i \bar q_i \gamma_{\mu}\gamma^5 q_i.\end{equation*}
\end{description}
}{
\begin{equation*}
\begin{aligned}
&\text{Spin-independent vector coupling:} && \frac{1}{\Lambda^2} \overline{ \chi} \gamma^{\mu} \chi ~\cdot~  \lambda_i \bar q_i \gamma_{\mu} q_i;\\
&\text{Spin-dependent axial-vector coupling:} && \frac{1}{\Lambda^2} \overline{ \chi} \gamma^{\mu}\gamma^5 \chi ~\cdot~ \lambda_i \bar q_i \gamma_{\mu}\gamma^5 q_i.
\end{aligned}
\end{equation*}
}
The model parameters are the scale of the effective interaction $\Lambda = M_\text{messenger}/\sqrt{g_\mathrm{DM}}$, which combines a heavy messenger particle with its
coupling constants to dark matter and to quarks $g_\mathrm{DM}$. The mass of the dark matter particle is denoted \Mchi and is included in the spinors $\chi$ and $\overline{\chi}$.
The parameter $\lambda_i$ introduces a relative coupling strength, which in general could be different for each quark flavor.

\begin{figure}[hbtp]
\centering
\includegraphics[width=\cmsFigWidth]{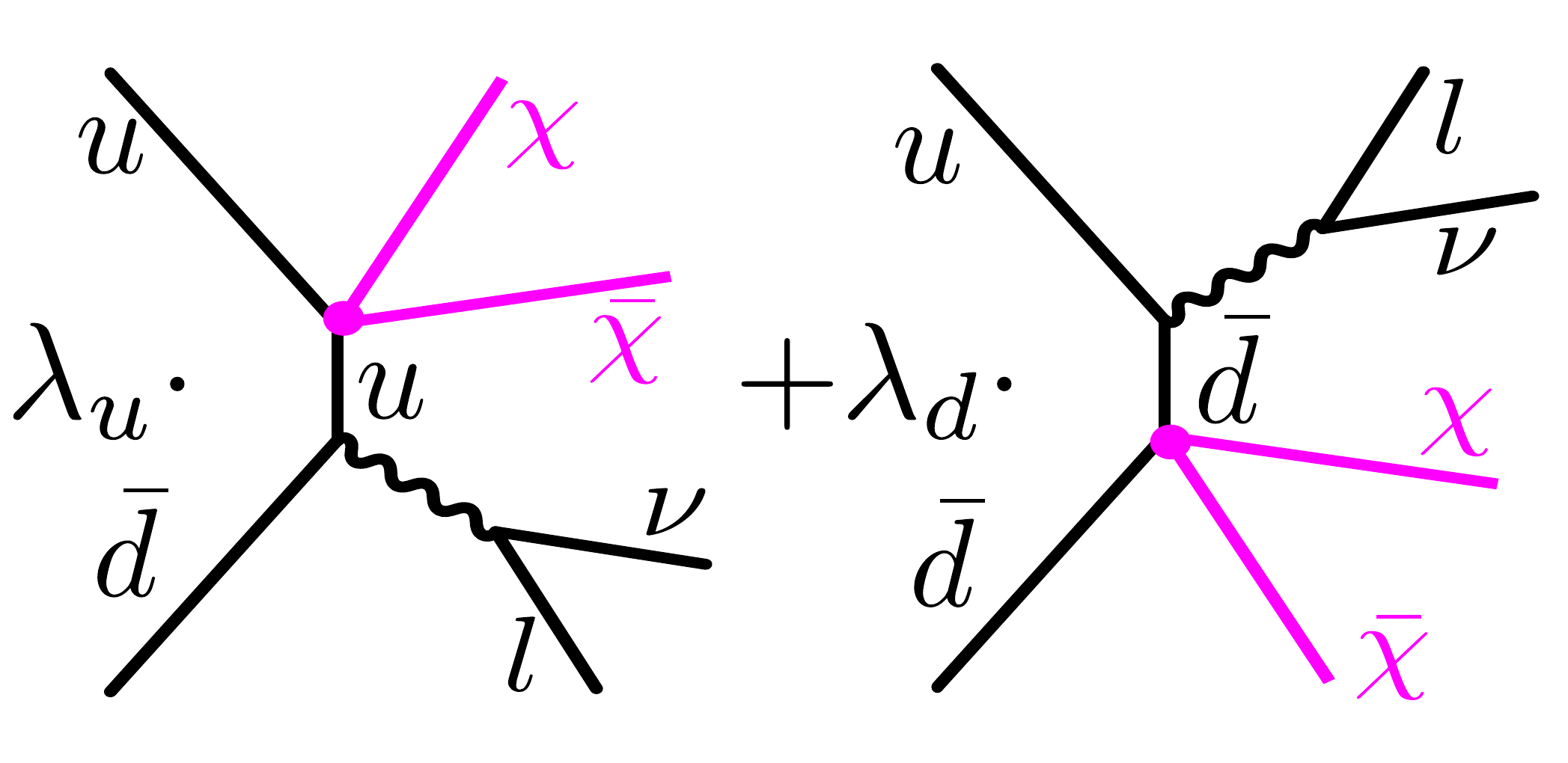}
\caption{Feynman diagrams for dark matter interference, shown as an example with an up and a down quark. The same initial and final state can have different particles coupling to the dark matter particles. }
\label{fig:DM-interference}
\end{figure}

\begin{figure*}[hbtp]
\centering
\includegraphics[width=\cmsFigWidthTwo]{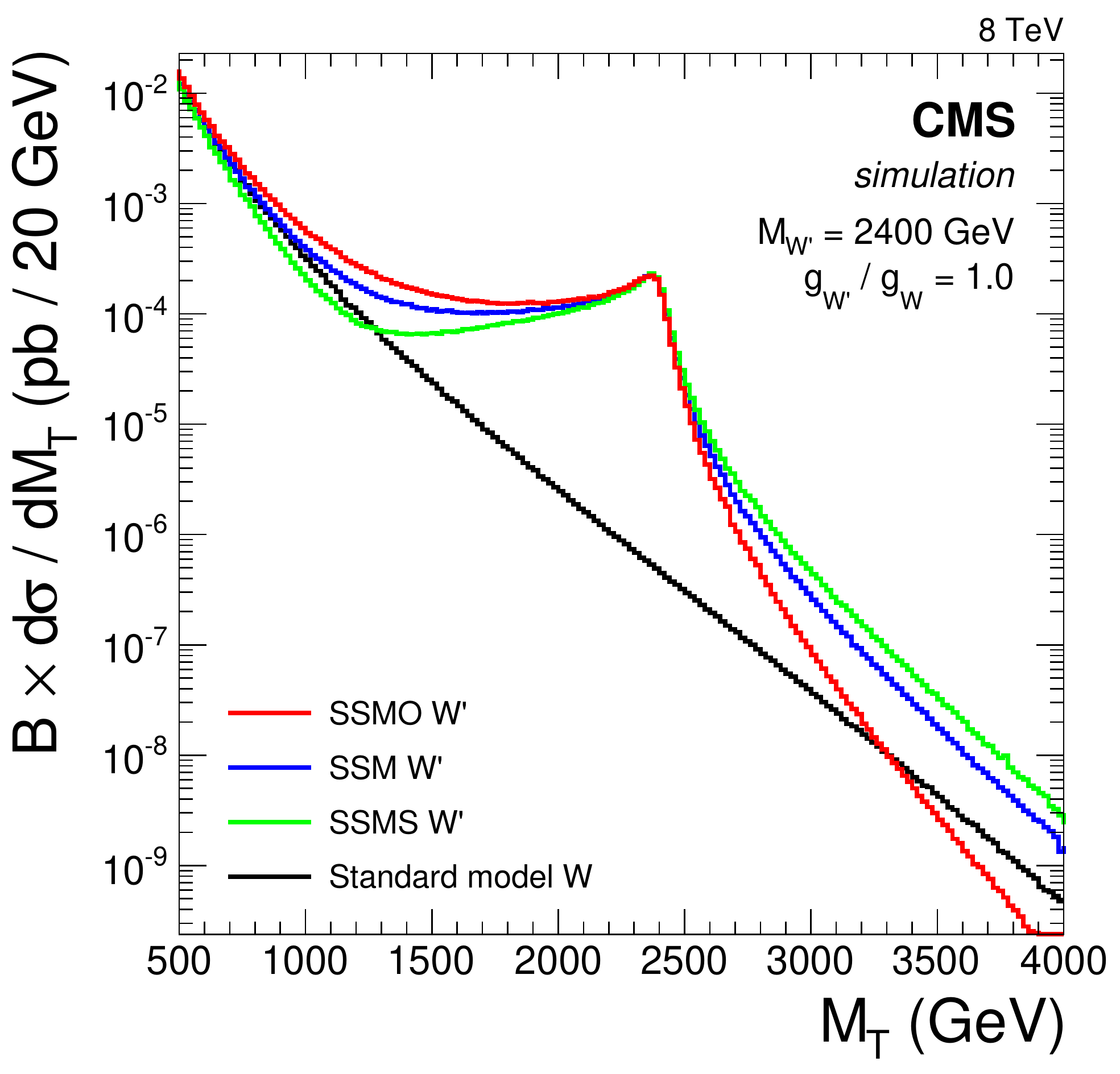}
\includegraphics[width=\cmsFigWidthTwo]{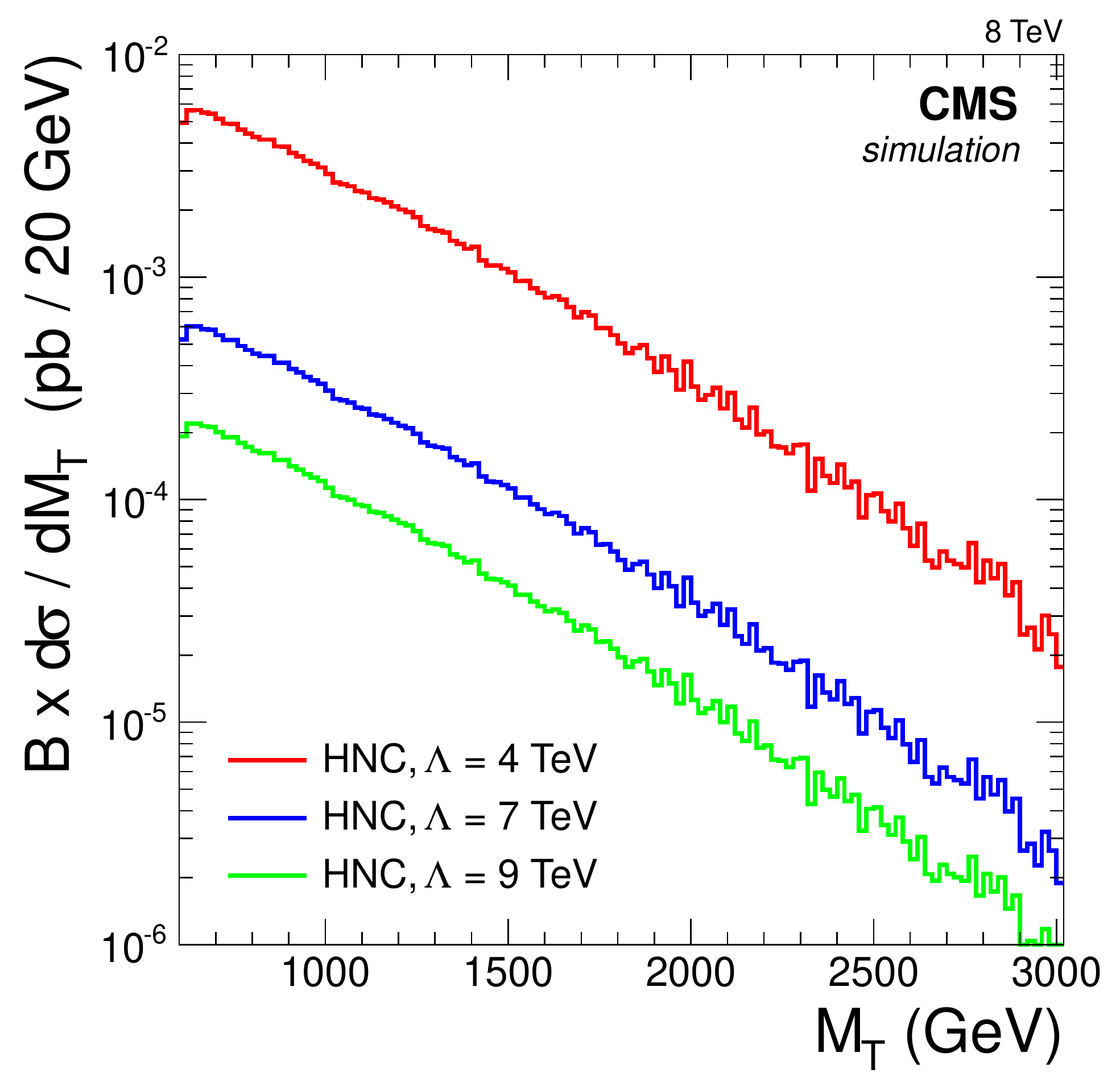}
\includegraphics[width=\cmsFigWidthTwo]{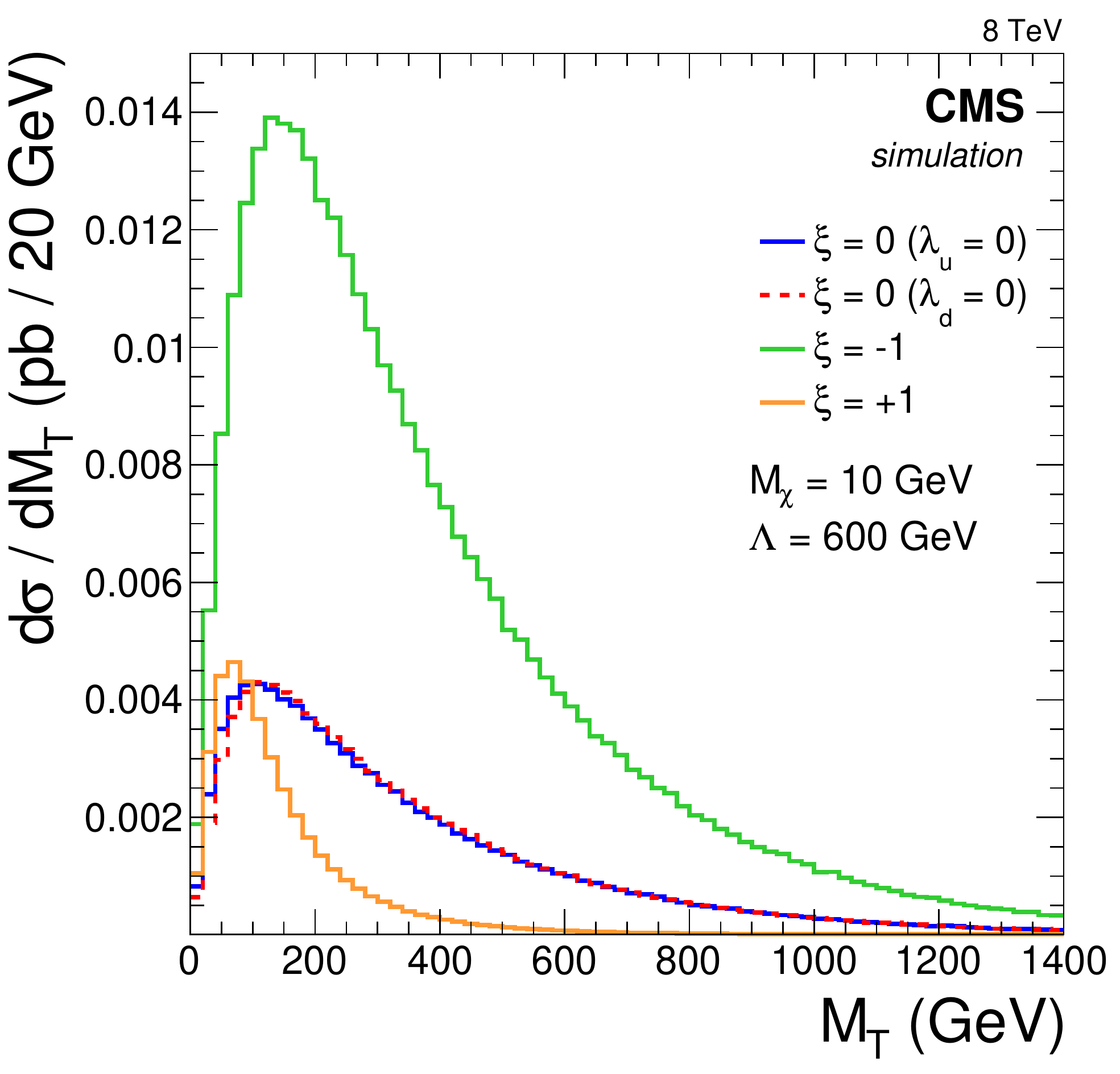}
\includegraphics[width=\cmsFigWidthTwo]{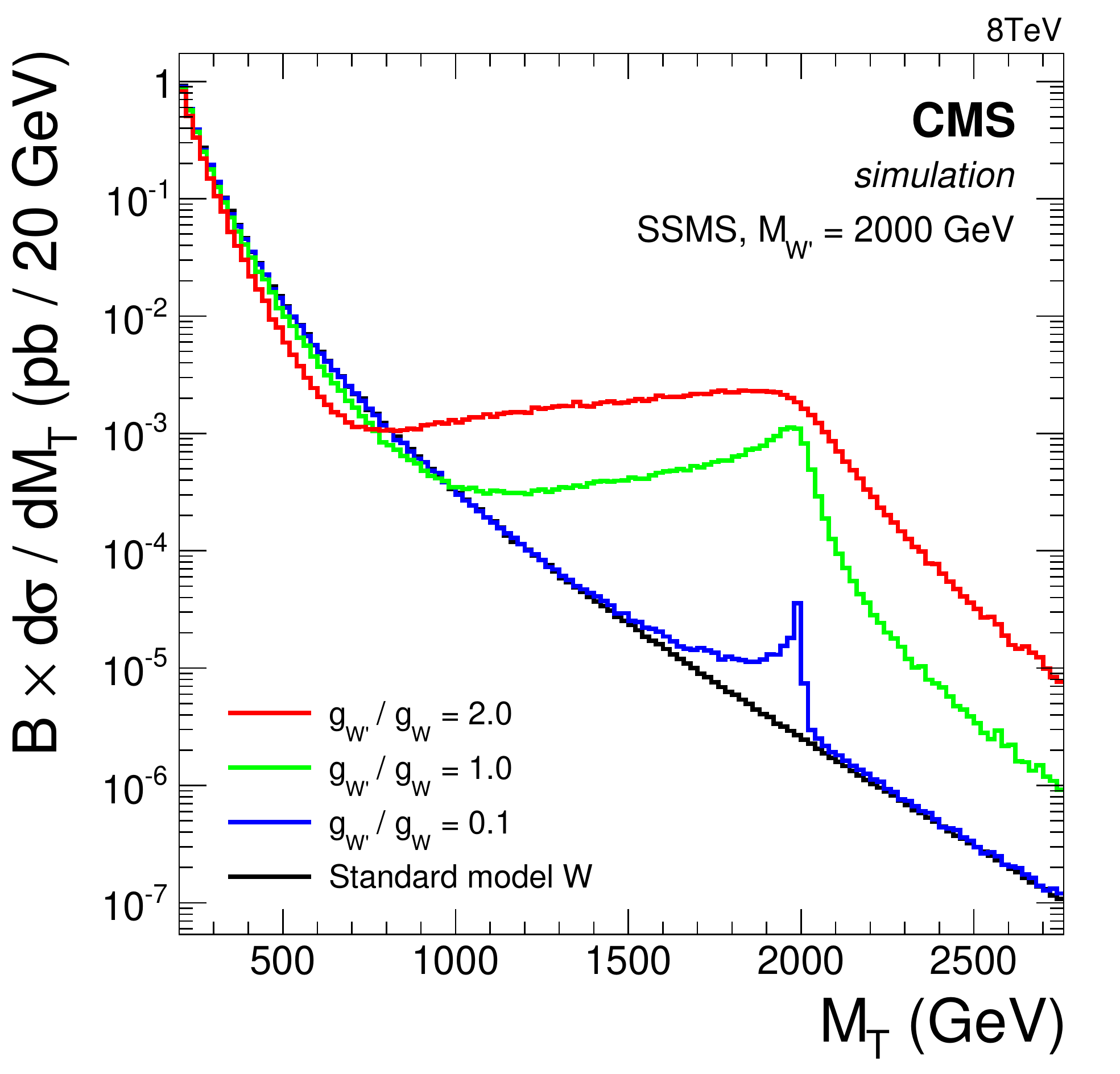}
\includegraphics[width=\cmsFigWidthTwo]{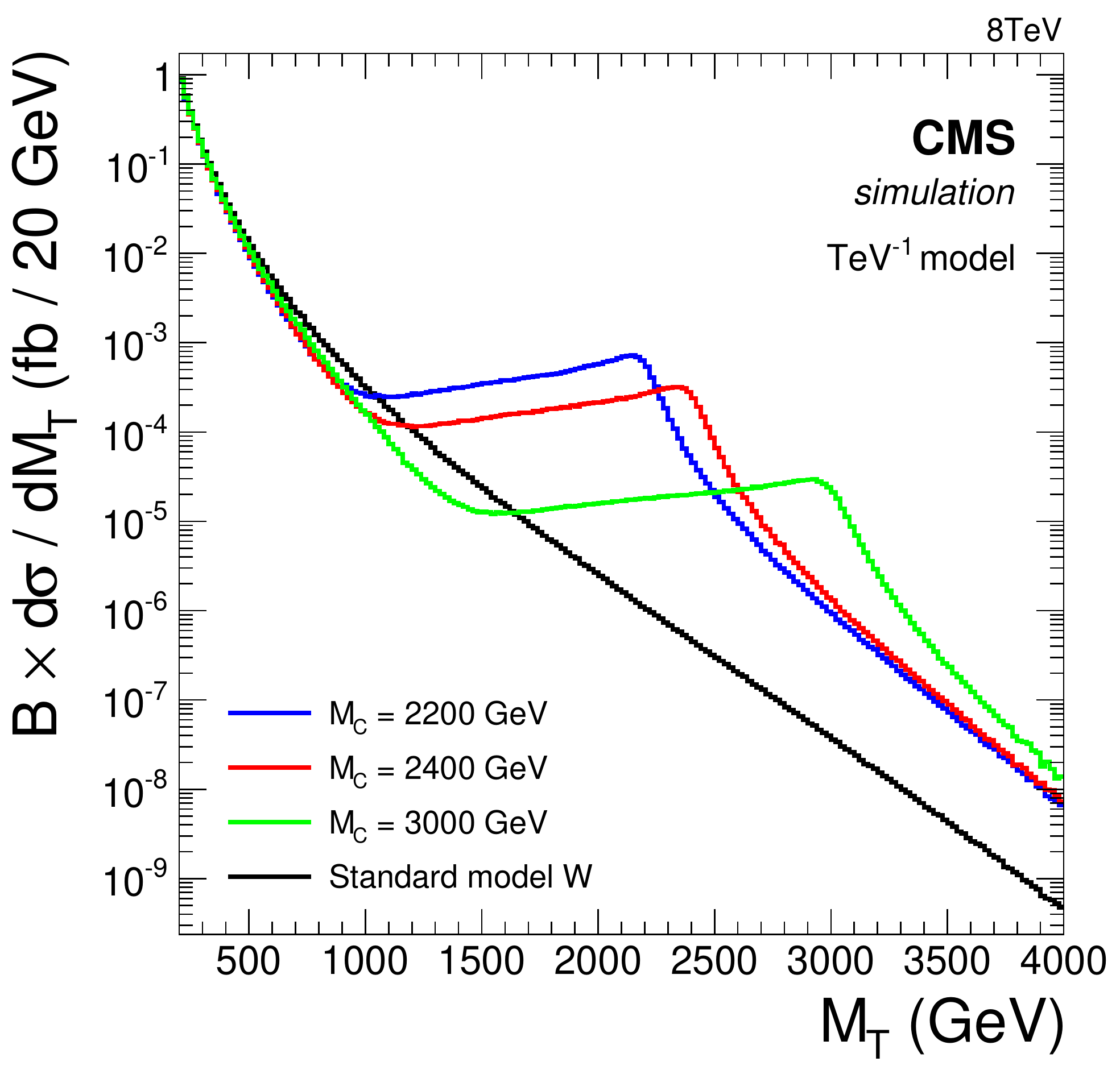}
\caption{Signal shapes at generator level: SSM model compared to the SSMS and SSMO models for $g_{\Wprime}/g_{\tPW}=1$ (top left); HNC-CI model for various values of $\Lambda$ (top right); DM for
various values of $\xi$ (middle left); \Wprime with \tPW boson interference and a varying coupling strength $g_{\PW}/g_{\Wprime}$ in the same sign scenario (middle right); and \TeVmOne model (bottom).}
\label{fig:signal}
\end{figure*}

Dark matter can couple to either up- or down-type quarks with the same initial and final state, as shown in Fig.~\ref{fig:DM-interference}.
Given that the couplings to up- or down-type quarks yield similar behavior,
the parameter $\xi=\lambda_u\lambda_d$
is most important for the phenomenology.
Following Ref.~\cite{monoLepton}, we consider three scenarios
with $|\lambda_i|$=1 or 0.
A value of $\xi = \pm1$ maximizes the effects of interference.
A choice of $\xi = 0$ can be assumed in two different ways, suppressing either the coupling to up- or to down-type quarks. Both cases are shown in Fig.~\ref{fig:signal} (middle left).
The difference between the two cases is small, therefore in the following we only consider the case with suppressed couplings to down-type quarks ($\lambda_d=0$) and denote it as $\xi = 0$.
The choice of the interference parameter changes the total cross section and the shape of the \MT spectrum, as shown in Fig.~\ref{fig:signal} (middle left) for $\Lambda = 600\GeV$, $\Mchi = 10 \GeV$.

In searches at proton-proton colliders, the difference between vector and axial-vector coupling is less important than in direct DM-nucleon interaction
experiments, as can be seen from Table~\ref{tab:xsec}. This is due to the large influence of the spin on the interaction at low $Q^2$ (of the order of 1 to 100\keV~\cite{pdg}), which is relevant for direct detection experiments, where coherent scattering at the nucleus is only possible for spin-independent (vector) couplings, but not for the spin-dependent (axial-vector) couplings.
At the LHC, half of the initial quarks originate from the quark-gluon sea, and all spin
configurations and light-quark flavors are available for production. For low \Mchi, no difference is observed between vector and axial-vector couplings, as shown in
Fig.~\ref{fig:signal} (middle left) for  $\Mchi = 10\GeV$. For masses above $100\GeV$, axial-vector cross sections  are lower than the vector cross sections, without a significant
shape difference.

The validity of this effective-theory model is limited.
For $\Lambda > M_\chi/2\pi$ the coupling is perturbative.
A more stringent criterion is $g_\mathrm{DM} = 1$, which constrains the model to $\Lambda > 2 M_\chi$.

Simulated signal samples are produced with \MADGRAPH 5.1.5~\cite{madgraph5} matched to \PYTHIA for showering and hadronization. The search is inclusive in terms of jet
multiplicity, and no constraints on the number of jets are applied. The samples are simulated for $\xi = +1$ and are rescaled on an event-by-event basis for $\xi = 0$ and
$-1$.

\subsection{\texorpdfstring{Interference of \tPW and \Wprime bosons with variable coupling strength}{Interference of W and W' bosons with variable coupling strength}}
\label{sec:models-interference}

If the \Wprime boson interacts with left-handed particles and right-handed antiparticles (V--A coupling), interference with the \tPW boson is
expected~\cite{criticalPaper,WprimeHelicity,dudkov}. The lowest-order effective Lagrangian for the interaction of two fermions and such a \Wprime boson is
\begin{equation}
\mathcal{L} = \frac{V^{ij}}{2\sqrt{2}}g^{ij}_{\Wprime} \bar f^i \gamma^\mu  (1-\gamma^5) \Wprime_\mu f^j +\text{h.c.},
\end{equation}
where h.c.\ is the hermitian conjugate, $V_{ij}$ is the CKM matrix element for quarks and unity for leptons, and $g_{\Wprime}$ is the coupling constant. The nature of the interference effects depends on
the ratio of the coupling $g_{\Wprime}$ to the SM weak coupling constant $g_{\tPW}=e/\sin{\theta_{\mathrm{W}}}$. Two different scenarios are considered: the same sign scenario (SSMS) with
$g^{\ell\nu}_{\Wprime}  g^{qq'}_{\Wprime}> 0$,
and the opposite sign scenario (SSMO) with $g^{\ell\nu}_{\Wprime}  g^{qq'}_{\Wprime} < 0$. The absolute value of the coupling $g_{\Wprime}$ is considered the same for quarks and leptons.

For the SSMS, the differential cross section for $\PW+\Wprime$ boson production can be smaller than the SM \tPW boson cross section, reflecting the effect of destructive interference. This effect is shown by
the green curve in Fig.~\ref{fig:signal} (top left) and is discussed in Ref.~\cite{criticalPaper}. In the SSMO, the cross sections exceed the SSM cross sections because of constructive interference, shown
by the red curve in the same figure.

The coupling strength $g_{\Wprime}$ affects both the total cross section and the width of the Jacobian peak.
It also influences the impact of the interference effects on the \MT spectrum. Assuming $g_{\Wprime}=g_{\tPW}$ the Jacobian peak partial widths of a \Wprime boson decaying to leptons can be seen in Fig.~\ref{fig:signal} (top left). For a larger coupling, the width increases, as shown in Fig.~\ref{fig:signal} (middle right).
This behavior is taken into account when deriving mass limits as a function of the coupling strength. A similar strategy was used in previous $\Wprime\to \cPqt\cPqb$ analysis~\cite{d0-limit,WprimeTBCMS}.

The contributions of the \tPW and \Wprime bosons to the overall cross section must be simulated simultaneously, as the final scattering amplitude is the sum of these two terms squared. An \MT requirement is used
at event generation to reduce the \tPW boson contribution resulting from the large \tPW boson production cross section, which is several orders of magnitude larger than that for the \Wprime boson production: $\sigma_\mathrm{LO}
(\tPW\to\ell\nu)$ falls from $9600\unit{pb}$ to $1.5\unit{fb}$ when the requirement $\MT > 500\GeV$ is imposed.

To provide an impression about the influence of the $\Wprime$ boson on the total cross section, effective cross sections are provided in Table~\ref{tab:xsec} for the SSMS and SSMO models.
They are derived from simulations of $\PW + \Wprime$ boson production with interference, with subsequent subtraction of the \tPW boson background.
For each mass point, an individual \MT cut-off is chosen and applied to both the \tPW boson sample and the $\PW + \Wprime$ sample before subtraction. This lowers the \tPW boson contribution and
therefore reduces the large uncertainty induced when subtracting two large and similar numbers. The cut-off is selected at the \MT value where the two distributions are compatible
within their
statistical uncertainties. For example, the cut-off for the \Wprime boson of 2\TeV in mass has been chosen to be 250\GeV. It is then varied up and down by 40\GeV to represent the interval in
which the two distributions approach each other and the procedure is repeated. The difference of the varied subtractions is provided as uncertainty in Table~\ref{tab:xsec}.
The resulting effective $\Wprime$ cross section can be compatible with $0$ or even less in case of destructive interference.
These effective cross sections are meant to provide additional information for the reader and not used in the analysis, which is based on the combined $\PW + \Wprime$ boson production
cross section.

A model of a \Wprime boson with an SM-like left-handed coupling has been implemented
within the \MADGRAPH 4.5.1 event generator~\cite{madgraph}. This model includes
spin correlations as well as finite-width effects. For each \Wprime-boson mass hypothesis, a sample with $g_{\Wprime} = g_{\PW}$ is processed by \PYTHIA with the Z2*~\cite{Field:2010bc}
tune, in order to simulate showering and hadronization. This sample is then reweighted using \MADGRAPH in order to simulate the \MT distribution for different values of the \Wprime boson
couplings. Both generators simulate at LO. The CTEQ6L1 PDF~\cite{CTEQ6L1} is employed.

To correct for higher-order effects in \tPW boson production, the difference between the next-to-leading order (NLO) and LO event yields are taken into account in the transverse mass distribution.
For the \Wprime boson contribution, the LO cross sections have been used.

\subsection{Split-UED model}
\label{sec:models-ued}

The leptonic final states under study may also be interpreted in the framework of universal extra dimensions with fermions propagating in the bulk, known as split-UED~\cite{PhysRevD.79.091702, JHEP04(2010)081}. This is a model based on an extended space-time with an additional compact fifth dimension of radius $R$. In this model all SM particles have corresponding Kaluza--Klein (KK) partners, for instance \WprimeKKn, where the superscript denotes the $n$th KK excitation mode. Only KK-even modes of \WprimeKKn couple to SM fermions, owing to
KK-parity conservation. Modes with $n \geq 4$ are not expected to be accessible under present LHC conditions, hence the only mode considered is $n = 2$. Interference with the SM is not considered in this model. Under this assumption, the decay to leptons is kinematically identical to the SSM \Wprime boson decay, and the observed limits obtained from the \WprimeENu and \WprimeMuNu searches can be reinterpreted directly in terms of the \WprimeKKn-boson mass, taking into account the difference in widths in the simulation. The LO and NNLO production cross sections for a \WprimeKKtwo boson are shown in Table~\ref{tab:xsec}.

The UED model is parameterized by the quantities $R$ and $\mu$, which are the radius of the extra dimension and the bulk mass parameter of the fermion field in five dimensions. In the split-UED model the parameter $\mu$ is assumed to be non-zero, following Refs.~\cite{PhysRevD.79.091702,JHEP04(2010)081}.

The mass of the \WprimeKKn is determined by $M_{\WprimeKKn}=\sqrt{\smash[b]{M_{\PW}^2+ ({n}/{R})^2}}$, \ie a larger radius corresponds to smaller KK masses. The mass of KK fermions depends on the
bulk mass parameter $\mu$. The cross section of the \WprimeKKn production times the branching fraction to standard model fermions
goes to zero as $\mu$ goes to zero.

\subsection{Model with a \texorpdfstring{\TeVmOne}{TeV-1} extra dimension }
\label{sec:models-tev-1}

Another extra dimensions model, the \TeVmOne model, has been proposed~\cite{TeVref1, TeVref2, TeVref3, TeVref4}, in which only the fermions are confined to ordinary three-dimensional space, with the SM gauge bosons and the Higgs field propagating in compactified extra dimensions. Under the assumption of a single extra dimension, the model is specified by one parameter $R = 1/M_\mathrm{C}$, the size of the compactified dimension, with $M_\mathrm{C}$ being the corresponding compactification scale. The \tPW bosons propagating in the compactified dimension are equivalent to the KK states \WprimeKKn with masses $M_{n} = \sqrt{\smash[b]{M_{0}^{2} + (n/R)^{2}} }$, where $M_{0}$ is the mass of the SM \tPW boson. The coupling constant of the KK states (for $n>0$) to fermions is $\sqrt{2}$ times larger than that of the SM \tPW boson.

The signal is similar in shape to an SSM \Wprime boson with destructive interference, as seen in Fig.~\ref{fig:signal} (bottom). This results in effective cross sections of $-0.966\pm0.025\fb$ for the $M_\mathrm{C} = 2.0\TeV$ case and $-0.079\pm0.014\fb$ for the $M_\mathrm{C} = 2.6\TeV$ case with respect to the \tPW boson.

Given the absence of higher-order calculations, the samples are generated at LO with \MADGRAPH 5.1.3.22 using the CTEQ6L1 parton distribution functions.
The cross sections with the \tPW boson cross section subtracted are summarized in Table~\ref{tab:xsec}.

\section{Object identification and event selection}
\label{sec:selection}

The models described in the previous section provide an event signature of a single high-\pt lepton (electron or muon) and one or more particles that cannot be detected directly (neutrino, dark matter particle), and so give rise to experimentally observed \MET.
This quantity is measured using a particle-flow technique~\cite{CMS-PAS-PFT-09-001,CMS-PAS-PFT-10-001,CMS-PAS-PFT-10-002}, an algorithm that combines measurements from all components of the CMS detector in order to
produce particle candidates. The modulus of the vector \pT sum of these candidates defines \MET, which is corrected for the jet energy
calibration~\cite{CMS-JME-10-011,Chatrchyan:2011tn}. At high \MT, the calculation of \MET is dominated by the high-\pT lepton in the event.

Candidate events with at least one high-\pT lepton are selected using single-muon (with $\pT > 40\GeV$)
and single-electron (with $\ET > 80\GeV$) triggers and loose electron identification criteria. The relatively high electron trigger threshold is required in order to suppress non-prompt electrons and jets. In the muon channel, the offline reconstructed \pT must be greater than 45\GeV, where the trigger is already fully efficient. This relatively low \pT requirement does not impair the search in the high-\MT region, while preserving an adequate number
of events in the low- and medium-\MT control regions. The requirement of $\ET > 100\GeV$ in the electron channel ensures a constant and high trigger efficiency. The trigger efficiency for single electrons has been determined with ``tag-and-probe'' methods~\cite{WZcrosssectionPaper} to be 99.1 (97.6)\% for the barrel (endcap) ECAL, with a
data-to-simulation scale factor of nearly one. The single-muon trigger efficiency varies from 94\% in the barrel to 82\% in the endcap regions, with data-to-simulation scale
factors of 0.98--0.96~\cite{CMS-DP-2013-009}.

Electrons are reconstructed as ECAL clusters that are matched to a tracker track and their identification has been optimized for high-\pT~\cite{EXO-12-061}.
They have to be sufficiently isolated, have an electron-like shape and be within the acceptance region of the barrel ($\abs{\eta} < 1.442$) or the endcaps ($1.56 < \abs{\eta} < 2.5$).
This acceptance region avoids the gap between barrel and endcap, where the misidentification probability is the highest.
Electron isolation in the tracker is ensured by requiring the \pT sum of all tracks that are in close proximity to the track of the electron candidate and originate from the same primary vertex, to be less than 5\GeV.
Here, only tracks that are within a cone of $\Delta R = 0.3$ around the electron candidate's track are considered.
The primary vertex is defined as the vertex with the highest $\sum \pT^2$ in the event, where the sum extends over the charged tracks associated with the vertex.
In the calorimeters, the \ET sum of energy deposits around the electron candidate is used as a measure of isolation.
It is corrected for the mean energy contribution from additional proton proton collisions during the same bunch crossing (pileup).
As in the tracker isolation calculation, contributions within a $\Delta R < 0.3$ cone around the electron candidate are considered.
To obtain sufficiently isolated electrons, this calorimeter isolation is required to be below a threshold of around 3\% of the electron's \ET.
Additionally, the energy deposits in the hadron calorimeter within a cone of $\Delta R = 0.15$ around the electron's direction must be less than 5\% of the electron's energy deposit in the ECAL.
In order to differentiate between electrons and photons, properties of the track matched to the calorimeter measurement must be consistent with those of a prompt electron.
Specifically, there must be $\leq 1$ hit missing in the innermost tracker layers, and the transverse distance to the primary vertex must be $<$0.02\unit{cm} (barrel) or $<$0.05\unit{cm} (endcap).
To reduce the Drell--Yan background, events with additional electrons of $\ET > 35\GeV$ are rejected.

The reconstruction of muons is optimized for high \pT. Information from the inner tracker and the outer muon system are used together. Each muon is required to have at least one hit in the pixel
detector, at least six tracker layer hits, and segments in two or more muon detector layers. Since segments are typically found in consecutive layers separated by thick layers of steel, the latter requirement significantly reduces the amount of hadronic punch-through~\cite{Chatrchyan:2012xi}.
To reduce background from cosmic ray muons, each muon is required to have a transverse impact parameter $\abs{d_0}$ of less than 0.02\unit{cm} and to have a longitudinal distance parameter $\abs{d_z}$ of the tracker track of less than 0.5\unit{cm}. Both parameters are defined with respect to the primary vertex. In order to suppress muons with mismeasured \pT, an additional requirement $\sigma_{\pT}/\pT < 0.3$ is applied, where
$\sigma_{\pT}$ is the uncertainty from the track reconstruction.
To match the trigger acceptance the muon must have $\abs{\eta} < 2.1$. Muon isolation requires that the scalar \pT sum of all tracks originating from the interaction
vertex within a $\Delta R = \sqrt{\smash[b]{(\Delta \phi)^2 + (\Delta \eta)^2}}< 0.3$ cone around its direction, excluding the muon itself, is less than 10\% of the muon's \pT. To further reduce
the Drell--Yan and cosmic ray backgrounds, the event must not have a second muon with $\pT > 25\GeV$.

The reconstruction efficiencies for both electrons and muons are measured using same-flavor dilepton events, up to the highest accessible \pT. Data and simulation agree within
statistical uncertainties for these high-energy events. For higher \pT, the flat efficiency is extrapolated and assigned an associated systematic uncertainty, as described in
Section~\ref{sec:uncertainties}.

In order to identify any differences in selection efficiency for observed and simulated data, efficiencies for both are determined using the ``tag-and-probe'' method. The total efficiency
in each case includes contributions from the trigger, lepton identification, and isolation criteria. The ratio of data to simulation efficiencies, denoted as the scaling factor
(SF), is determined to be $0.975\pm0.023$ ($0.970\pm0.042$) for barrel (endcap) in the electron channel. For the muon channel, the SFs are $0.967 \pm 0.026$ for $\abs{\eta} < 0.9$
(barrel), $0.948 \pm 0.026$ for $0.9 < \abs{\eta} < 1.2$ (barrel-endcap interface), and $0.979 \pm 0.026$ for $1.2 < \abs{\eta} < 2.1$ (forward region)~\cite{CMS-DP-2013-009}.

\begin{figure}[hbtp]
\centering
\includegraphics[width=\cmsFigWidth]{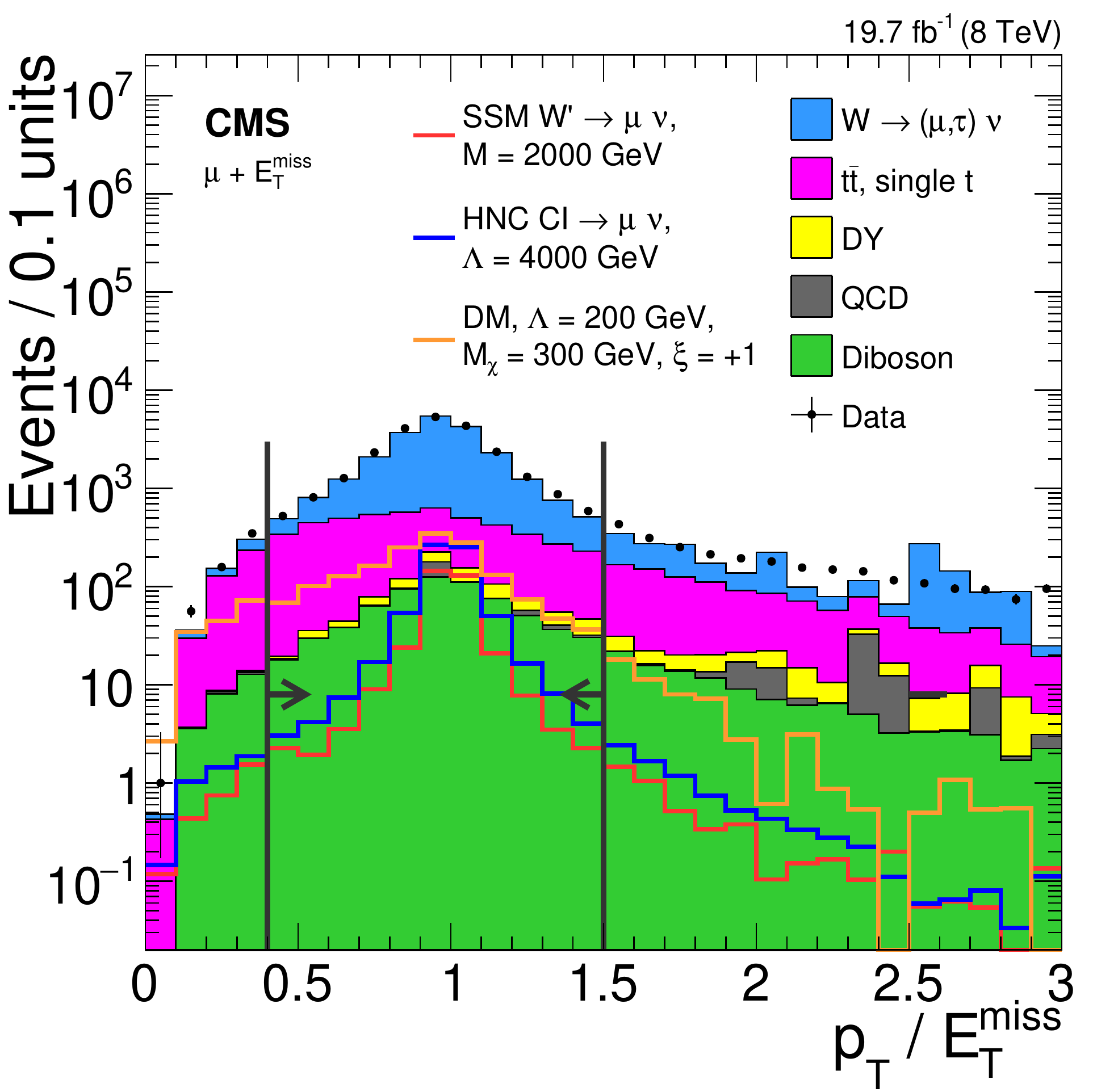}
\includegraphics[width=\cmsFigWidth]{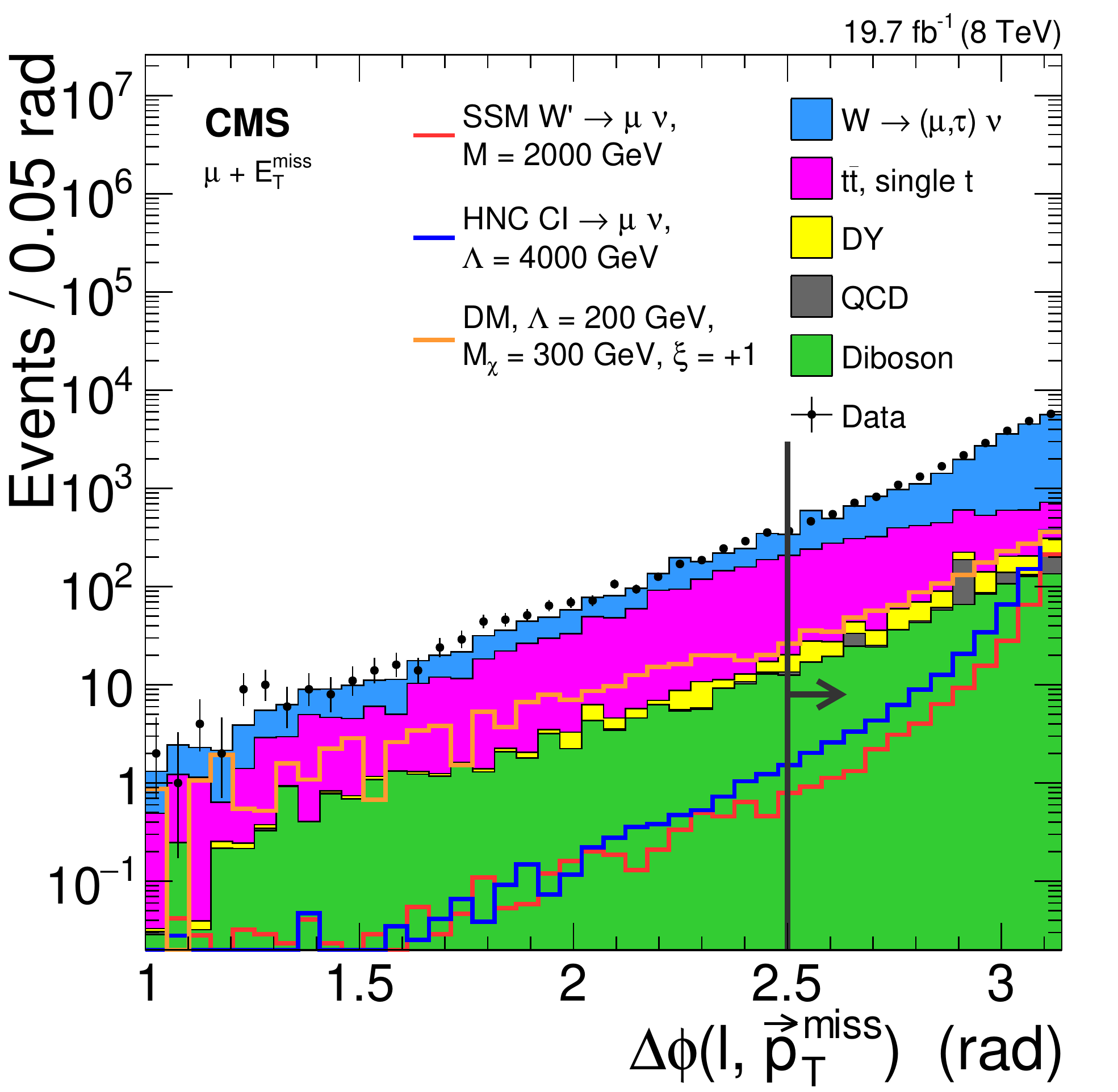}
\caption{The distribution in \pt/\MET (\cmsLeft) and $\Delta \phi(\ell,\MPT)$ (\cmsRight), for data, background, and some signals in the muon channel with an \MT threshold of 220\GeV. The simulated background labeled as `diboson' includes $\PW\PW$, $\PZ\PZ$ and $\PW\tPZ$ contributions, while `DY' denotes the Drell--Yan process.
}
\label{fig:kinematics}
\end{figure}

In the models considered, the lepton and \MPT are expected to be nearly back-to-back in the transverse plane, and balanced in transverse energy. This is illustrated in
Fig.~\ref{fig:kinematics} for the muon channel for three example signals (SSM, HNC-CI, and DM) with an \MT threshold of 220\GeV.
To incorporate these characteristics in the analysis, additional kinematic criteria select events based on the ratio of the lepton \pT to \ETmiss, requiring $0.4 < \pT/{\ETmiss} < 1.5$, and on the angular
difference between the lepton  and \MPT, with $\Delta \phi(\ell,\MPT) > 2.5 \approx 0.8\pi$.

Signal efficiencies are model-dependent, determined by the signal shape in the distribution of \MT. For simulated events passing all the selection criteria, the average signal efficiencies for a given parameter
are summarized in Table~\ref{tab:signal-eff}.

\begin{table}[htb]
\topcaption{Signal efficiencies for the various models.}
\label{tab:signal-eff}
\centering
\begin{scotch}{llcc}
Model       & Parameter  & Electron channel   & Muon channel  \\ \hline
SSM \Wprime & M$_{\Wprime}$ = 0.5\TeV     &  62$\pm$3\%   & 64$\pm$3\%             \\
SSM \Wprime & M$_{\Wprime}$ = 1.5\TeV     &  74$\pm$6\%   & 71$\pm$7\%             \\
SSM \Wprime & M$_{\Wprime}$ = 4\TeV       &  50$\pm$5\%   & 55$\pm$4\%             \\
HNC-CI      & $\Lambda$ independent     &  80$\pm$6\%   & 80$\pm$6\%        \\
DM & $\xi$ = -1 & 42$\pm$4\%       & 42$\pm$4\%        \\
DM & $\xi$ =  0 & 39$\pm$4\%       & 39$\pm$4\%        \\
DM & $\xi$ = +1 & 12$\pm$2\%       & 13$\pm$2\%        \\
\end{scotch}
\end{table}

The SSM \Wprime has maximal signal efficiency at the mass of 1.5\TeV, decreasing gradually for larger and smaller masses. The uncertainties quoted in Table~\ref{tab:signal-eff} are explained in Section~\ref{sec:uncertainties}.
The geometrical acceptance is roughly 90\% for both electron and muon channels. For the higher \Wprime-boson masses up to 4\TeV the signal efficiencies slowly decrease to 50\%, because of an increasing fraction of off-shell \Wprime bosons.

For the HNC-CI model, the signal efficiency is independent of the interaction scale $\Lambda$ and has been determined from simulation to be 80\% with 6\% uncertainty for both the $e + \MET$ signal and the $\mu + \MET$ signals.

For the DM models the signal efficiency depends on the steepness of the \MT distribution and the total cross section, both of which are sensitive to the interference parameter $\xi$.
For $\xi = +1$, the spectrum falls more rapidly, and the search region corresponds to the low-to-medium part of the \MT spectrum, resulting in a rather low signal efficiency of $(13\pm2)\%$. For the other two interference cases, $\xi = 0$ and $-1$, the spectrum extends to very high \MT, where the expected background is negligible and the electron and muon channel signal efficiencies are as high as $(39\pm4)\%$ for $\xi = 0$ and $(42\pm4)\%$ for $\xi = -1$.
As expected, no difference in efficiency is observed between the vector and the axial-vector couplings.

\section{Distribution in \texorpdfstring{\MT}{invariant mass}}
\label{sec:results}

\begin{figure}[hbtp]
\centering
\includegraphics[width=\cmsFigWidth]{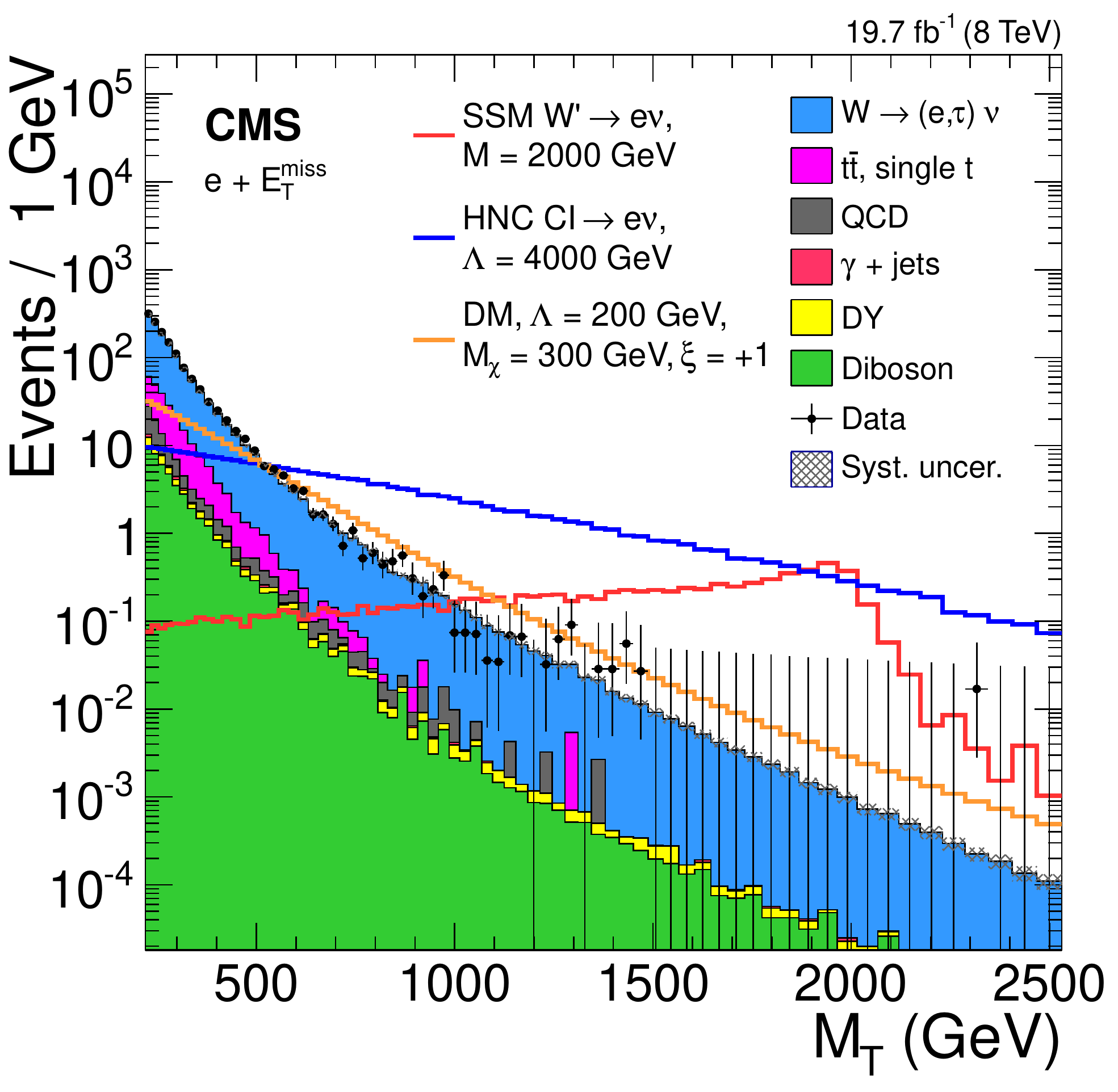}
\includegraphics[width=\cmsFigWidth]{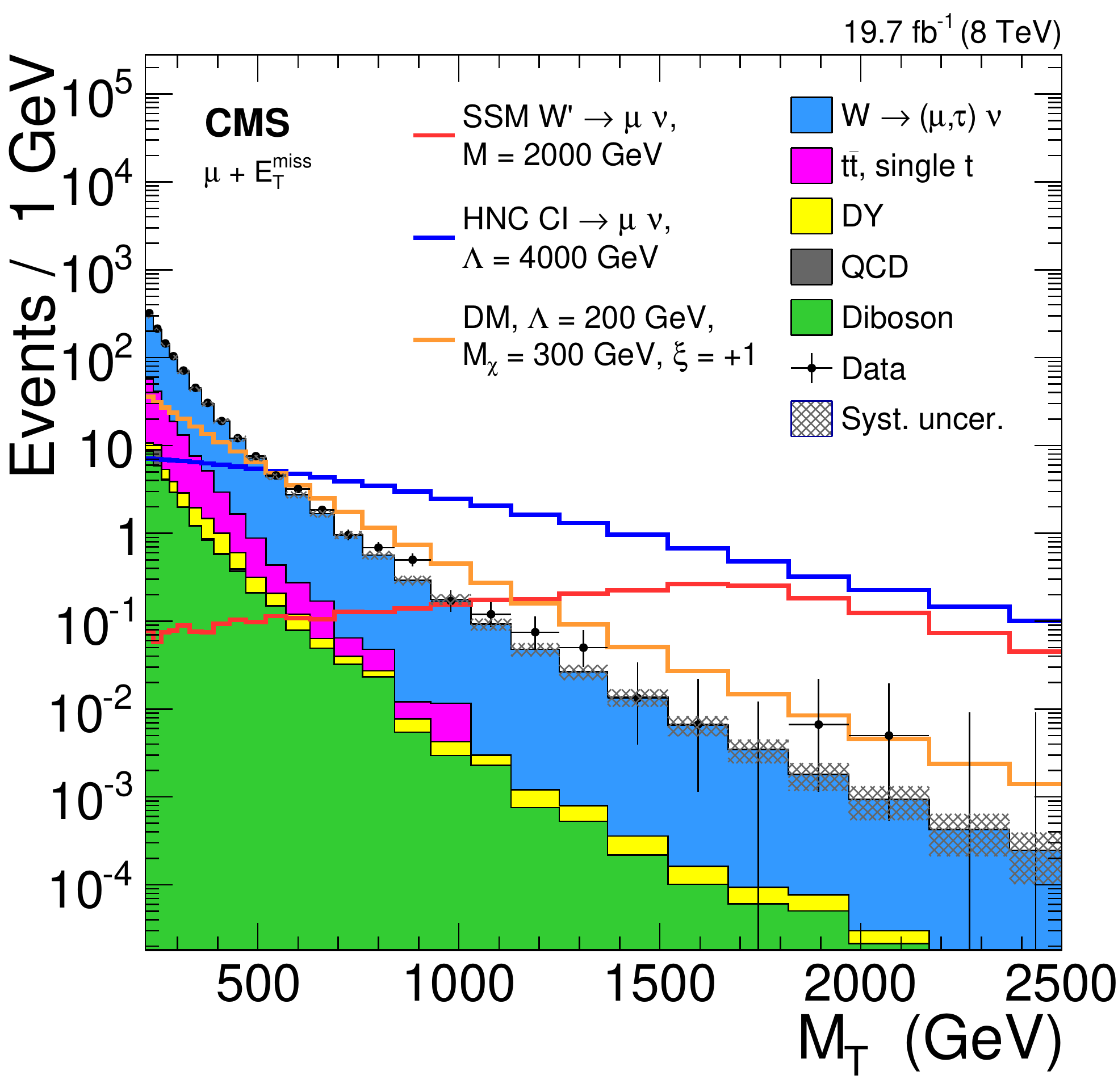}
\caption{Observed \MT distributions for the electron (\cmsLeft) and muon (\cmsRight) channels. The horizontal bars on the data points indicate
the widths of the bins. The asymmetric error bars indicate the central confidence intervals for Poisson-distributed data and are obtained from the Neyman construction as described in Ref.~\cite{Garwood1936}.}
\label{fig:MT}
\end{figure}

The observed \MT distributions for the analyzed data sets are shown in Fig.~\ref{fig:MT} for the electron and muon channels. Included in the same figure are the predicted \MT distributions
for the accepted SM events, separated into contributions from each background process, along with example signal distributions for DM, SSM \Wprime, and HNC-CI models. For both channels, a variable binning commensurate with the energy-dependent \MT-resolution is used. The expected systematic uncertainty in the \MT distribution is also shown.

\begin{figure}[hbtp]
\centering
\includegraphics[width=\cmsFigWidth]{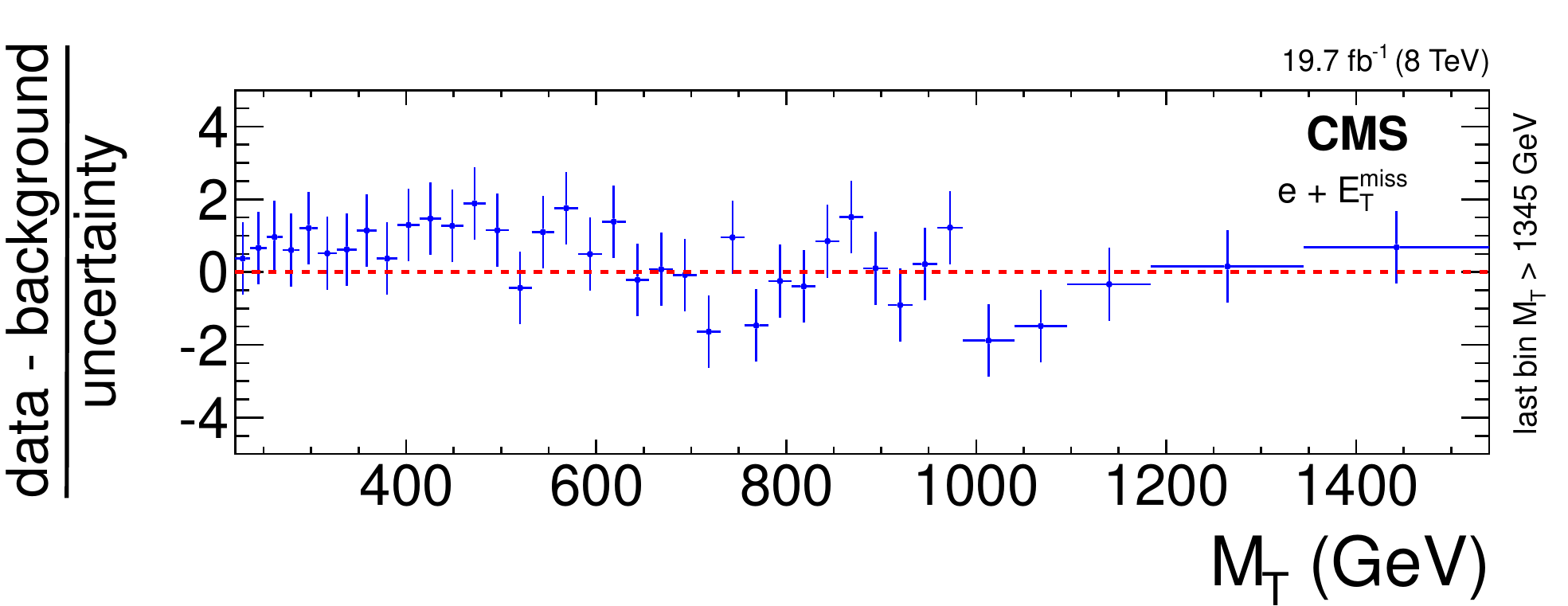}
\includegraphics[width=\cmsFigWidth]{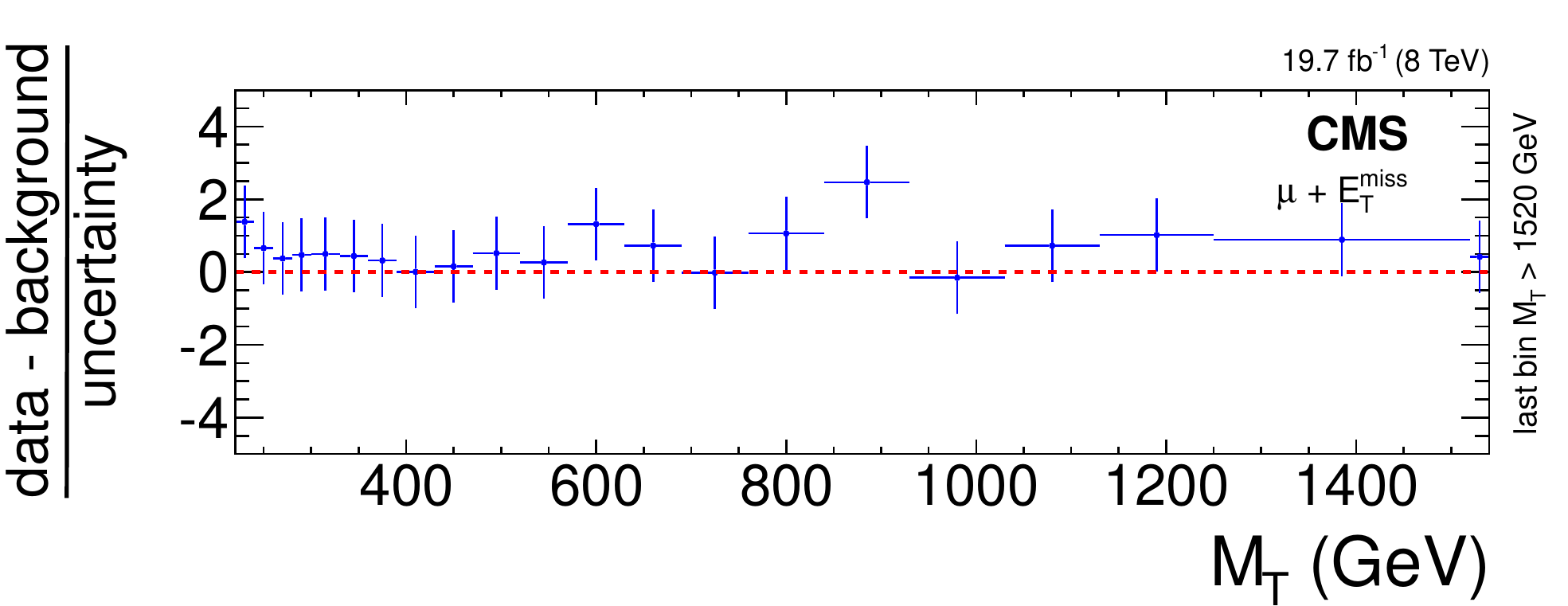}
\caption{Deviation of the number of observed events from the expectation,
as shown in Fig.~\ref{fig:MT},
in units of standard deviations, binned in \MT for the electron (\cmsLeft) and muon (\cmsRight) channels.
Both the systematic and statistical uncertainties are included.
The horizontal bars on the data points indicate the widths of the bins, which are chosen as in Fig~\ref{fig:MT} with the extension to
include always 5 background events in a bin.}
\label{fig:ratioplots}
\end{figure}

The deviation of the data from the standard model prediction is shown in Fig.~\ref{fig:ratioplots}. No significant deviation from the predicted background is observed in the \MT spectrum. The highest transverse mass events observed have $\MT = 2.3\TeV$ in the electron channel and $\MT = 2.1\TeV$ in the muon channel.
Both events have a well-reconstructed high-\pt lepton and very
little hadronic activity.

\section{Background}
\label{sec:background}

\subsection{Sources of background}
\label{sec:background_sources}

The primary source of background for all signals is the presence of off-peak, high-transverse-mass tails of the SM $\PW \to \ell\nu$ decays. Other important backgrounds arise from QCD multijet, \ttbar,
and Drell--Yan events. Contributions from dibosons ($\PW\PW$, $\PW\PZ$, $\PZ\PZ$) decaying to $\Pe$, $\mu$, or $\tau$-leptons are also considered. The following background sources  are considered in this
analysis. They are listed in the order of Fig.~\ref{fig:MT}, where their distribution in \MT can be seen.

\begin{figure}[hbtp]
\centering
  \includegraphics[width=\cmsFigWidth]{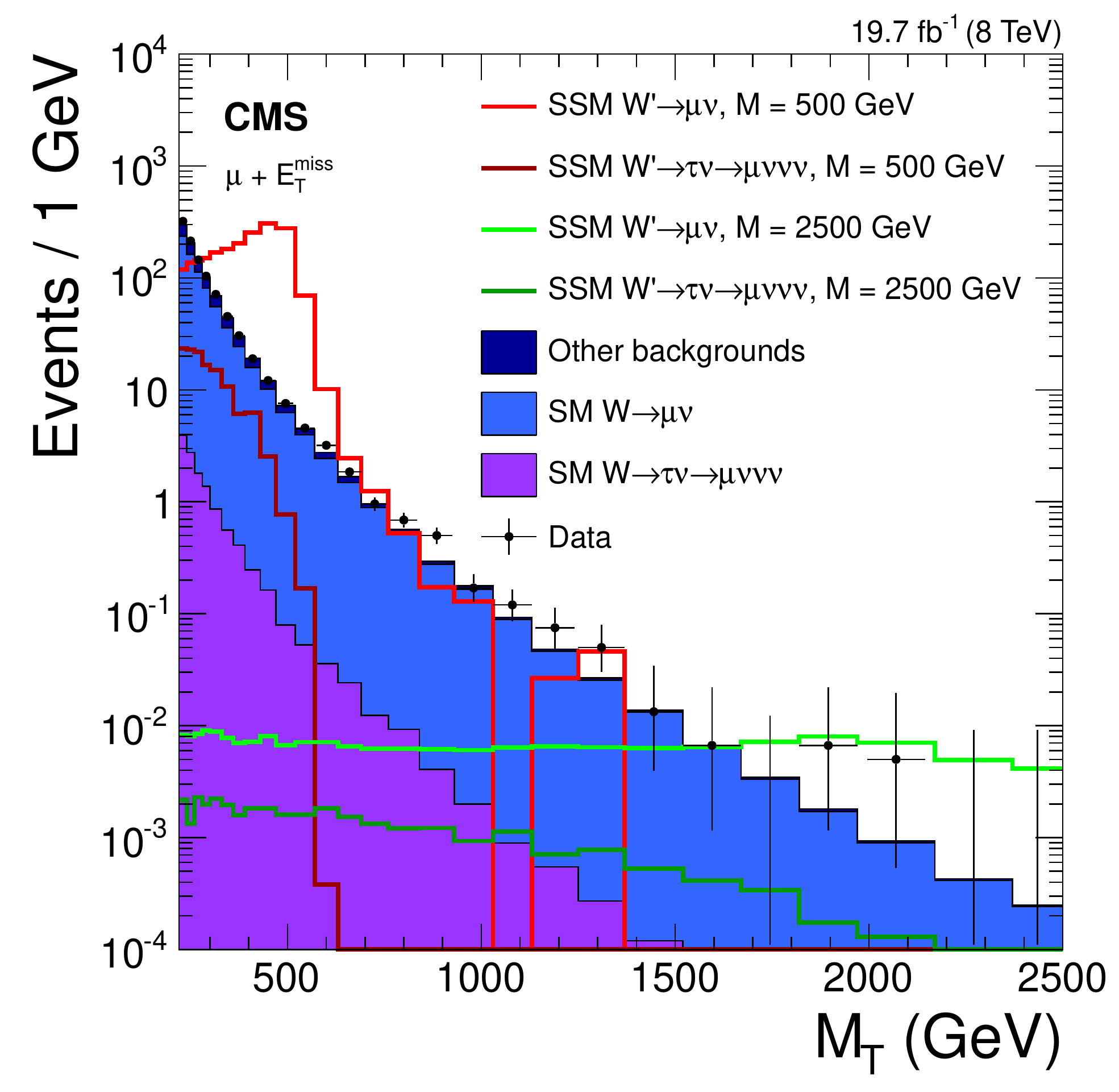}
  \caption{Contributions from \tPW and \Wprime bosons decaying to $\tau\nu\to\mu\nu\nu\nu$ compared to the prompt $\mu\nu$ decay channel. Also shown are the contributions from  the other backgrounds and the number of data events. The horizontal bars on the data points indicate
the widths of the bins.}
  \label{fig:tau-contribution}
\end{figure}

\begin{enumerate}

\item A $\PW\to \ell \nu$ with $\ell = \Pe, \mu$ sample is simulated at LO with \PYTHIA. To ensure a good description of the considered phase space, two samples with lepton \pt ranges 100--500\GeV
and $\geq$500\GeV, respectively, are used. A transverse-mass-dependent K-factor is calculated, including NLO QCD and electroweak corrections (see Section~\ref{sec:W-kfactor}).

\item A $\PW\to \tau \nu$ sample, where the $\tau$-lepton decays to an electron or a muon and the two corresponding neutrinos, is generated with \PYTHIA using the same K-factors as above. Electrons and
muons from these decays have a small impact parameter with respect to the primary vertex and are not separable from prompt leptons. In addition, the \pt/\MET ratio is approximately one, despite the
presence of three neutrinos in the event, since the vector sum of the neutrinos' \pt balances the charged lepton \pt for small \tPW boson \pt. These features prevent an efficient rejection of $\PW\to
\tau \nu$ events, which therefore contribute to the background, albeit at low \MT (see Fig.~\ref{fig:tau-contribution}) and with little contribution in the high-\MT region.

\item Top-quark pair and single top-quark production are other sources of high-\pt leptons and \MET, and these are generated with \MCATNLO~\cite{mcatnlo,mcatnloheavyflavour} in combination with \HERWIG~\cite{herwig}, and \POWHEG~\cite{powheg_1,powheg_2,powheg_3,Re:2010bp} in combination with \PYTHIA, respectively. A newly calculated NNLO cross section~\cite{Czakon:2013goa} is used to rescale the NLO predictions. These events are largely rejected by the requirement of two-body decay kinematics (see Section~\ref{sec:selection}) but can extend into high \MT as seen in Fig.~\ref{fig:MT}.

\item Multijet background (QCD), enriched in electrons/photons and muons, is generated with \PYTHIA. Although this process has by far the largest cross section, it is efficiently rejected by the isolation requirements imposed to select the lepton candidates as well as the requirement on the ratio of $\pt/\MET$ (see Section~\ref{sec:selection}). Despite the large suppression of these events, the misidentification of jets as leptons (especially as electrons) still occurs. The contribution of QCD multijet events to the electron channel is derived from data as explained in Section~\ref{sec:qcdfromdata}.

\item Drell--Yan production of dileptons ($\ell = \Pe, \mu$) constitutes a background when one lepton escapes detection. The samples are generated with \POWHEG~\cite{powheg_DY}. Contributions from Drell--Yan production of $\tau\bar\tau$ are simulated using \PYTHIA, applying a uniform QCD K-factor of 1.26 (calculated with \FEWZ~\cite{fewz,Gavin:2012kw}).

\item Contributions from dibosons ($\PW\PW$, $\PW\PZ$, $\PZ\PZ$) decaying to a state with at least one lepton, are generated with \PYTHIA and scaled to NLO cross sections.

\item In the electron channel, a $\gamma$+jet event sample, generated with \PYTHIA, is used to estimate the effects of photons misidentified as electrons.

\end{enumerate}

Background selection efficiencies for $\MT>220\GeV$ are summarized in Table~\ref{tab:bkgr}. The total background predictions, listed in Table~\ref{tab:exampleEventYields}, are comparable in the electron
and muon channels. Resolution effects, reconstruction efficiencies, and statistical fluctuations are taken into account. Systematic uncertainties are discussed in Section~\ref{sec:uncertainties}.

\begin{table*}[htb]
\topcaption{Background processes and their number of events $N_\text{events}$ and selection efficiencies after the full selection and the requirement $\MT > 220 \GeV$.}
\label{tab:bkgr}
\centering
\begin{scotch}{lcccc}
Process  &  \multicolumn{2}{c}{$\Pe$ channel }  & \multicolumn{2}{c}{$\mu$ channel }  \\
       & $N_\text{events}$    &  selection efficiency  & $N_\text{events}$    &  selection efficiency  \\
\hline
$\PW \to \ell \nu$ with $\ell = \Pe, \mu$           &  $18390 \pm 170$ & $(745.8 \pm 7.0)\times10^{-7}$   & $17261 \pm 151$  & $(697.5 \pm 6.1)\times10^{-7}$ \\
$\PW \to \tau \nu$, $\tau \to (\Pe, \mu) \nu \nu$     &  $491 \pm 74$    & $(19.8 \pm 3.0)\times10^{-7}$   & $281.3 \pm 3.4$  & $(113.7 \pm 1.7)\times10^{-8}$ \\
Top-quark pair and                       & \multirow{2}{*}{$2831 \pm 28 $} & \multirow{2}{*}{$(398.4 \pm 3.9)\times10^{-6}$}     & \multirow{2}{*}{$3132 \pm 30$}        & \multirow{2}{*}{$(486.8 \pm 4.6)\times10^{-6}$}           \\
single top-quark & & & & \\
Diboson                                                & $783.8 \pm 8.3$ & $(376.5 \pm 4.0)\times10^{-6}$     & $629.6 \pm 7.7 $         & $(325.8 \pm 4.0)\times10^{-6}$           \\
Multijet (QCD)                                          & $705 \pm 19$  & data driven & $16.5 \pm 7.3$   & $(11.1 \pm 4.9)\times10^{-12}$ \\
DY $\to \ell \ell$ with $\ell = \Pe, \mu, \tau$       & $159 \pm 19$  & $(20.8 \pm 2.5)\times10^{-7}$  & $216.1 \pm 5.9$   & $(280.8 \pm 7.6)\times10^{-8}$ \\
$\gamma$+jet                                            & $56.7 \pm 8.4$& $(31.7 \pm 4.7)\times10^{-7}$    & \NA & \NA                 \\
\end{scotch}
\end{table*}

All simulated event samples are normalized to the integrated luminosity of the recorded data, using calculated NNLO cross sections. The only exceptions are the diboson and QCD samples, for which the NLO and LO cross sections are used, respectively. The simulation of pileup is included in all event samples by superimposing minimum bias interactions onto all simulated events. For the data set used, the average number of interactions per bunch-crossing is 21.

\subsection{Prediction of the expected background}

Searching for deviations from the steeply falling \tPW-boson \MT spectrum requires an accurate background estimate at very high transverse masses. Several methods are used to evaluate the expected background in the signal region, based either on control samples in data or on a fully simulation-based approach.

For the majority of background sources, the estimate is determined from simulation, based on samples with large event counts at high \MT. However, to avoid bin-to-bin statistical fluctuations that may still occur at very high \MT, the total background prediction is parameterized with the empirical function given in Eq. (\ref{mc_parametrisation}).
\begin{equation}
\label{mc_parametrisation}
f(\MT)=\re^{a+b \MT+c \MT^2} \MT^d,
\end{equation}
where $a$, $b$, $c$, and $d$ are the fit parameters. This function was found to provide a good description of the steeply falling SM background up to high \MT. Based on this parameterization, the
expected number of SM background events for all transverse mass bins can be predicted, as shown for three typical thresholds \MTlower in Table~\ref{tab:exampleEventYields}. Contributions from SM
processes drop quickly with increasing \MT, leading to an expected yield of less than 0.5 events in each channel,
for $\MT > 2\TeV$.

\begin{table*}[hbtp]
\renewcommand{\arraystretch}{1.3}
\topcaption{Event yields observed in data, and expected from background and signal, for different transverse mass thresholds. The quoted uncertainties are the combined uncertainties assuming a log-normal distribution, not including a 2.6\% integrated luminosity uncertainty~\cite{LUM-13-001}.}
\label{tab:exampleEventYields}
\centering
\begin{scotch}{llccc}
                                                                     &                          & $\MT > 1.0\TeV$           & $\MT > 1.5\TeV$              & $\MT > 2.0\TeV$\\ \hline
\multicolumn{5}{c}{Electron channel}\\\hline
Data                                                                   &                          & $24$                      & $1$                          & $1$ \\
SM Background                                                          &                          & $26.0^{+2.5}_{-2.5}$      & $2.02^{+0.26}_{-0.25}$       & $0.207^{+0.036}_{-0.033}$ \\
\multirow{2}{*}{$\Wprime$}	                                           & $M_{\Wprime} = 2.5 \TeV$ & $50.5^{+7.5}_{-7.5}$      & $38.8^{+6.1}_{-6.1}$         & $24.0^{+3.9}_{-3.9}$ \\
                                                                     & $M_{\Wprime} = 3 \TeV$   & $10.3^{+2.1}_{-2.1}$      & $7.8^{+1.9}_{-1.9}$          & $5.8^{+1.5}_{-1.5}$ \\
\multirow{2}{*}{HNC-CI}                                                & $\Lambda = 4 \TeV$       & $1120^{+110}_{-110}$      & $368^{+47}_{-47}$            & $105^{+19}_{-19}$ \\
                                                                     & $\Lambda = 9 \TeV$       & $43.4^{+4.3}_{-4.3}$      & $14.3^{+1.8}_{-1.8}$         & $4.08^{+0.75}_{-0.75}$ \\
\multirow{3}{4cm}{DM vector-coupling, $M_\chi = 50\GeV$, $\Lambda=300$}& $\xi = +1$               & $0.402^{+0.050}_{-0.050}$ & $0.0346^{+0.0072}_{-0.0070}$ & $0.0033^{+0.0010}_{-0.0010}$ \\
                                                                     & $\xi = 0$                & $6.8^{+1.5}_{-1.5}$       & $1.25^{+0.42}_{-0.42}$       & $0.22^{+0.11}_{-0.11}$ \\
                                                                     & $\xi = -1$               & $27.4^{+5.9}_{-5.9}$      & $5.0^{+1.7}_{-1.7}$          & $0.89^{+0.44}_{-0.43}$ \\
\hline
\multicolumn{5}{c}{Muon channel}\\ \hline
Data                                                                   &                          & $35$                      & $3$                          & $1$ \\
SM Background                                                          &                          & $26.1^{+4.4}_{-4.3}$      & $2.35^{+0.70}_{-0.60}$       & $0.33^{+0.16}_{-0.12}$ \\
\multirow{2}{*}{$\Wprime$}                                             & $M_{\Wprime} = 2.5 \TeV$ & $48.7^{+4.1}_{-4.1}$      & $36.1^{+2.8}_{-3.1}$         & $20.3^{+3.0}_{-3.4}$ \\
                                                                     & $M_{\Wprime} = 3 \TeV$   & $9.88^{+0.99}_{-0.98}$    & $7.33^{+0.64}_{-0.65}$       & $5.00^{+0.16}_{-0.39}$ \\
\multirow{2}{*}{HNC-CI}                                                & $\Lambda = 9 \TeV$       & $42.4^{+3.8}_{-3.8}$      & $13.8^{+2.0}_{-2.0}$         & $4.47^{+0.90}_{-0.94}$ \\
                                                                     & $\Lambda = 4 \TeV$       & $1091^{+97}_{-98}$        & $356^{+50}_{-52}$            & $115^{+23}_{-24}$ \\
\multirow{3}{4cm}{DM vector-coupling, $M_\chi = 50\GeV$, $\Lambda=300$}& $\xi = +1$               & $0.271^{+0.070}_{-0.067}$ & $0.0151^{+0.0061}_{-0.0056}$ & $0.00088^{+0.00051}_{-0.00043}$ \\
                                                                     & $\xi = 0$                & $6.7^{+1.6}_{-1.6}$       & $1.43^{+0.54}_{-0.51}$       & $0.31^{+0.17}_{-0.15}$ \\
                                                                     & $\xi = -1$               & $27.1^{+6.6}_{-6.5}$      & $5.8^{+2.2}_{-2.1}$          & $1.25^{+0.68}_{-0.60}$ \\

\end{scotch}
\end{table*}

\subsection{Higher-order corrections for SM \texorpdfstring{\tPW}{W}\-boson background}
\label{sec:W-kfactor}

\begin{figure*}[hbtp]
\centering
\includegraphics[width=\cmsFigWidth]{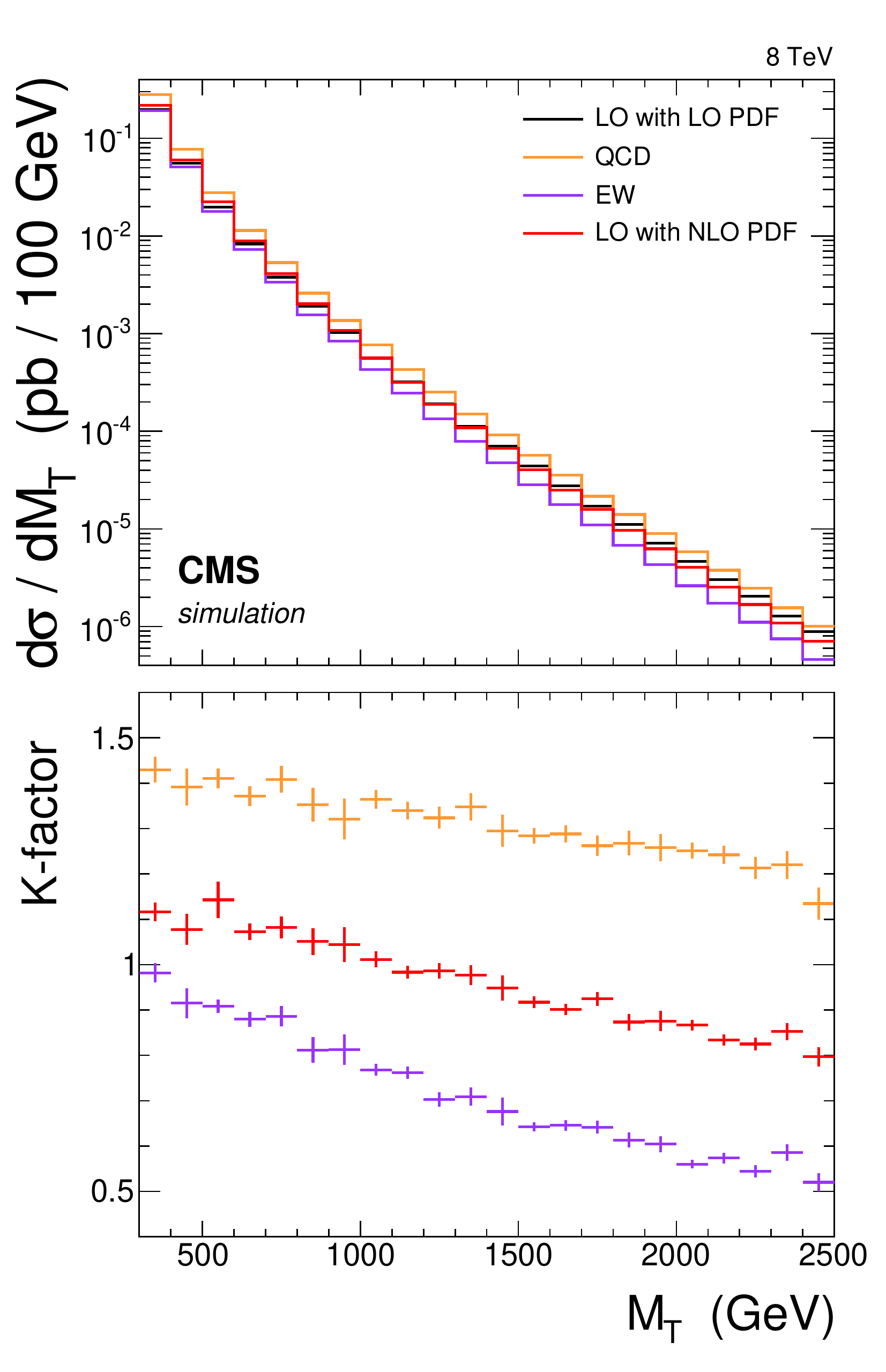}
\includegraphics[width=\cmsFigWidth]{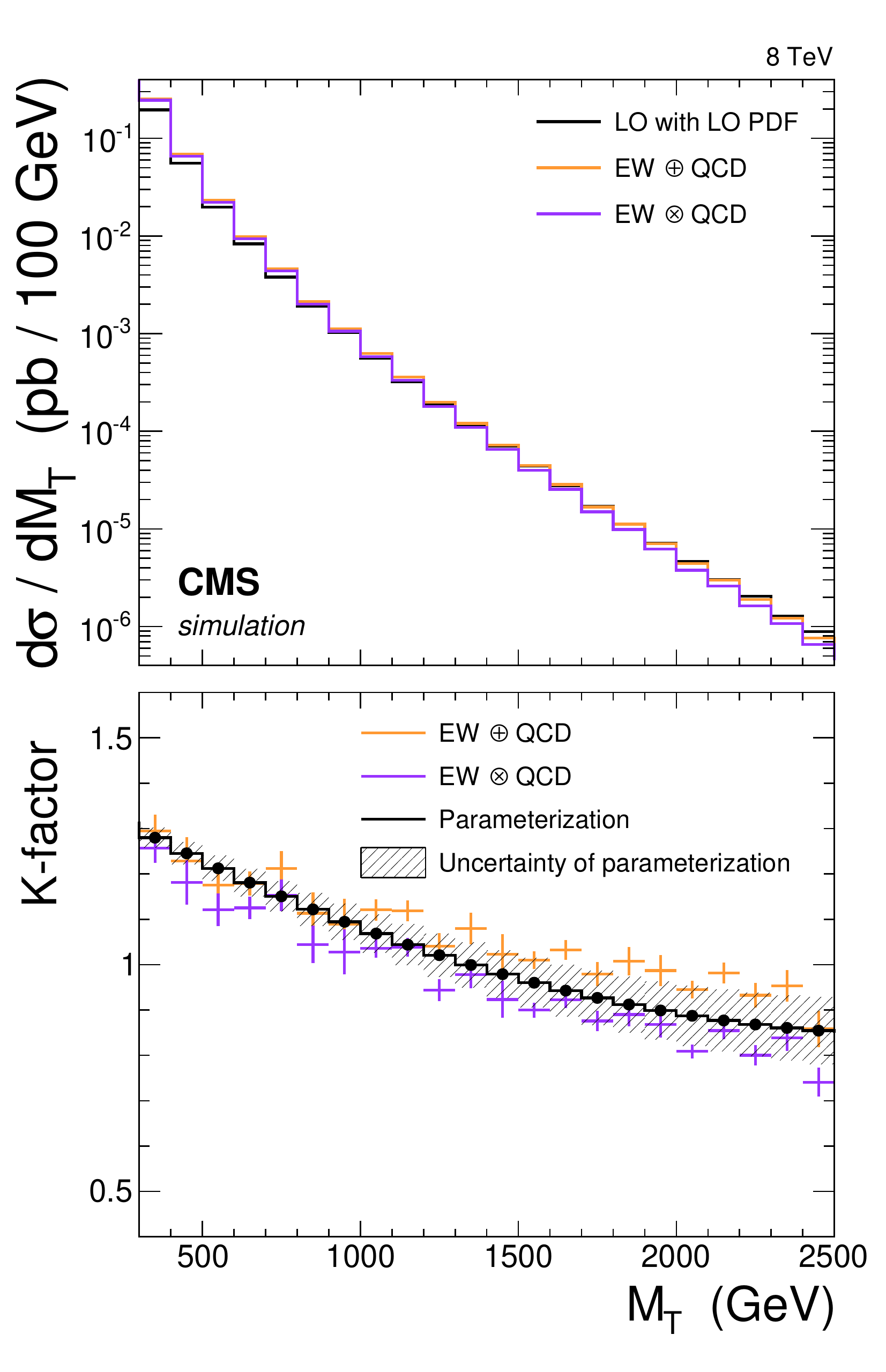}
\caption{Electroweak and QCD corrections to the SM \tPW-boson background prediction. Left: QCD and electroweak contributions compared to the LO calculation with LO PDF. Right: Combination of higher-order
corrections with an additive and a multiplicative approach compared to the LO calculation with LO PDF, as well as a parameterization of the mean of the values obtained by the two methods.
To derive the K-factor data on the right plot, one divides the sum or product of the K-factors on the left plot by the NLO/LO PDF K-factor.
}
\label{fig:Wkfactor}
\end{figure*}

The \tPW boson, particularly through the off-shell tail of its \MT distribution, contributes an important and irreducible background in this analysis. An accurate prediction of
the \tPW boson's off-shell tail is also needed
to establish the effects of interference with the signal.
Therefore, higher-order electroweak (EW) and QCD corrections are evaluated, binned in \MT:
\begin{equation}
K(\MT) = \frac{\Delta\sigma (\text{NLO})/\Delta \MT}{\Delta\sigma(\mathrm{LO})/\Delta \MT}.
\end{equation}
The NLO EW corrections, calculated with the \textsc{horace}~\cite{horace} event generator using the CT10~\cite{CT10} PDF set, depend strongly on \MT. While the corresponding K-factor is around 1.0 for
transverse masses of 300\GeV, it decreases to around 0.5 for \MT = 2.5\TeV. The QCD corrections are calculated with \MCATNLO, also using the CT10 PDF set, and are less \MT dependent, leading to a K-factor
ranging from around 1.4 to 1.2. The impact of both these corrections is illustrated in Fig.~\ref{fig:Wkfactor}. Also shown is the influence of the PDF set using the LO generator \PYTHIA comparing CT10 and CTEQ6L1~\cite{CTEQ6L1}.

To combine the EW and QCD differential cross sections, two different approaches have been used~\cite{Wkfactors}: an additive or a multiplicative combination.
As both, the EW and QCD K-factor include the NLO PDF information, a correction must be made to not account for the NLO PDF twice. The combined K-factors are calculated by:
\newcommand{\dsdo}{\left[\frac{\Delta\sigma}{\Delta\MT}\right]}
\ifthenelse{\boolean{cms@external}}{
\begin{equation}
\begin{aligned}
K_\text{QCD $\oplus$ EW}^\mathrm{CT10} = &\frac{\dsdo_\mathrm{QCD}^\mathrm{CT10} + \dsdo_\mathrm{EW}^\mathrm{CT10} - \dsdo_\mathrm{LO}^\mathrm{CT10}}{\dsdo_\mathrm{LO}^\mathrm{CTEQ6L1}},\\
K_\text{QCD $\otimes$ EW}^\mathrm{CT10} = &\frac{\dsdo_\mathrm{QCD}^\mathrm{CT10}}{\dsdo_\mathrm{LO}^\mathrm{CTEQ6L1}}\times \frac{\dsdo_\mathrm{EW}^\mathrm{CT10}}{\dsdo_\mathrm{LO}^\mathrm{CTEQ6L1}} \\
&\times \frac{\dsdo_\mathrm{LO}^\mathrm{CTEQ6L1}}{\dsdo_\mathrm{LO}^\mathrm{CT10}},
\end{aligned}
\end{equation}
}{
\begin{equation}
\begin{aligned}
K_\text{QCD $\oplus$ EW}^\mathrm{CT10} &= \frac{\dsdo_\mathrm{QCD}^\mathrm{CT10} + \dsdo_\mathrm{EW}^\mathrm{CT10} - \dsdo_\mathrm{LO}^\mathrm{CT10}}{\dsdo_\mathrm{LO}^\mathrm{CTEQ6L1}},\\
K_\text{QCD $\otimes$ EW}^\mathrm{CT10} &= \frac{\dsdo_\mathrm{QCD}^\mathrm{CT10}}{\dsdo_\mathrm{LO}^\mathrm{CTEQ6L1}}\times \frac{\dsdo_\mathrm{EW}^\mathrm{CT10}}{\dsdo_\mathrm{LO}^\mathrm{CTEQ6L1}} \times \frac{\dsdo_\mathrm{LO}^\mathrm{CTEQ6L1}}{\dsdo_\mathrm{LO}^\mathrm{CT10}},
\end{aligned}
\end{equation}
}
where $\dsdo$ is the differential cross section, the upper index gives the PDF set name and the lower index gives the type of correction applied.

Their effects differ by around 10\%. The K-factor assumed in this analysis is obtained by taking the average of the two approaches, and treating half of their difference as a systematic uncertainty.

The means of the K-factors resulting from these two approaches are shown in Fig.~\ref{fig:Wkfactor} (right), with the distribution parametrized using a second-order polynomial. These higher-order corrections have a significant influence on the final result.
This treatment represents an improvement over previous analyses~\cite{2010cms-electron-limit,2010cms-muon-limit,2011limit}, in which a constant K-factor of 1.3 was used across the whole \MT spectrum. For $\MT \geq 1.5\TeV$ the value used in this analysis is closer to $\sim$0.9, as shown in the figure.
This reduces the expected background, while leaving the signal unchanged since corresponding EW corrections have not been calculated.

\subsection{Multijet background estimation from data}
\label{sec:qcdfromdata}

As stated in Section \ref{sec:background_sources}, the misidentification of jets as leptons, which is more likely for electrons than for muons, is a possible source of background for this search.
While the contribution of QCD multijet events to the muon channel is negligible, a small contribution to the electron channel remains.
In the latter channel, the shape and normalization of the QCD multijet background, as shown in Fig.~\ref{fig:MT} and used for the final results, is derived from data.

A QCD template is obtained from the events in which the electron candidate fails the isolation requirement but where all other event requirements are met.
QCD template events are scaled with normalization factors from an independent control region, which is defined by the requirement $1.5<\ET/\MET<10$.
In this region, the ratio $r_{ttl}$ of `tight' events (electron candidate passes all requirements of a well-isolated electron) to `loose' events (all events in the region) is measured as a function of \ET and $\eta$.
The resulting normalization factor for QCD template events is $r_{ttl}/(1-r_{ttl})$.
Contributions from processes with genuine electrons or photons are estimated via simulation and are subtracted.
They amount to 4--13\% of the loose event counts, the most important contributions being
$\PW$+jets and $\gamma$+jets events, along with small contributions from \ttbar, single top-quark, Drell--Yan, and dibosons.
This results in ratios of `tight' to `loose' event counts varying from 7\% in the barrel to 25\% in the very forward region, for electrons with $\ET>200\GeV$.
Based on a set of cross-checks, a total uncertainty of 40\% is assigned to the multijet background.

\section{Systematic uncertainties}
\label{sec:uncertainties}

Mismeasurement of lepton energy or momentum, resulting from both detector resolution and imperfect scale calibration, will result in a smearing of the \MT spectrum.
For each source of uncertainty, shifts of ${\pm}1\sigma$ are applied, the kinematics of the objects ($\Pe$, $\mu$, \MET) are recalculated, as well as \MT, and the kinematic selection is reapplied.
The resulting distribution is parameterized, and the difference with respect to the original parameterization is used to estimate the systematic uncertainty in the number of background events.
The total uncertainty of the expected background is indicated in Fig.~\ref{fig:MT} and specified in Table~\ref{tab:exampleEventYields}.

The systematic uncertainty in the electron energy scale is estimated to be 0.4\,(0.8)\% in the barrel (endcaps).
For the electron energy resolution uncertainty, an additional Gaussian smearing of 1.2\,(2.4)\% for the barrel (endcap) region is applied to the MC simulation~\cite{Chatrchyan:2013dga}.

The muon transverse momentum scale uncertainty is estimated as $5\% \times\pT / \TeV$. The momentum resolution uncertainty is taken into account by applying an additional smearing of 3.2\% to the MC simulation. Both, scale and resolution uncertainty were estimated from the measurement of cosmic muons~\cite{Chatrchyan:2012xi}.
The uncertainty of the muon momentum measurement relates to the smallness of curvature of tracks for high-\pT muons, while the energy of the electrons is measured in the crystal calorimeter and the uncertainty is smaller.

As explained in Section~\ref{sec:W-kfactor} the difference between the two ways of combining the EW and QCD corrections is treated as the systematic uncertainty in the K-factor for $\PW$-boson
production. The effect of even higher order corrections like Sudakov corrections is expected to be small and therefore not considered.

The ratio of data to MC efficiencies are the scale factors (SFs) defined in Section~\ref{sec:selection}. The uncertainty due to the determination method and the extrapolation to high \MT is taken as
the systematic uncertainty in the SFs.

The simulated distribution of $\Pp\Pp$ collision vertices per bunch-crossing has to be reweighted to the distribution measured in data. The uncertainty due to this reweighting method is treated as the
systematic uncertainty of the pileup simulation. The effect on the background event yield due to this uncertainty is smaller than 1\%.

\begin{figure}[hbtp]
\centering
\includegraphics[width=\cmsFigWidth]{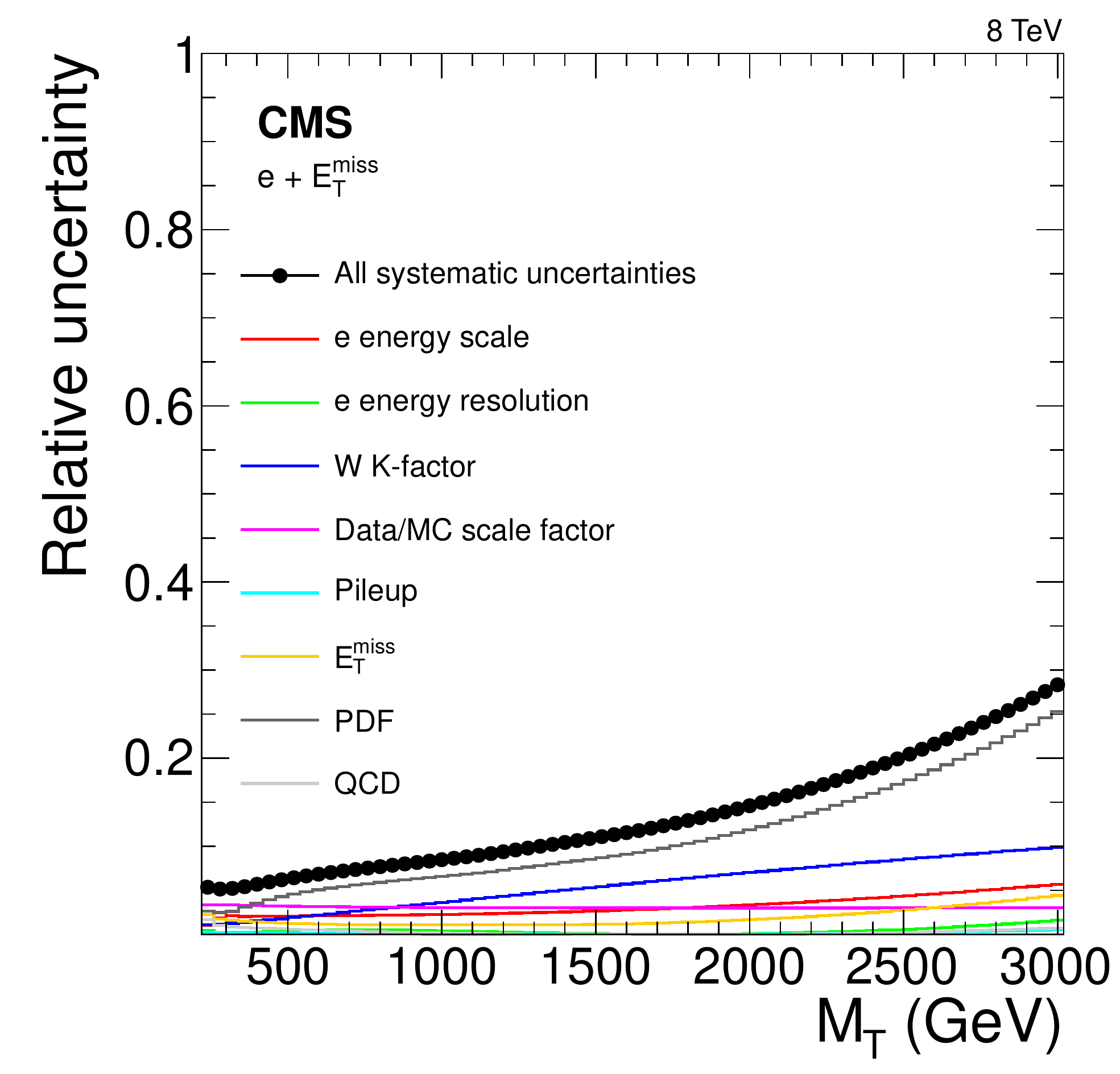}
\includegraphics[width=\cmsFigWidth]{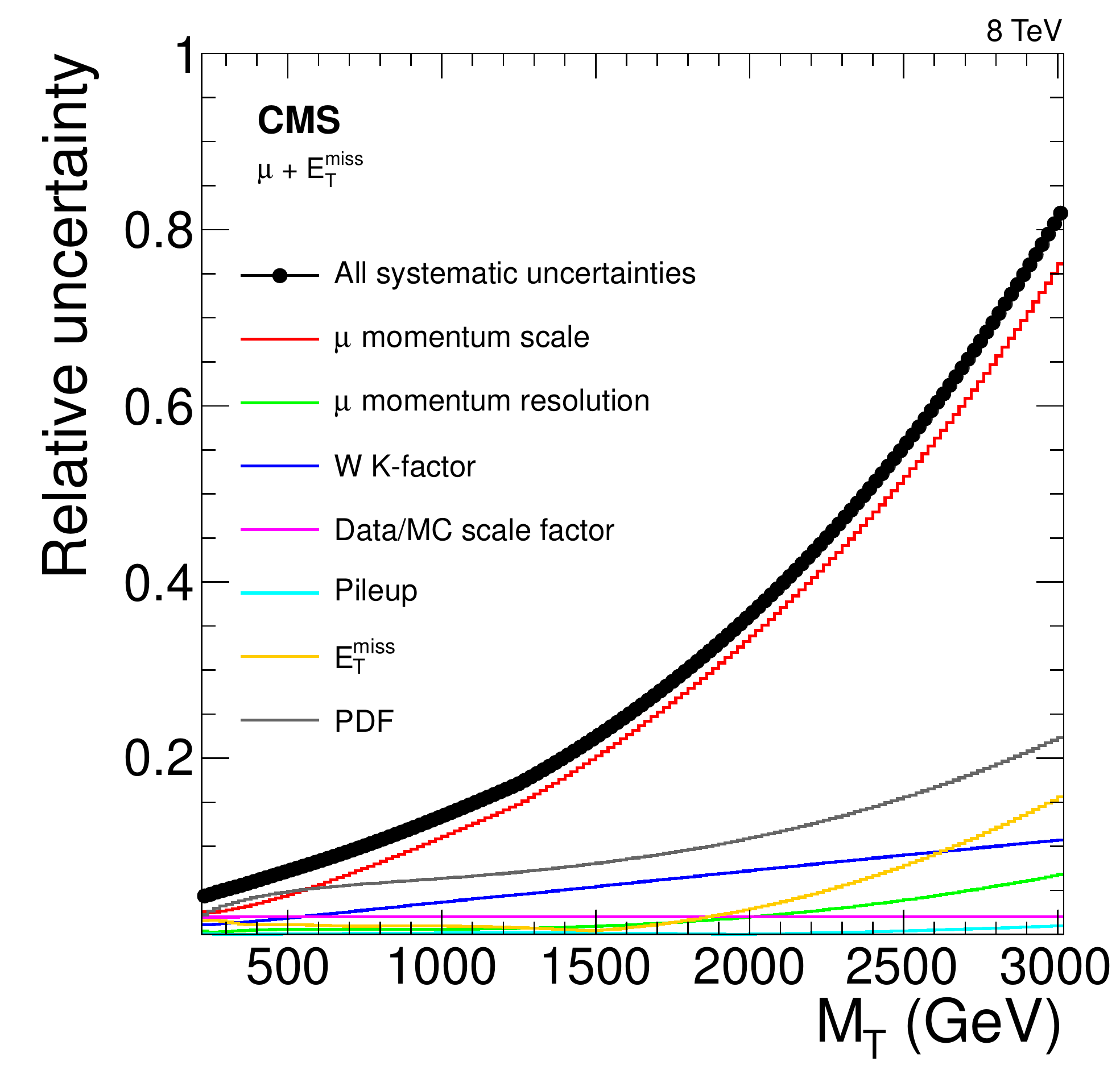}
\caption{Individual contributions of relative systematic uncertainties on the background event yields in the electron (\cmsLeft) and muon (\cmsRight) channels.}
\label{fig:uncertainties}
\end{figure}

The overall uncertainty in the determination of \MET in each event is derived from the individual uncertainties assigned to the objects (jets, $\Pe$, $\mu$, $\tau$, $\gamma$, and unclustered energy) used
by the particle-flow algorithm. The contribution of each object type is varied according to its uncertainty. This uncertainty is propagated to the particle-flow \MET~\cite{Chatrchyan:2011tn}. The
quadratic sum of the individual uncertainties gives the overall uncertainty in the particle-flow \MET.
To account for correlations, the \MET uncertainty components that are also used for the lepton (\eg electron energy scale and resolution in the electron channel) are not included in the
\MET distributions of Fig.~\ref{fig:uncertainties} but are included in their respective distributions.

The theoretical uncertainty related to the choice of the PDF set was estimated using the PDF4LHC recommendation~\cite{Alekhin:2011sk,Botje:2011sn}, reweighting the background samples with three
different PDF sets: NNPDF2.3~\cite{NNPDF2.3}, MWST2008~\cite{MSTW2008}, CT10~\cite{CT10}. For each central PDF set an uncertainty band is derived from the different error PDF sets. The error PDF sets
describe the uncertainties of the PDF set including uncertainties due to $\alpha_{S}$ variation. The envelope of these three error bands is then taken as the uncertainty due to the choice of PDF set.
The same procedure was used for the signal cross section predictions.

For the multijet background prediction, the uncertainties described in Section~\ref{sec:qcdfromdata} are used.

The accuracy of the integrated luminosity estimate is 2.6\%~\cite{LUM-13-001}.

An estimate of the uncertainty in the number of background events in the $M_T$ spectrum arising from the uncertainties described above, but not including the luminosity uncertainty, is shown in
Fig.~\ref{fig:uncertainties}. The dominant systematic uncertainties are due to the PDF sets in the electron channel and due to momentum scale uncertainty in the muon channel.

\section{Limit-setting procedures}
\label{sec:limittools}

As no significant deviation from predictions is seen in the \MT distribution, exclusion limits on new signals can be set.
Upper limits on the production cross section times branching fraction $\sigma_{\Wprime} \,\mathcal{B}(\Wprime \to \ell \cPgn)$, with $\ell = \Pe$ or $\mu$, are determined using a Bayesian method~\cite{pdg} with a uniform positive prior probability distribution for the signal cross section. Systematic uncertainties on the expected signal and background yields are included via nuisance parameters with log-normal prior distributions.

To determine a model independent upper limit on the cross section times branching fraction,
all events above a threshold \MTlower are summed.
From the number of background events, signal events, and observed data events, the cross section limit can be calculated.
No assumptions on the shape of the signal \MT distribution have to be made.
This method has a good sensitivity, comparable to a multi-bin approach, when the background is low.

For the limits on the SSM \Wprime, HNC-CI, DM, split-UED, and \TeVmOne models, the entire \MT spectrum as displayed in Fig.~\ref{fig:MT} with $\MT>220\GeV$ selection is considered, taking the shape of the distribution into account using a binned likelihood (multi-bin counting). This is performed for different values of the model parameters of each signal, resulting in limits in terms of these model parameters, such as the \Wprime boson mass or the interaction scale $\Lambda$.

The analyses of the SSMS and SSMO hypotheses are technically challenging, as the number of events in an \MT region can be larger or smaller than predicted by the standard model.
For the SSMS model, the \PW-\Wprime-boson production cross section would be reduced with respect to the standard model $\PW$ boson, as seen in Table~\ref{tab:xsec}, affecting the \MT range below the Jacobian peak as shown in Fig.~\ref{fig:signal} (top left).
An assumption of the overall cross section always influences the whole signal distribution.
However, the effect of the SSMS and SSMO signal compared to the SM expectation is to reduce the observed event count in one region, while increasing it in another.
This makes the analysis less sensitive in setting a cross section limit. The \Wprime boson coupling strength $g_{\Wprime}$ has therefore been chosen as a free parameter for the models including \PW-\Wprime interference (SSMS and SSMO).
A smaller coupling will result in a narrower Jacobian peak, as well as less modulation in the interference region.

The limits on the \Wprime boson coupling strength in the SSMS and SSMO scenarios have been determined using the modified-frequentist  \CLs method~\cite{CLSRead,Junk:1999kv}. The test statistic used is
\begin{equation}
q_\mu = -2 \ln \frac{\mathcal{L}(\text{data}|\mu,\hat\theta_\mu)}{\mathcal{L}(\text{data}|0,\hat\theta_0)},
\end{equation}
where $\mathcal{L}$ is the likelihood, $\mu$ is the parameter of interest (here, the coupling),
and $\hat\theta$ is the set of nuisance parameters maximizing the likelihood.

\section{Interpretation of the results}
\label{sec:limits}

This section discusses the limits established for the various models summarized in Table~\ref{tab:models}. All limits presented here are at 95\% \CL unless stated otherwise. Note that all presented
mass limits come with a lower bound of $\MT = 220\GeV$, below which background dominates the \MT distribution.

\subsection{Model-independent cross section limit}
\label{sec:modelindependentlimit}

\begin{figure}[hbtp]
\centering
\includegraphics[width=\cmsFigWidth]{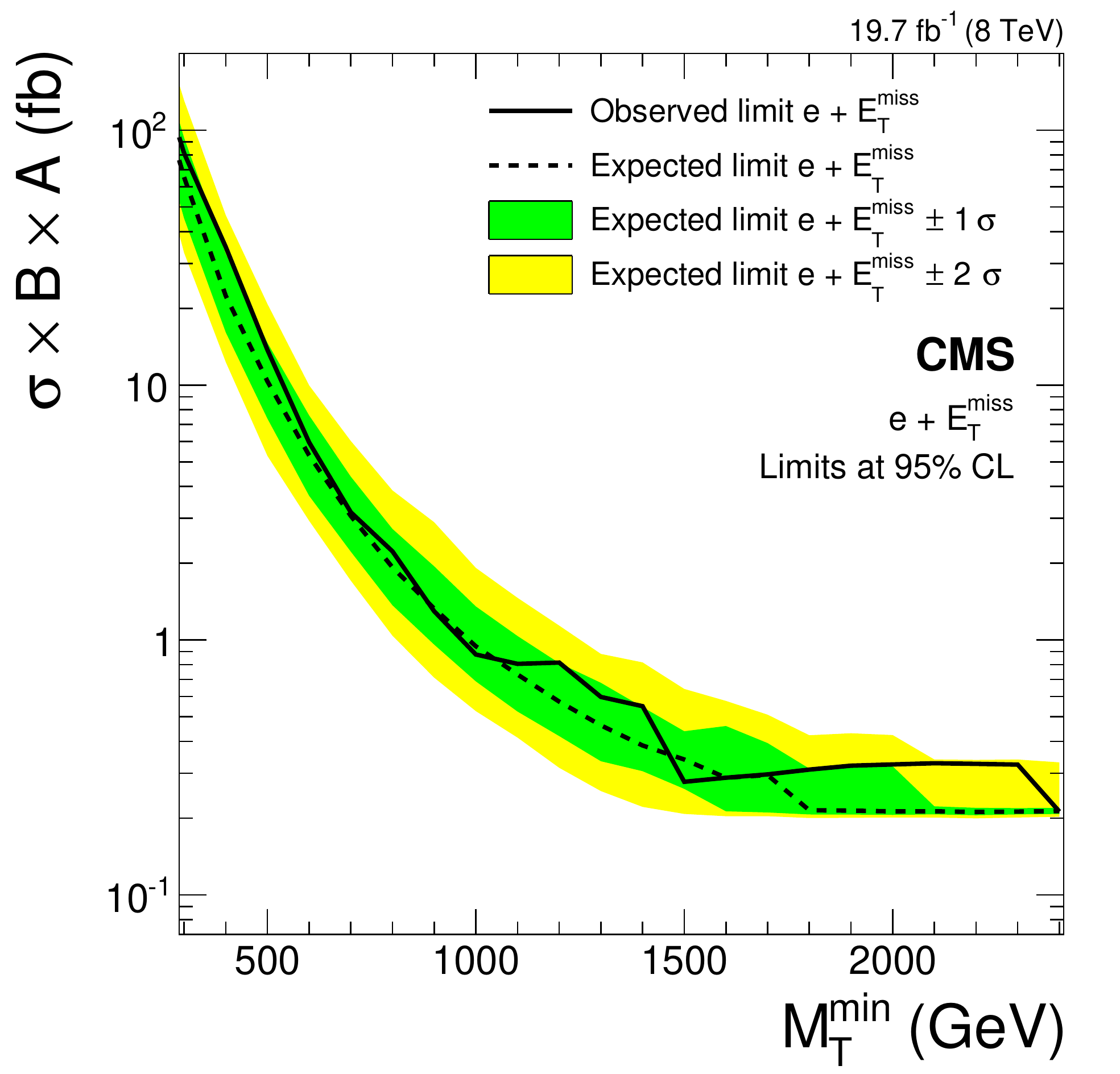}
\includegraphics[width=\cmsFigWidth]{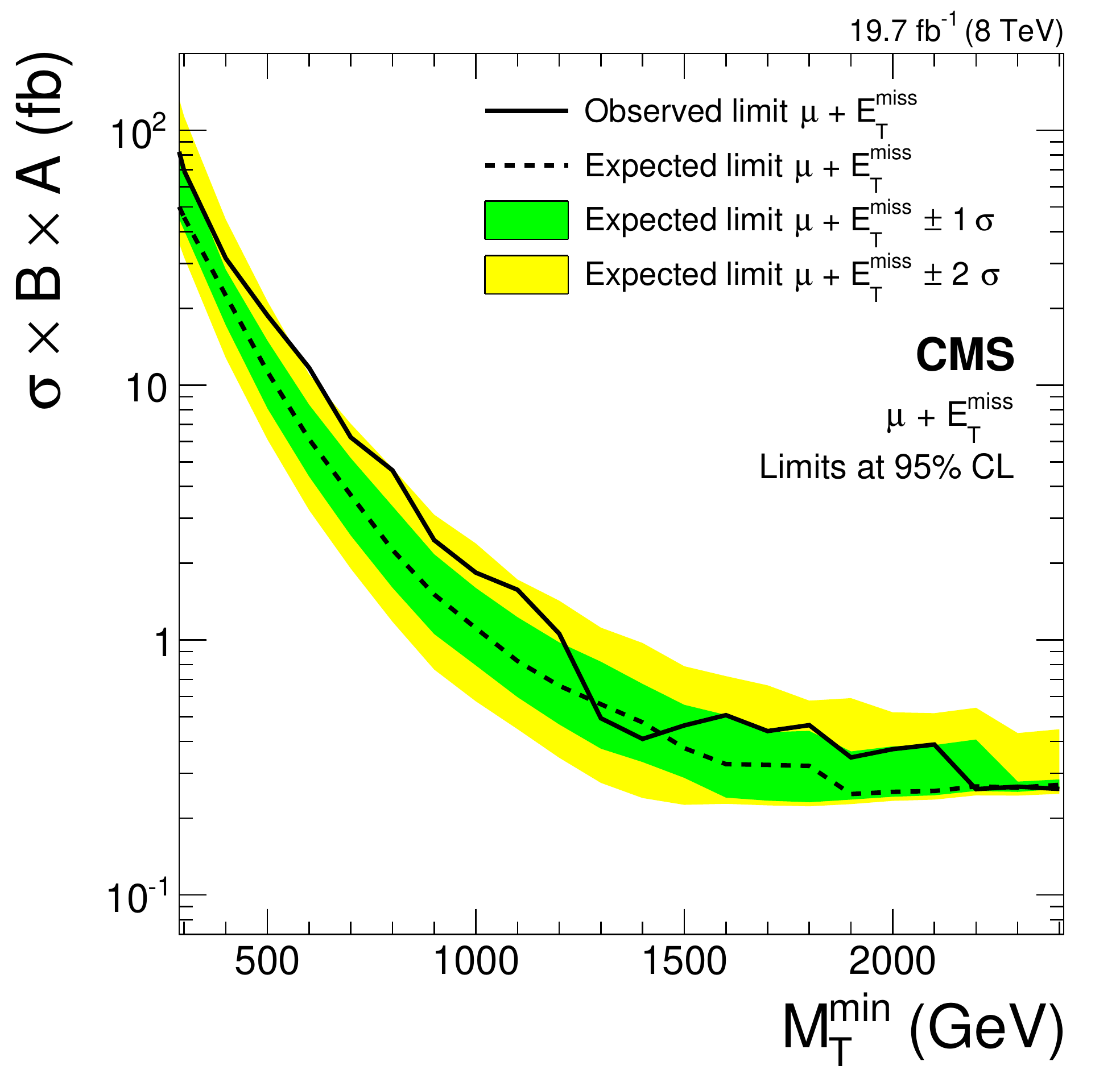}
\caption{Cross section upper limits at 95\% \CL on the effective cross section $\sigma(\Wprime)\,\mathcal{B}(\Wprime \to \ell \nu)A$ above a threshold $\MTlower$ for the individual electron and muon
channels. Shown are the observed limit, expected limit,
and the expected limit $1\sigma$ and $2\sigma$ intervals. Only detector acceptance is taken into account for the signal. The parameter $A$ describes the efficiency derived from the kinematic
selection criteria and the \MTlower threshold. }
\label{fig:LimitMI_singlechannel}
\end{figure}

\begin{figure}[hbtp]
\centering
\includegraphics[width=\cmsFigWidth]{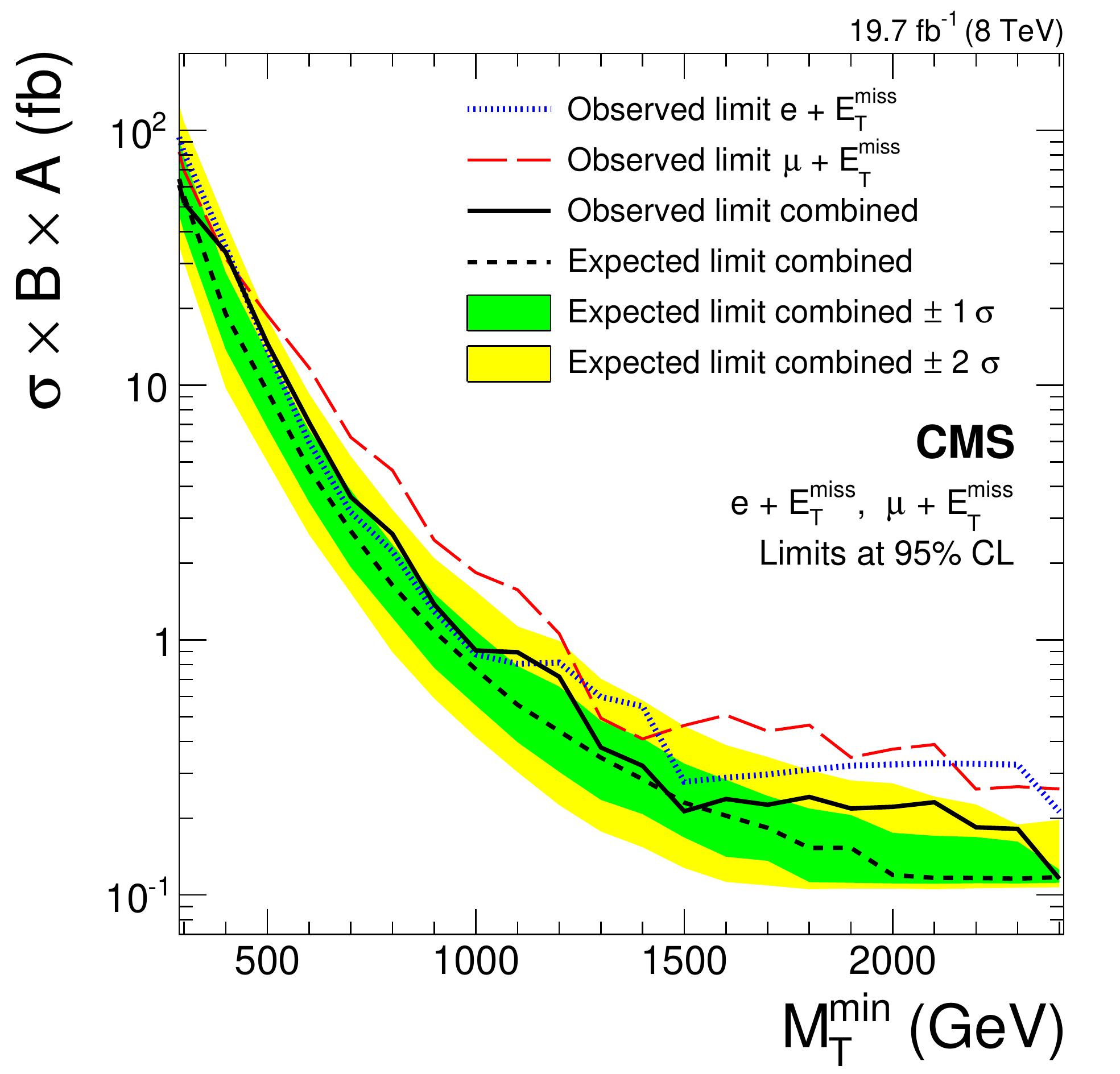}
\caption{Cross section upper limits at 95\% \CL on the effective cross section $\sigma(\Wprime)\,\mathcal{B}(\Wprime \to \ell \nu)A$ above a threshold $\MTlower$ for the combination of the electron
and muon channels. Shown are the observed limits of the electron channel, muon channel, and the combination of both channels
and the combined expected limit, together with the combined expected
limit $1\sigma$ and $2\sigma$ intervals. Only detector acceptance is taken into account for the signal. The parameter $A$ describes the efficiency derived from the kinematic selection criteria and the
\MTlower threshold. }
\label{fig:LimitMI}
\end{figure}

Apart from the model-dependent multi-bin limits, a model-independent cross section limit is determined using a single bin ranging from a lower threshold on $\MT$ to infinity, with the results shown in Fig.~\ref{fig:LimitMI_singlechannel} for the individual electron and muon channels and Fig.~\ref{fig:LimitMI} for the combination of both channels. Only model-independent contributions to signal efficiencies, \eg, the lepton reconstruction efficiency including detector acceptance, are considered, derived using the simulated $\tPW \to \ell \nu$ sample. The signal efficiencies are estimated to be 86\% in the muon channel and 83\% in the electron channel.

In order to determine any limit for a specific model from the model-independent limit shown here, only the model-dependent part of the efficiency must be taken into account. This means an efficiency $A$
describing the effect on the signal of the $\MTlower$ threshold and of the two kinematic selection criteria, $0.4 < \pT/{\ETmiss} < 1.5$ and $\Delta \phi(\ell, \MPT) > 2.5$ (see
Section~\ref{sec:selection}), must be determined.
The \pt and \ETmiss measurements are affected dominantly by the lepton resolutions, as described in Section~\ref{sec:detector}.
Multiplying $A$ with the theoretical cross section $\sigma$ and the branching fraction $\mathcal{B}$, the result can be compared with an exclusion limit from Figs.~\ref{fig:LimitMI_singlechannel}
or~\ref{fig:LimitMI}. Values of $\sigma \times \mathcal{B}\times A$ larger than the limit indicated by the solid line can be excluded. To find the best value of $\MTlower$, the threshold should be
optimized with respect to the expected limit.

The electron channel is more sensitive than the muon channel, as the energy resolution is superior. In the muon channel, shown in Fig.~\ref{fig:LimitMI_singlechannel} (\cmsRight), a small excess of events yields a larger and therefore worse observed limit for most values of $\MTlower$ compared to the expectation.
The step-like structure for high transverse mass thresholds corresponds to the small discrete numbers of events in these regions.

\subsection{Limits on an SSM \texorpdfstring{\Wprime}{W'} boson}
\label{sec:SSMlimit}

\begin{figure}[hbtp]
\centering
\includegraphics[width=\cmsFigWidth]{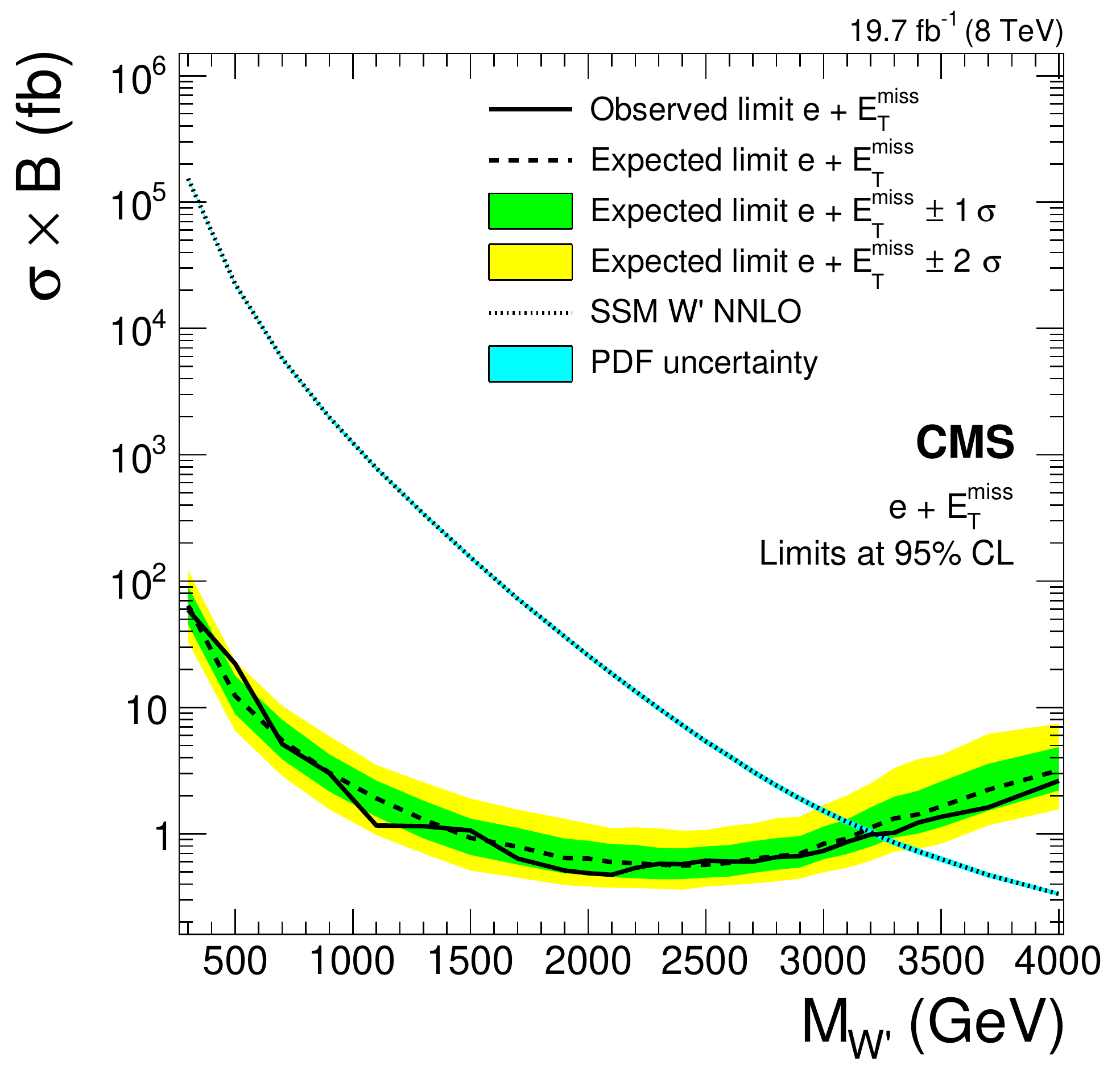}
\includegraphics[width=\cmsFigWidth]{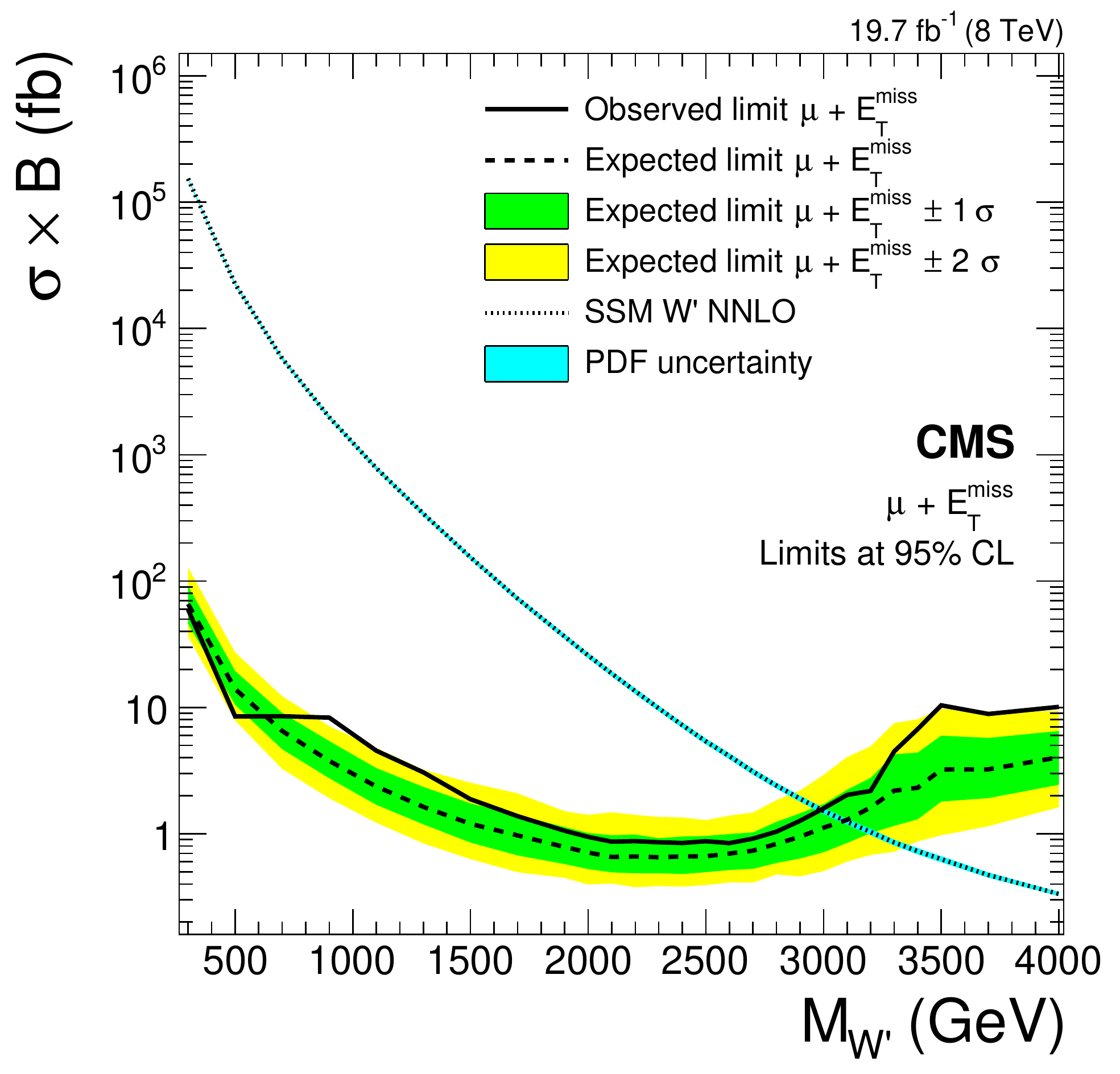}
\caption{Upper limits at 95\% \CL on $\sigma(\Wprime)\,\mathcal{B}(\Wprime \to \ell \nu)$ with $\ell=\Pe$ (\cmsLeft) and $\ell= \mu$ (\cmsRight). Shown are the theoretical cross section, the observed
limit, the expected limit, and the expected limit $1\sigma$ and $2\sigma$ intervals. The theoretical cross section incorporates a mass-dependent NNLO K-factor.}
\label{fig:Limit2012}
\end{figure}

The search for an SSM \Wprime boson yields limits on the cross section times branching fraction for the electron and muon channels. The multi-bin method is used to determine the 95\% \CL upper cross section limits, as shown in Fig.~\ref{fig:Limit2012}. The indicated theoretical cross sections are the NNLO values for lepton+\MET channel, as detailed in Section~\ref{sec:models}, and are the same for both channels. The PDF uncertainties are shown as a thin band around the NNLO cross section. The central value of the theoretical cross section times branching fraction is used for deriving the mass limit.
The existence of an SSM \Wprime boson of mass less than 3.22\TeV (compared with an expected limit of 3.18\TeV) in the electron channel, and 2.99\TeV (compared with an expected 3.09\TeV) in the
muon channel, is excluded. The electron channel has a slightly higher expected sensitivity because of its better resolution.

\begin{figure}[hbtp]
\centering
\includegraphics[width=\cmsFigWidth]{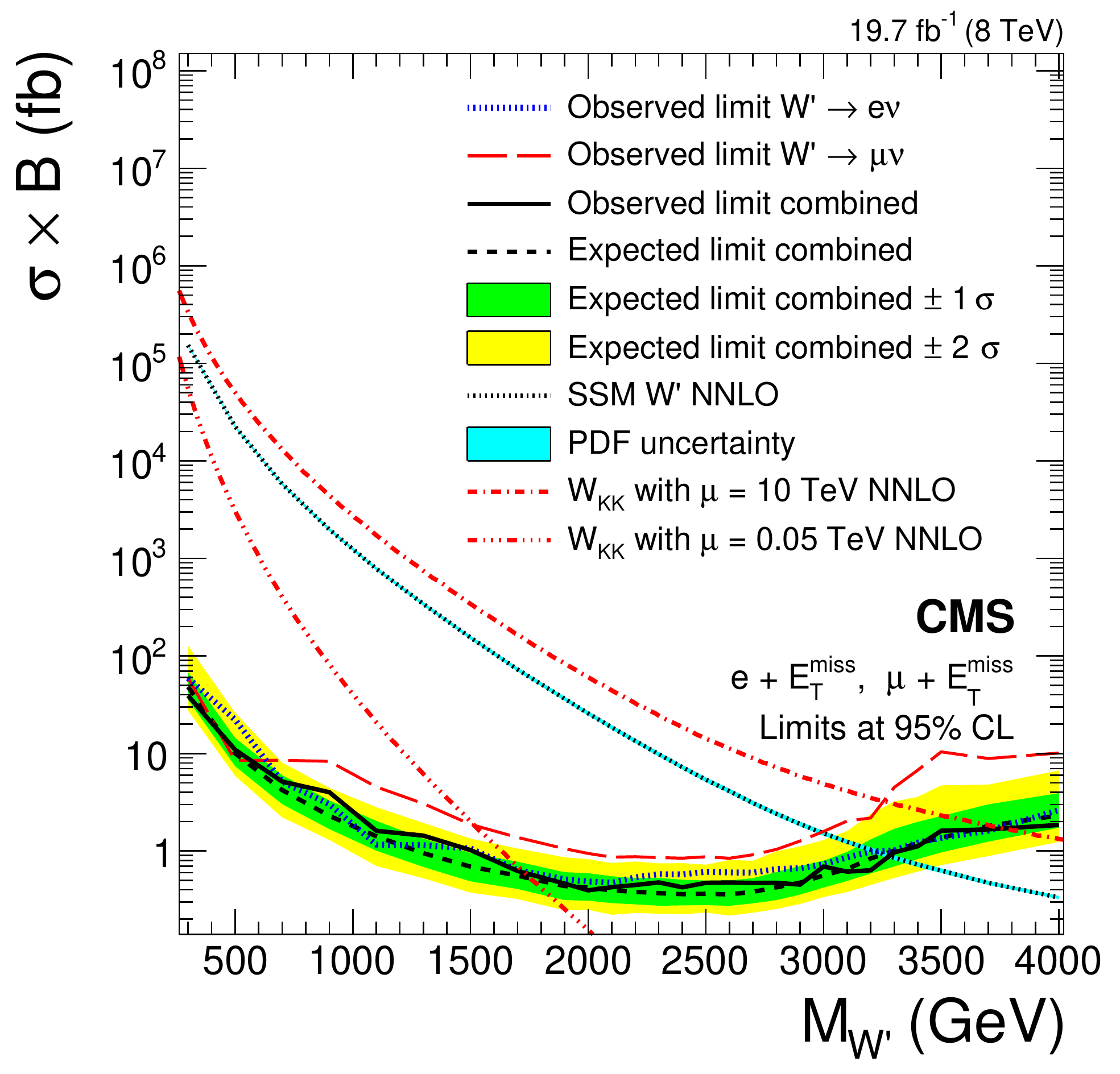}
\caption{Limits for heavy \Wprime bosons for the electron and the muon channels, and for the two channels combined. Shown are the theoretical cross section for the SSM and two split-UED scenarios, the
observed limit, the expected limit, and the expected limit $1\sigma$ and $2\sigma$ intervals. The theoretical cross sections incorporate a mass-dependent NNLO K-factor. The PDF uncertainties for the
SSM are shown as a light blue band around the cross section curve. For the split-UED scenarios, the PDF uncertainties are expected to be small, similar to the uncertainty in the SSM scenario.}
\label{fig:CombLimit2012}
\end{figure}

 Limits can also be obtained for the combined electron and muon channels. Uncertainties deriving from the lepton identification efficiencies for each channel are assumed to be independent. Uncertainties due to the \MET determination, pileup estimate, and luminosity measurement are each assumed to be fully correlated between the channels. Combining both channels, which corresponds to doubling the event count, increases the mass limit to 3.28\TeV (compared with an expected limit of 3.26\TeV). This compares with the previously established combined limit of 2.5\TeV~\cite{2011limit}, which is based on an integrated luminosity of 5\invfb at $\sqrt{s}=7\TeV$. Figure~\ref{fig:CombLimit2012} displays the excluded \Wprime cross section times branching ratio as a function of the \Wprime-boson mass. The corresponding values are summarized in Table~\ref{tab:CombinedLimit2012}.

If the cross section limits are compared to the LO cross section, the SSM \Wprime mass limits change slightly to 3.16\TeV (compared with an expected limit of 3.14\TeV) for the electron channel, 2.96\TeV (compared with an expected limit of 3.04\TeV) for the muon channel, and 3.25\TeV (compared with an expected limit of 3.21\TeV) for the combination of both channels.

\begin{table}[htb]
\topcaption{Upper cross section limits at 95\% \CL for various SSM \Wprime boson masses, based on the combination of the electron and muon channels.}
\label{tab:CombinedLimit2012}
\centering
\begin{scotch}{lccccccc}
$M_{\Wprime}$ (\GeVns)  & 300 & 1100 & 1500 & 2000 & 3000 & 3500 & 4000\\
\hline
Expected limit (fb) & 48 & 1.4 & 0.69 & 0.42 & 0.56 & 1.4 & 2.3    \\
Observed limit (fb) & 39 & 1.6 & 1 & 0.4 & 0.69 & 1.6 & 1.9        \\
\end{scotch}
\end{table}

\subsection{Interpretation in the HNC-CI model}

\begin{figure}[hbtp]
\centering
\includegraphics[width=\cmsFigWidth]{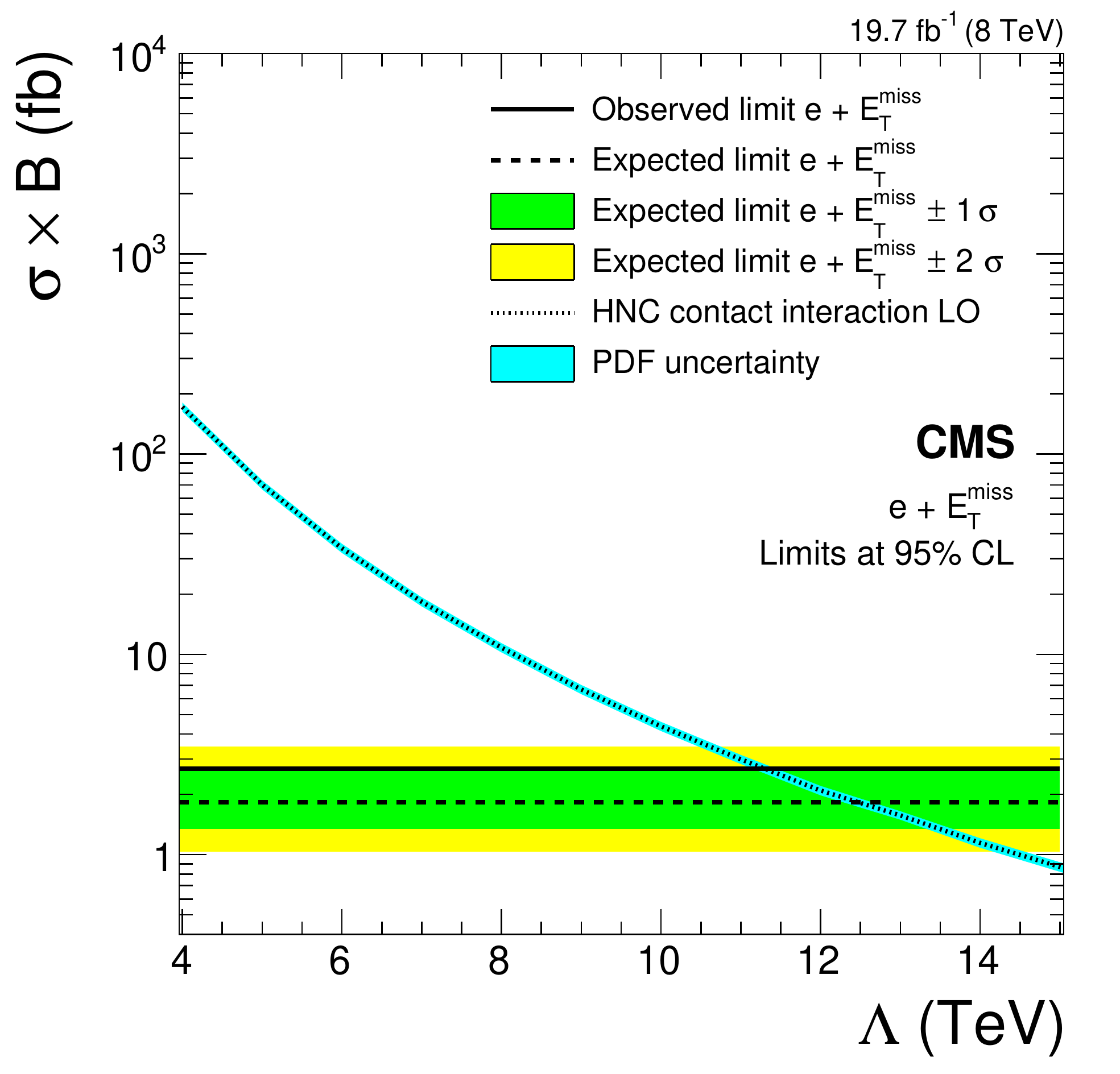}
\includegraphics[width=\cmsFigWidth]{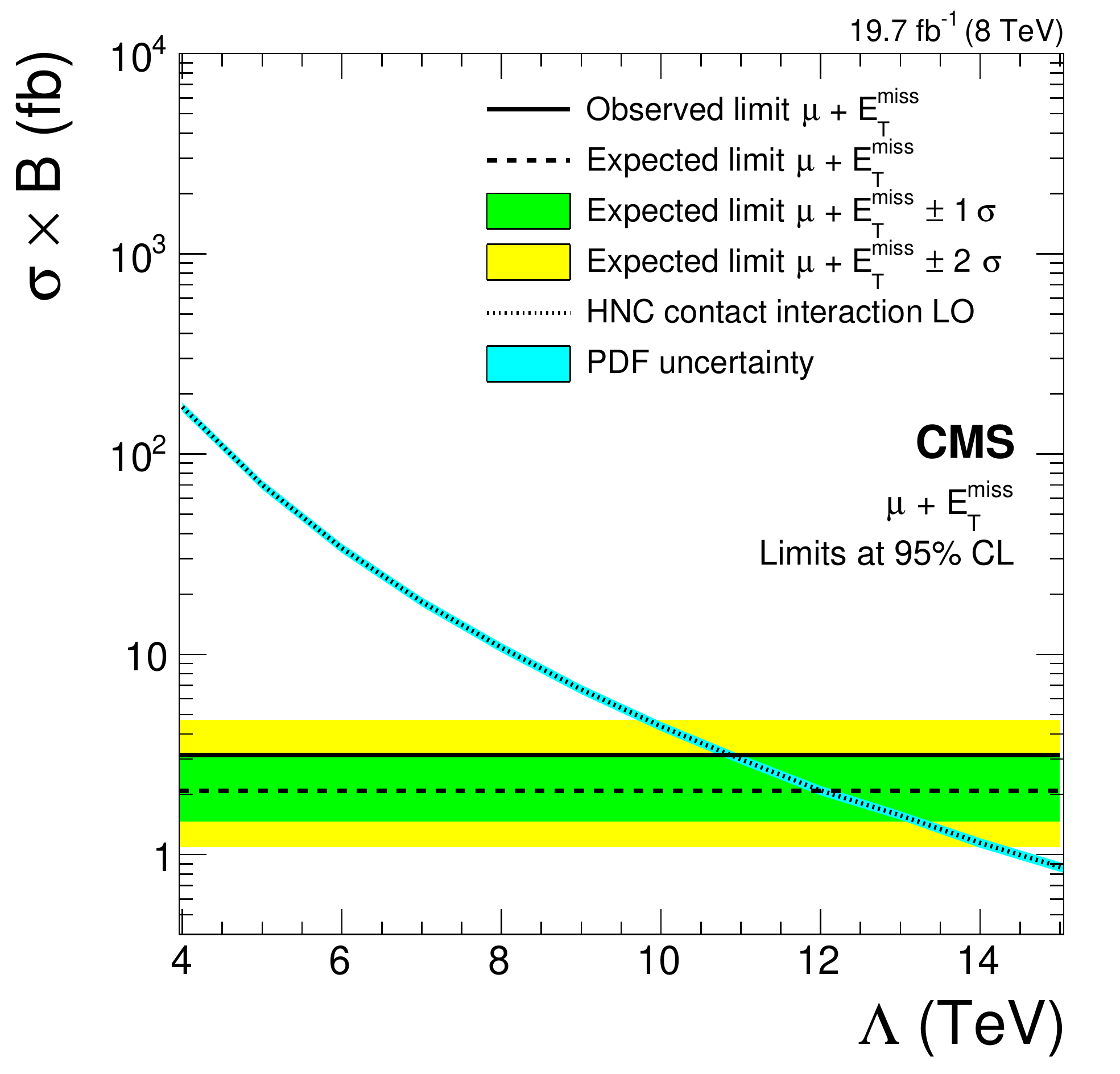}
\caption{Upper limits at 95\% \CL on $\sigma \mathcal{B}(\Pp\Pp \to \ell \nu)$ with $\ell=\Pe$ (\cmsLeft) and $\ell= \mu$ (\cmsRight) in terms of the contact interaction scale $\Lambda$ in the HNC-CI model.}
\label{fig:CILimit}
\end{figure}

Another interpretation of the observed data can be made in the framework of the HNC-CI model, providing a limit on the contact interaction scale $\Lambda$. The statistical interpretation is identical
to that for an SSM \Wprime boson, using a Bayesian multi-bin approach with a uniform prior for the signal cross section~\cite{pdg}. The difference in shape with respect to the \Wprime
Jacobian peak does not affect the limit-setting procedure. The cross section scales as $\Lambda^{-4}$. The shape and signal efficiency, however, are independent of $\Lambda$, leading
to the constant expected and observed limits shown in Fig.~\ref{fig:CILimit}. The limit on $\Lambda$ is calculated to be 11.3\TeV in the electron and 10.9\TeV in
the muon channel.
In the considered model the contact interaction must be flavor symmetric for $\Lambda <$500\TeV \cite{HNCM-CI}.
Combining both channels, a limit on $\Lambda$ of 12.4\TeV is observed and 13.6\TeV is expected.

\subsection{Dark matter interpretation}

The data may be interpreted in the context of an effective dark matter theory. The search presented is inclusive, meaning that it includes final states of lepton+\MET and well as lepton+\MET+jet, since
no event selection criteria based on jets are applied. The statistical interpretation is a Bayesian approach with a uniform prior~\cite{pdg}, based on the multi-bin approach. In order to be comparable
to other dark matter searches, the limits in this model are determined at 90\% \CLend
Electron and muon channels are combined, since the recoiling \tPW boson is a  standard model boson, for which the
decay channel should not depend on the new physics model. Vector-like (spin-independent) and axial-vector-like (spin-dependent) couplings are considered.

\begin{figure*}[htpb]
\centering
\includegraphics[width=\cmsFigWidthTwo]{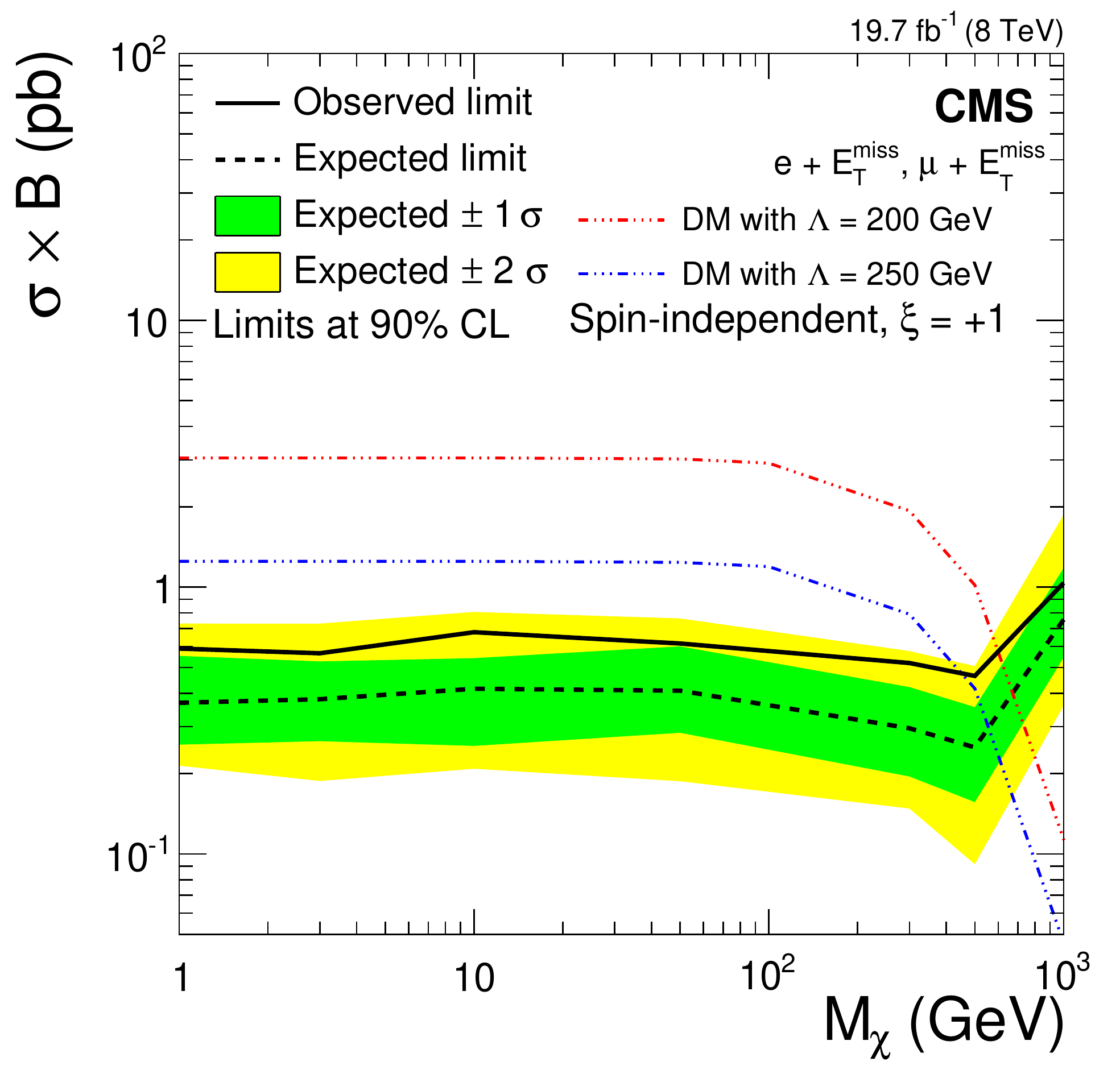}
\includegraphics[width=\cmsFigWidthTwo]{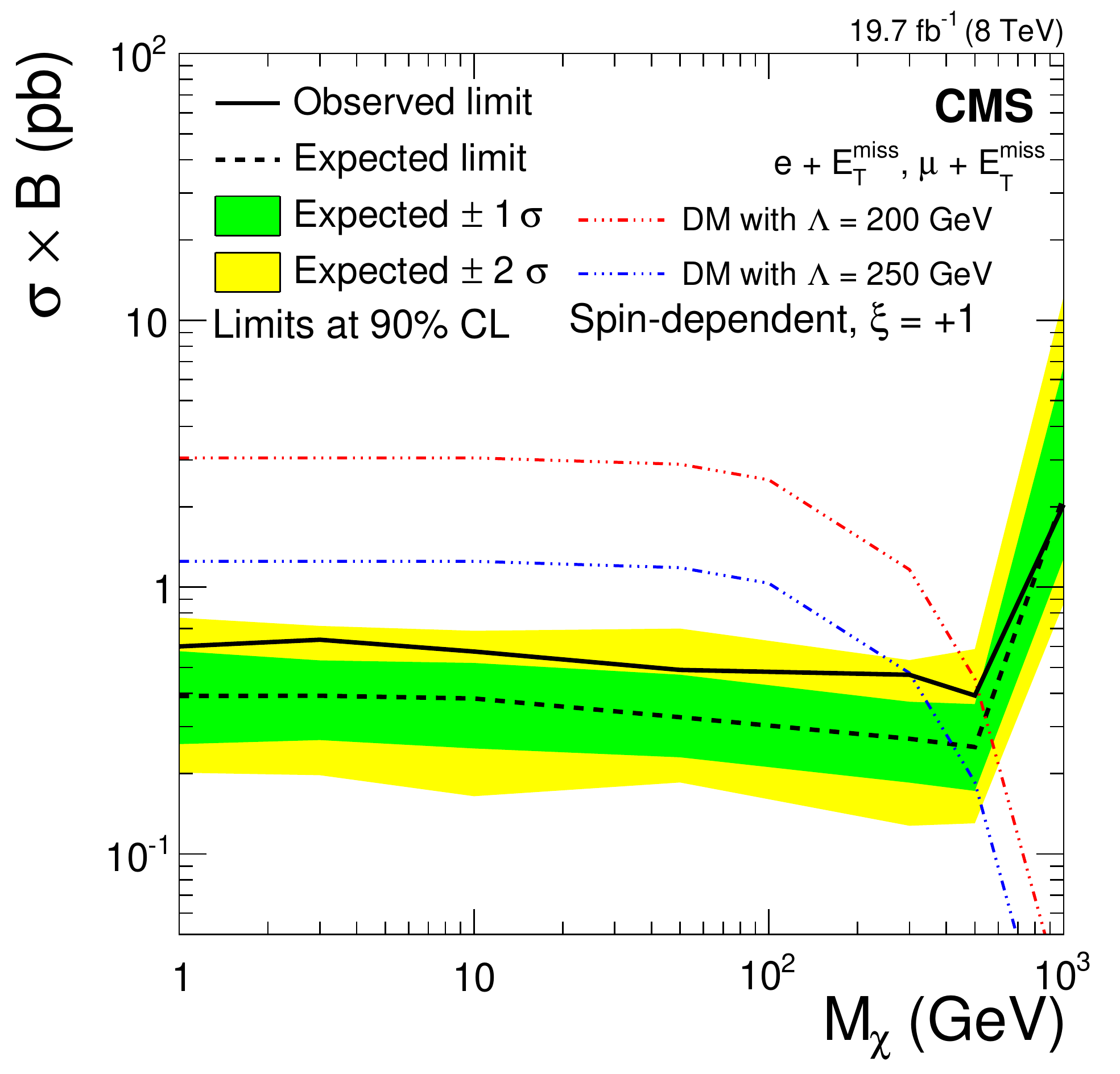}\\
\includegraphics[width=\cmsFigWidthTwo]{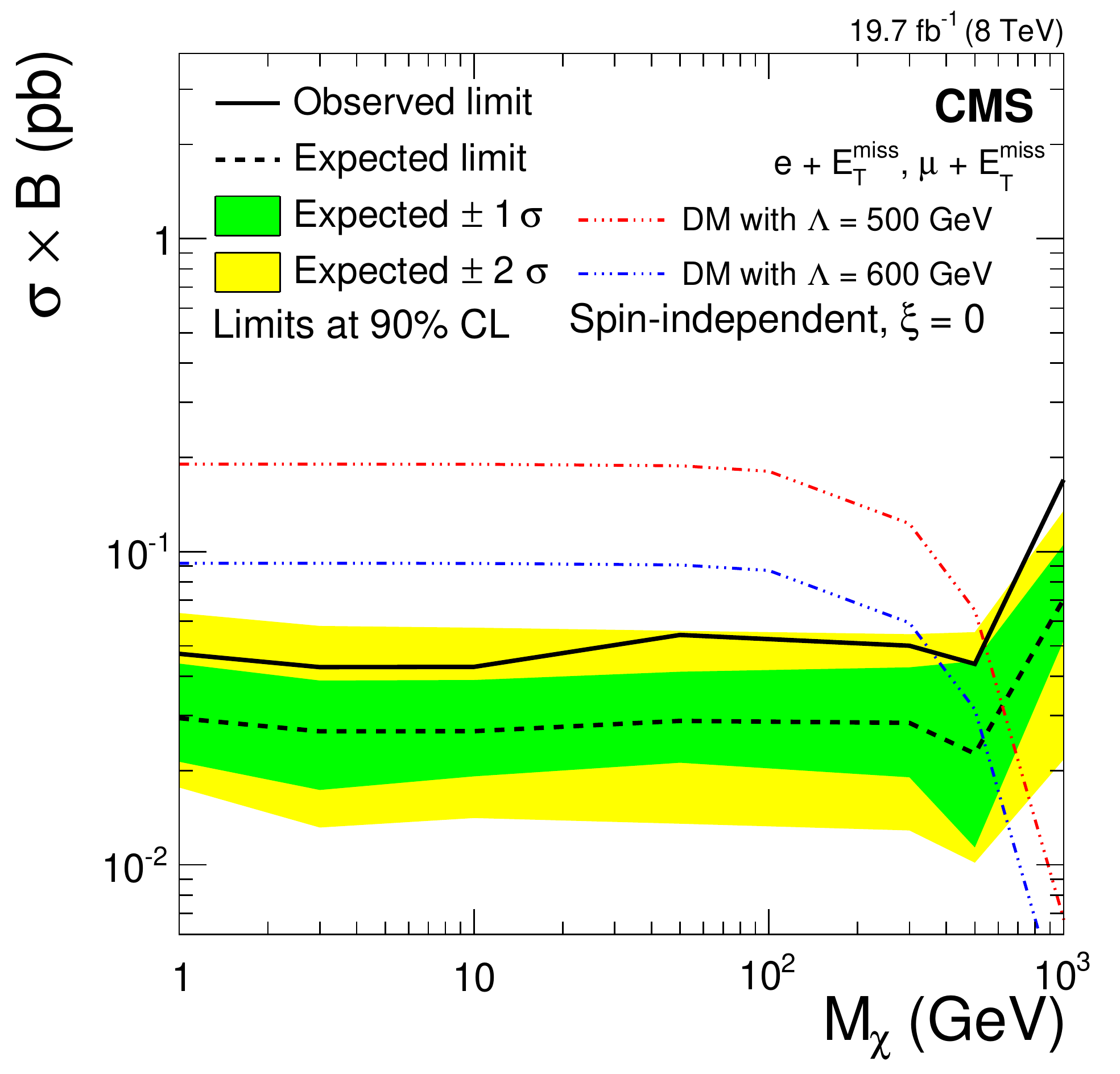}
\includegraphics[width=\cmsFigWidthTwo]{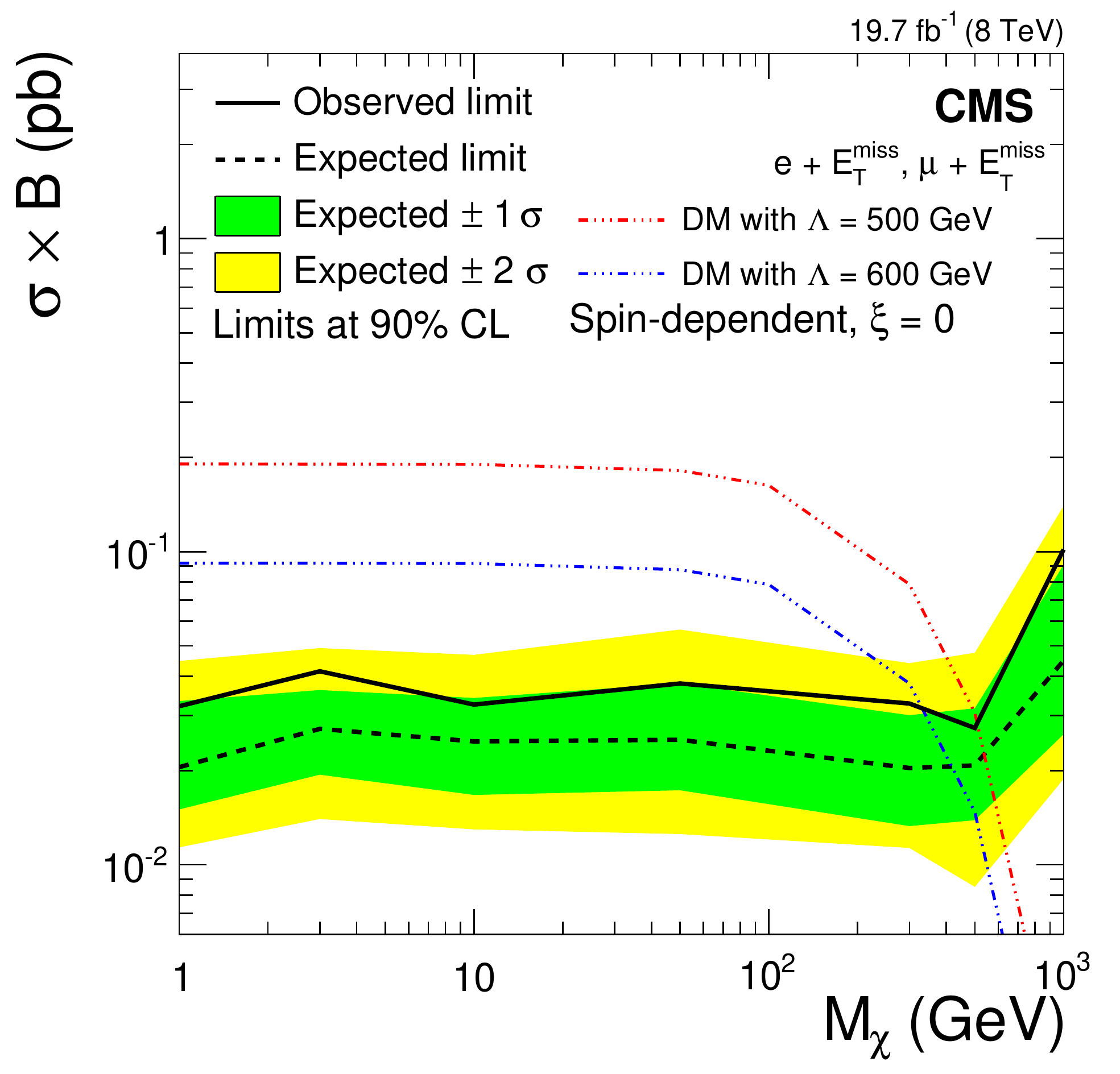}\\
\includegraphics[width=\cmsFigWidthTwo]{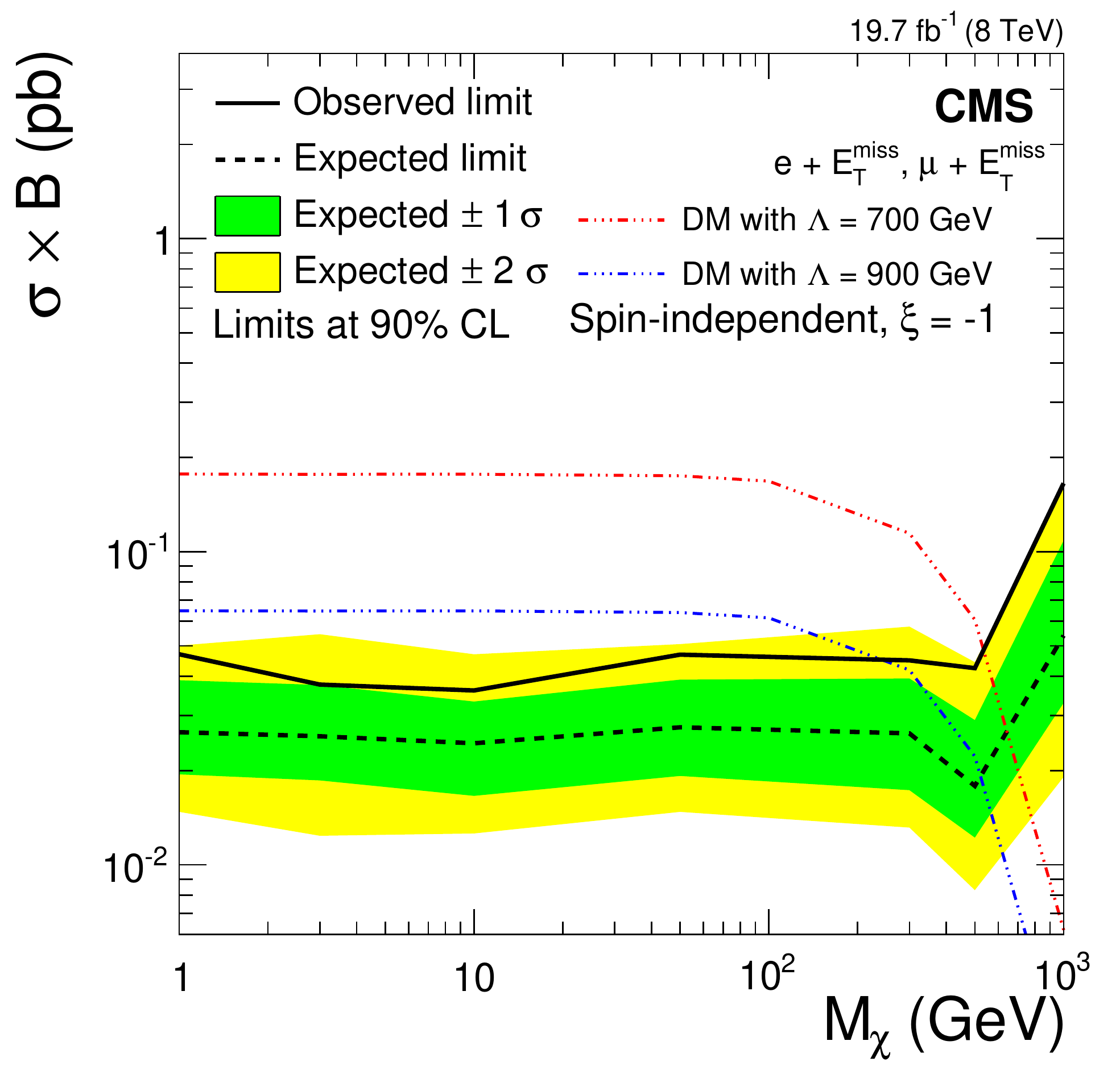}
\includegraphics[width=\cmsFigWidthTwo]{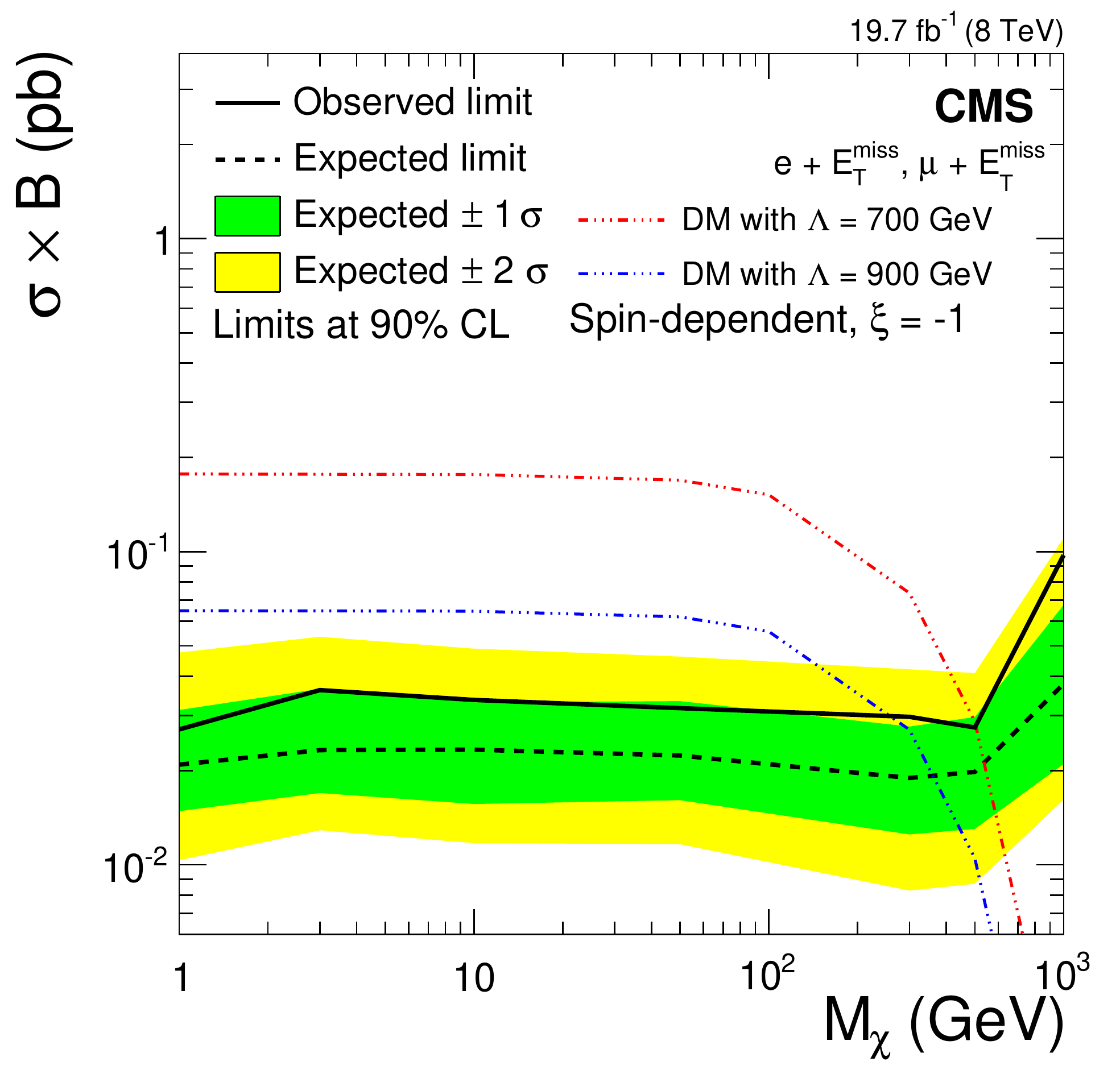}

\caption{Upper limits on $\sigma \,\mathcal{B}(\Pp\Pp \to \chi\chi\ell\nu)$ for vector-like (spin-independent, left column) and axial-vector-like (spin-dependent, right column) couplings. Limits are calculated for the cases (from top to bottom) $\xi = +1,$ 0 and $-1$. Note the different vertical scales.}
\label{fig:LimitDM-collider}
\end{figure*}

\begin{figure}[htp]
\centering
\includegraphics[width=\cmsFigWidth]{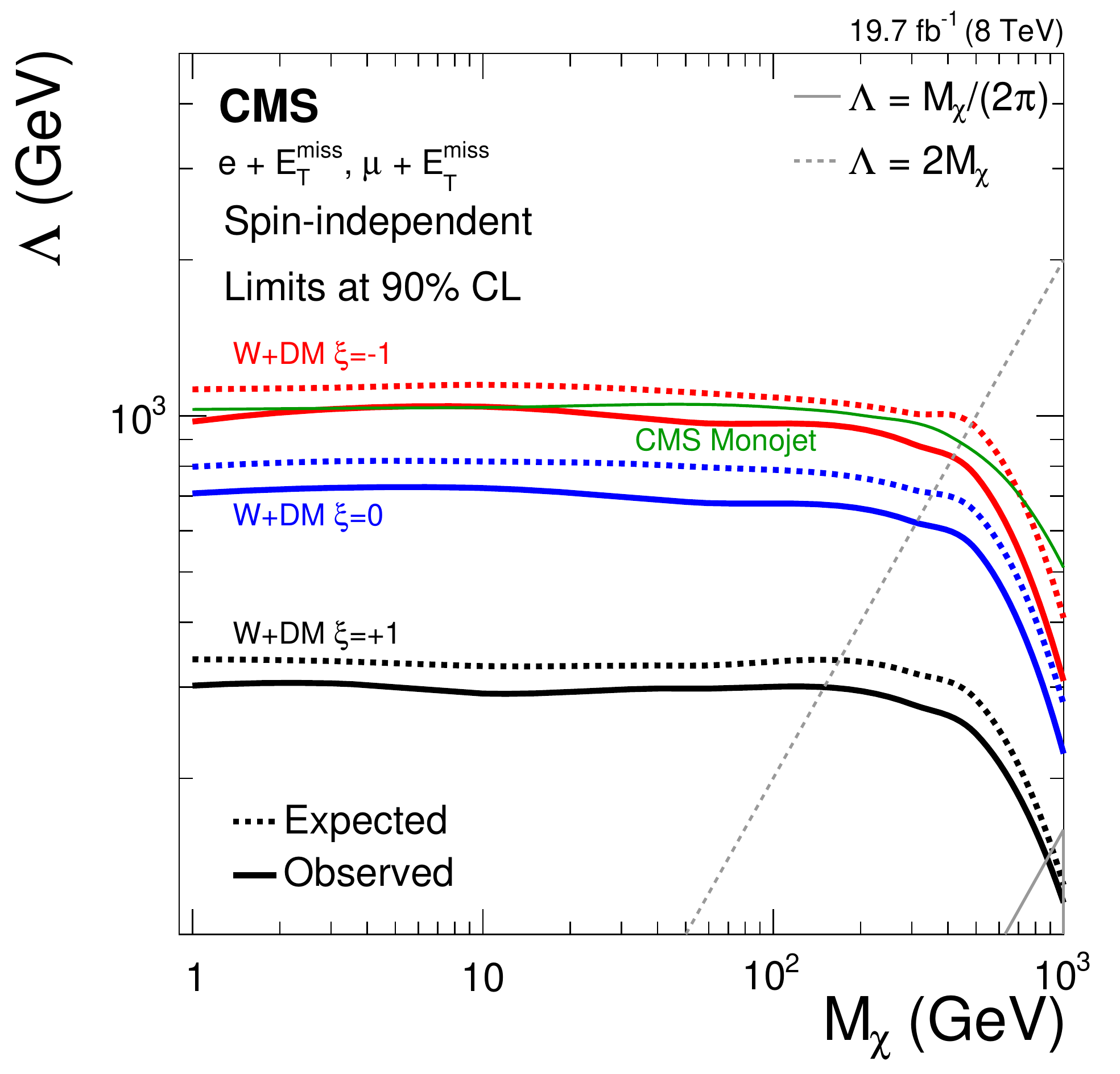}
\includegraphics[width=\cmsFigWidth]{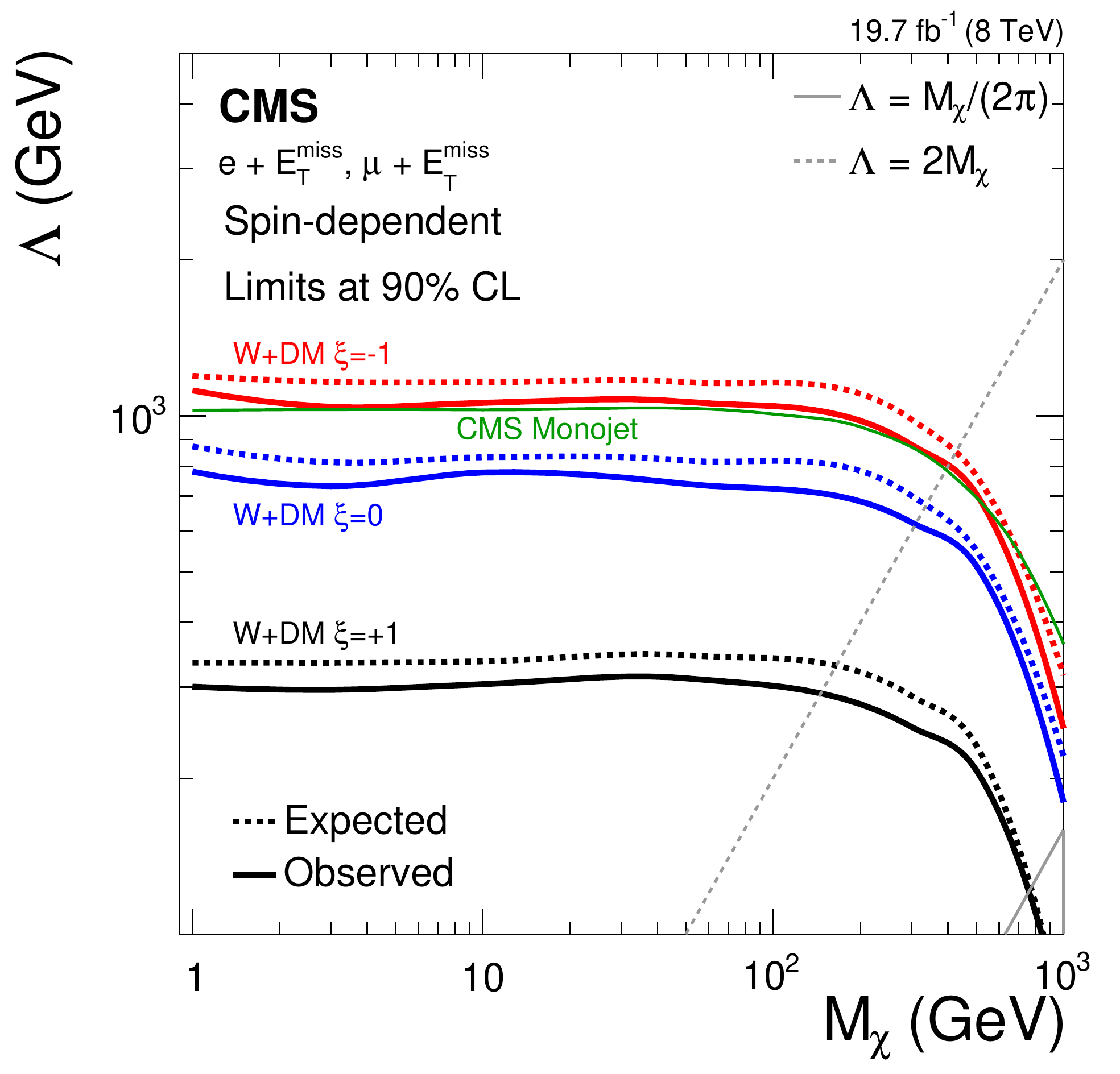}
\caption{Exclusion plane in $\Lambda$--\Mchi, for the combination of the electron and muon channels.
Vector-like (\cmsLeft) and axial-vector-like (\cmsRight) couplings are shown. The two gray lines indicate
where the coupling becomes non-perturbative and ($g_\mathrm{DM}$) is equal to 1, as described in Section~\ref{sec:models-dm}.
The green line shows the limit in the monojet final state~\cite{CMS-PAS-EXO-12-048}, which is independent of $\xi$ for the limit on $\Lambda$.}
\label{fig:LimitDM-lambda}
\end{figure}

\begin{figure}[htp]
\centering
\includegraphics[width=\cmsFigWidth]{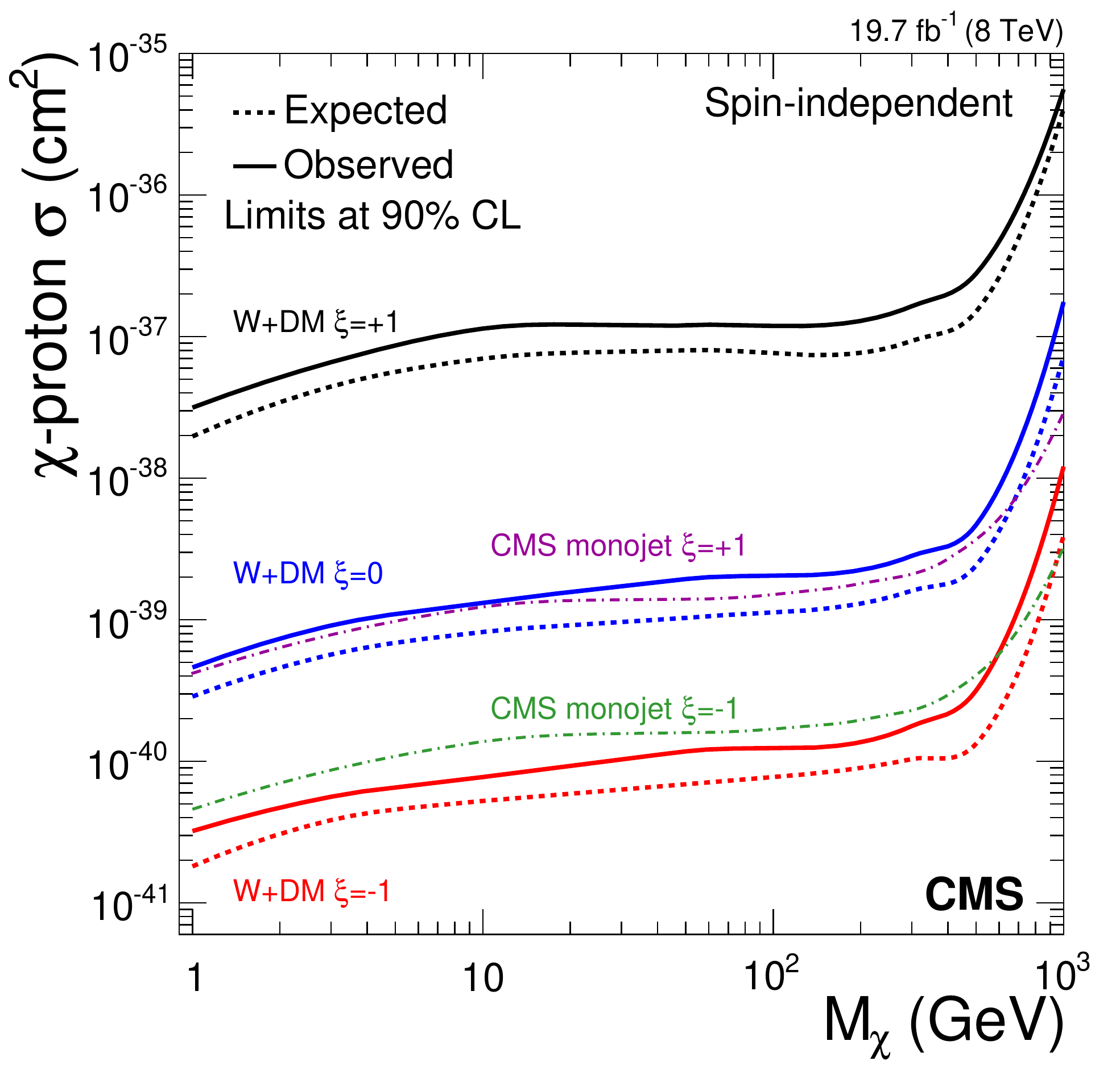}
\includegraphics[width=\cmsFigWidth]{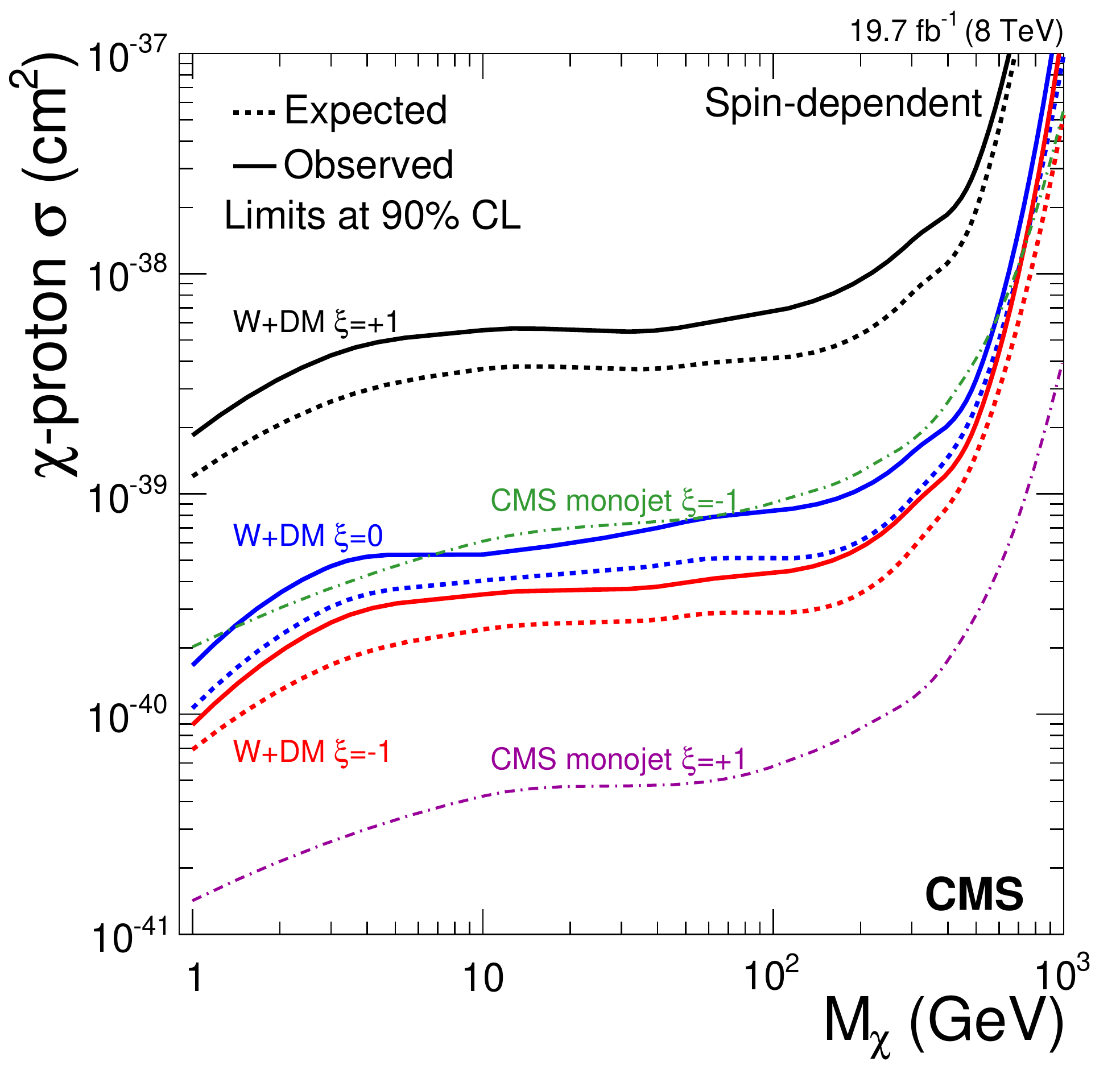}
\caption{Excluded proton-dark matter cross section for vector-like (\cmsLeft) and axial-vector-like
(\cmsRight) couplings, for the combination of the electron and muon channels.
For comparison the result from the monojet DM search~\cite{CMS-PAS-EXO-12-048}
is also shown.
}
\label{fig:LimitDM-direct-interaction}
\end{figure}

The exclusion limits are determined as cross section limits (see Fig.~\ref{fig:LimitDM-collider}) which are subsequently transformed into limits on the effective scale parameter $\Lambda$ as a function of $M_\chi$, as shown in Fig.~\ref{fig:LimitDM-lambda}. In order to compare these collider limits with results from direct detection experiments, they are translated into limits on the DM-proton cross section, shown in Fig.~\ref{fig:LimitDM-direct-interaction}.
We recalculate the excluded nucleon cross section for a given $\Lambda$ and \Mchi using the conversion formula with interference from Ref.~\cite{TevatronDMFrontier}.

The collider cross section limits in Fig.~\ref{fig:LimitDM-collider} show that the excluded cross section is flat as a function of \Mchi, as expected since the signal kinematics do not change
appreciably for different \Mchi. The coupling does not have a large effect on the excluded cross section. The different interference scenarios have a visible influence on the limit. In the case of
$\xi=+1$, a cross section greater than 0.6\pb is excluded, whereas for $\xi=0$ and for $\xi=-1$ the cross section limit is 0.05\pb.
For high \Mchi
the phase space to produce two heavy particles and a $\PW$ boson is small, therefore the signal cross section is reduced and its shape more consistent with the background. These effects yield a
reduced sensitivity at this center-of-mass energy.

As the \MT distributions for vector-like (spin-independent) and axial-vector-like (spin-dependent) couplings are very similar, the derived limits for the two cases do not differ substantially.

For lower masses a constant $\Lambda$ exclusion is obtained for $\Mchi \leq 100\GeV$ of $\Lambda < 300\GeV$ for $\xi = +1$, $\Lambda < 700\GeV$ for $\xi = 0$, and $\Lambda < 1000\GeV$ for $\xi = -1$. The difference between vector-like and axial-vector-like couplings is small for low \Mchi for all three values of $\xi$, but a difference is observed in the high-\Mchi region, above 100\GeV. An overview is given in Fig.~\ref{fig:LimitDM-lambda}.

For comparison the limit from the monojet final state~\cite{CMS-PAS-EXO-12-048} is shown. The conversion to $\chi$-proton cross section depends on the coupling parameter $\xi$, although the monojet analysis is not sensitive to $\xi$.
The limits determined from the direct detection experiments depend on different model assumptions~\cite{BuchmuellerDarkMatterAssumptions}. Therefore we do not give a direct comparison here.

The $\chi$-proton cross section upper limits at 90\% \CL for $\Mchi = 10\GeV$ are presented in Table~\ref{tab:protoncslimits} and Fig.~\ref{fig:LimitDM-direct-interaction}.

\begin{table}[htpb]
\centering
\topcaption{The $\chi$-proton cross section upper limits at 90\% \CL for \Mchi = 10\GeV.}
\label{tab:protoncslimits}
\begin{scotch}{rcc}
\multicolumn{1}{c}{$\xi$} & Vector coupling     & Axial-vector coupling\\
&($\cm^{2}$)&($\cm^{2}$)\\
\hline
$-1$   &  $4  \times 10^{-41}$ & $1 \times 10^{-40}$\\
$0$    &  $6  \times 10^{-40}$ & $2 \times 10^{-40}$\\
$+1$   &  $3  \times 10^{-38}$ & $2 \times 10^{-39}$\\

\end{scotch}
\end{table}

\subsection{Limits on coupling strength in models with interference}

Limits on the \Wprime boson models, taking into account interference effects with the $\tPW$ boson, are set on the \Wprime coupling $g_{\Wprime}$ in terms of the SM \tPW-boson coupling strength $g_{\PW}$.
The corresponding impact on the observable distribution is modeled using a reweighting technique.
Thus, effects such as the influence on the decay width and the impact on interference by the altered coupling are taken into account, as shown in Fig.~\ref{fig:signal} (middle right).
The following interference occurs in the mass range between $M_\tPW$ and $M_{\Wprime}$.
If the coupling of the \Wprime boson has the same sign (SSMS) with respect to the \tPW-boson coupling to left-handed fermions, the interference effect is destructive; in case of opposite sign (SSMO) coupling, it is constructive. For $\MT>M_{\Wprime}$ the effect is vice-versa.

\begin{figure*}[hbtp]
\centering
\includegraphics[width=\cmsFigWidthTwo]{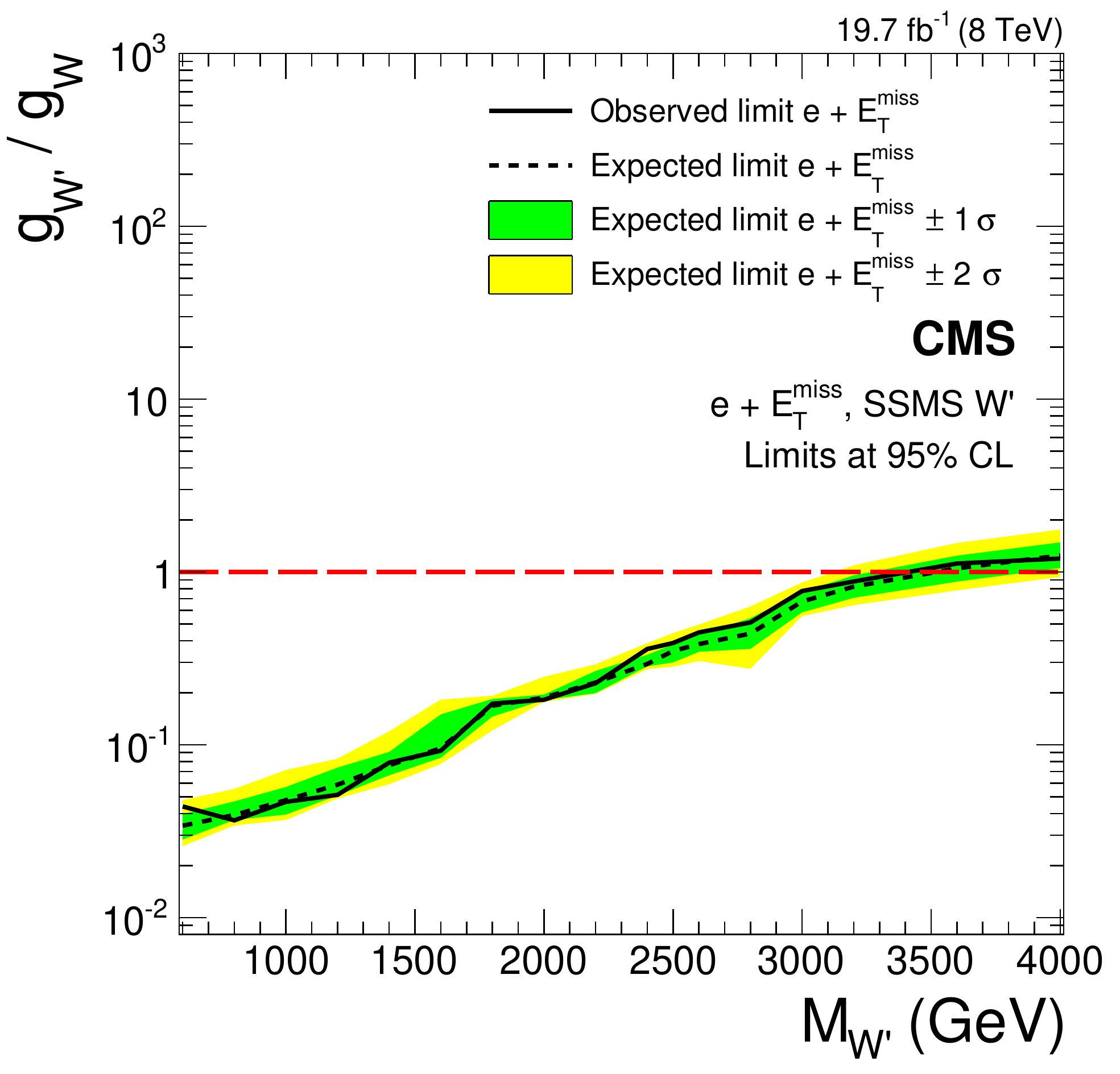}
\includegraphics[width=\cmsFigWidthTwo]{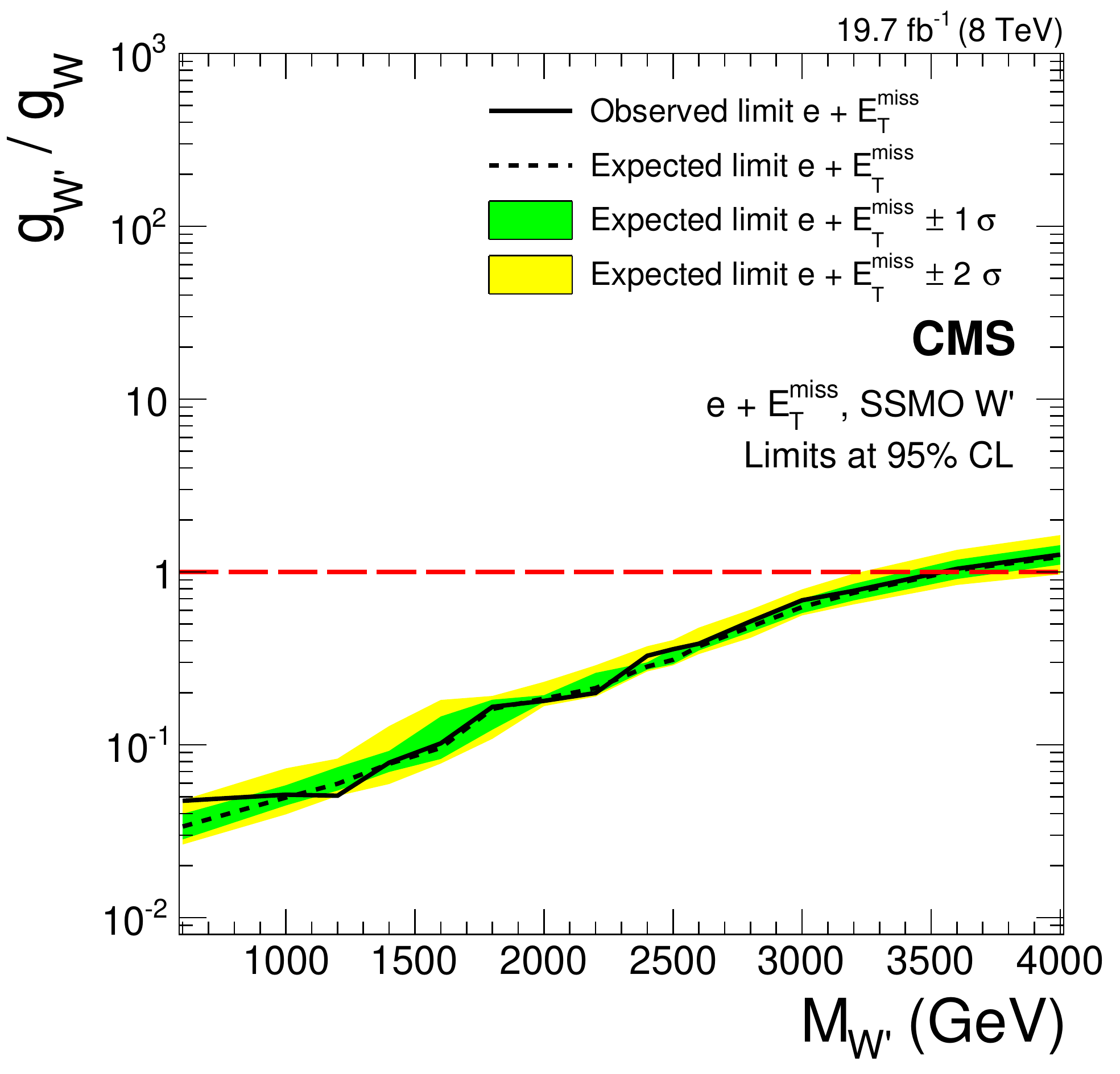}\\
\includegraphics[width=\cmsFigWidthTwo]{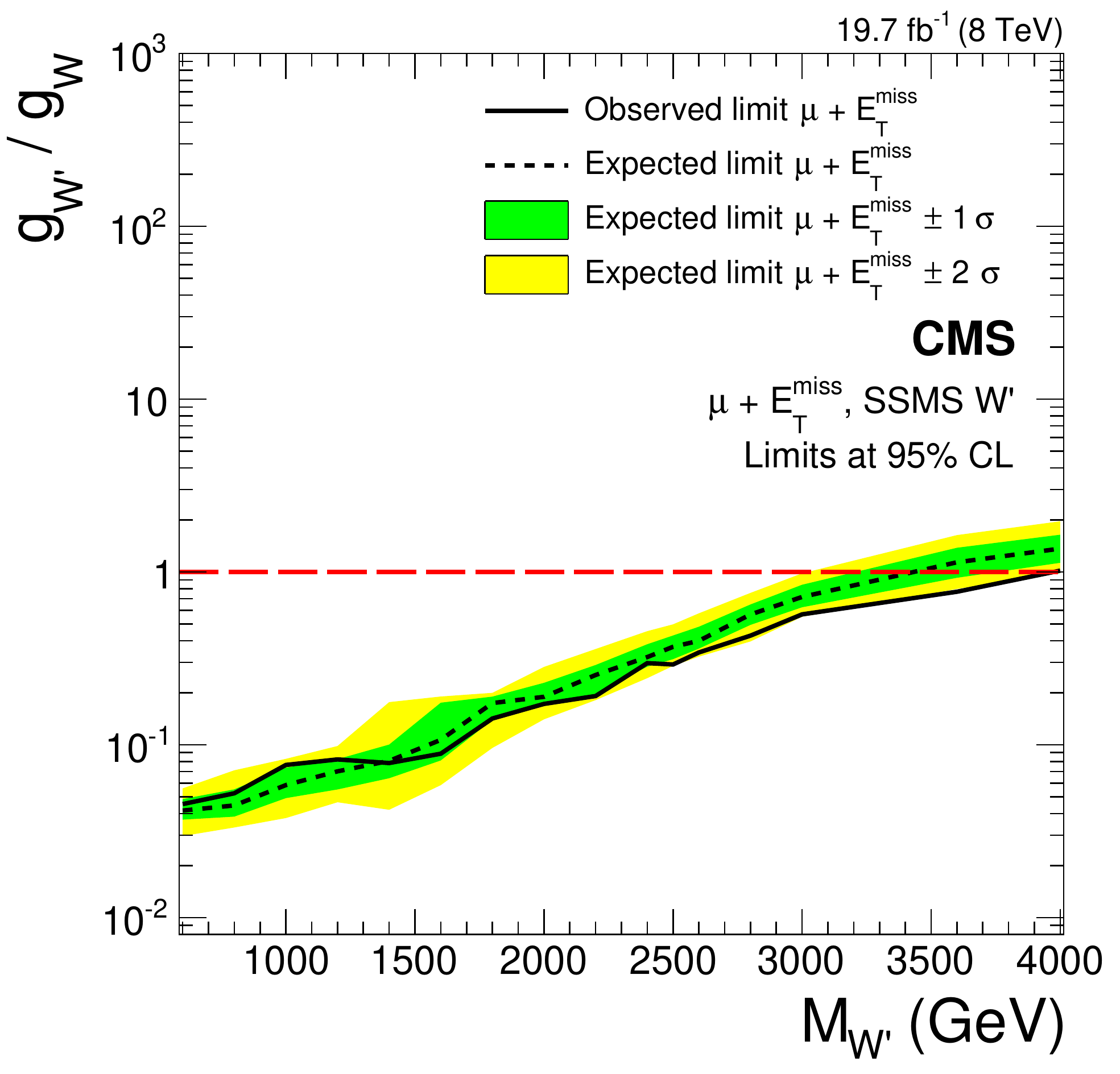}
\includegraphics[width=\cmsFigWidthTwo]{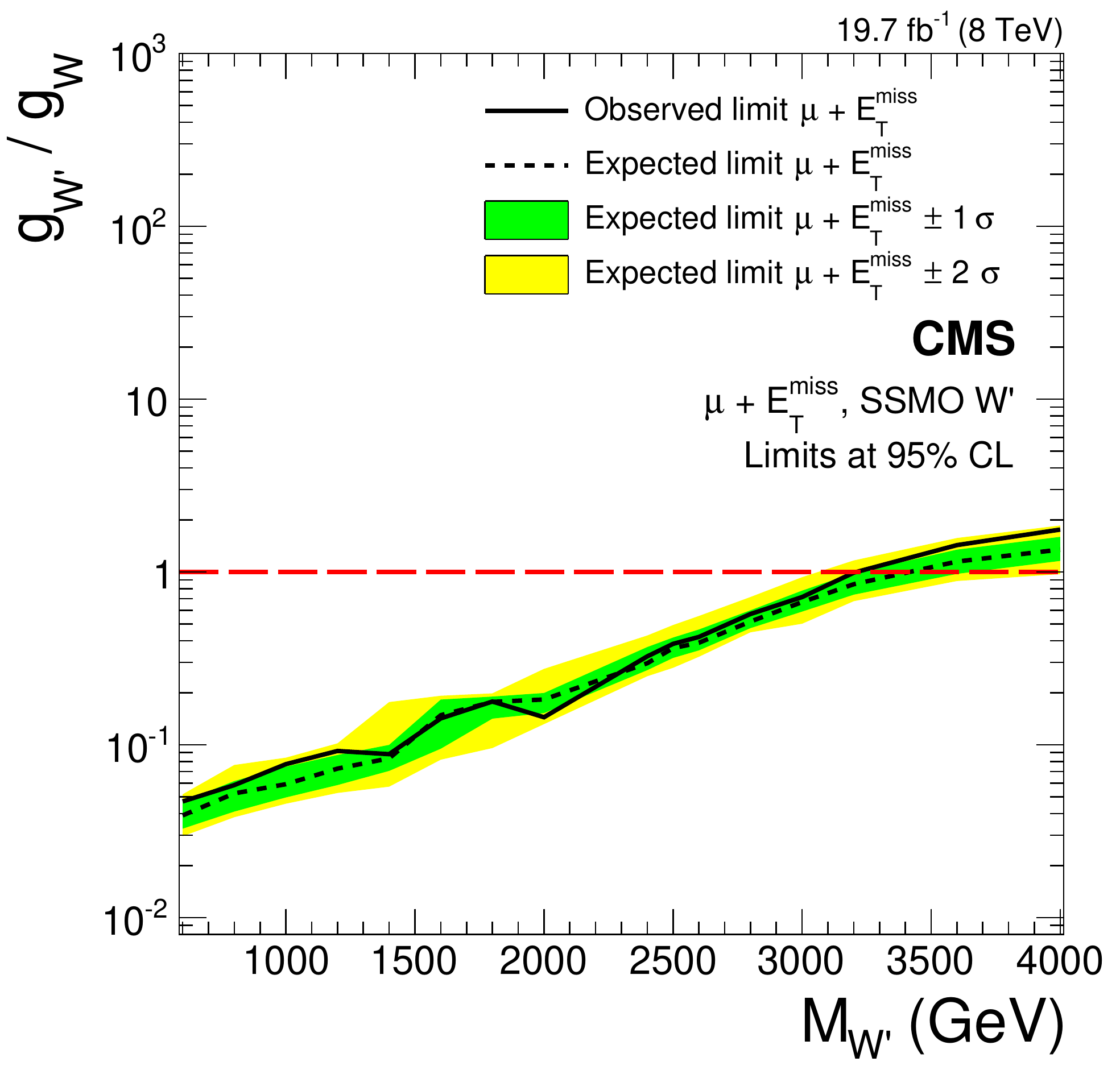}\\
\includegraphics[width=\cmsFigWidthTwo]{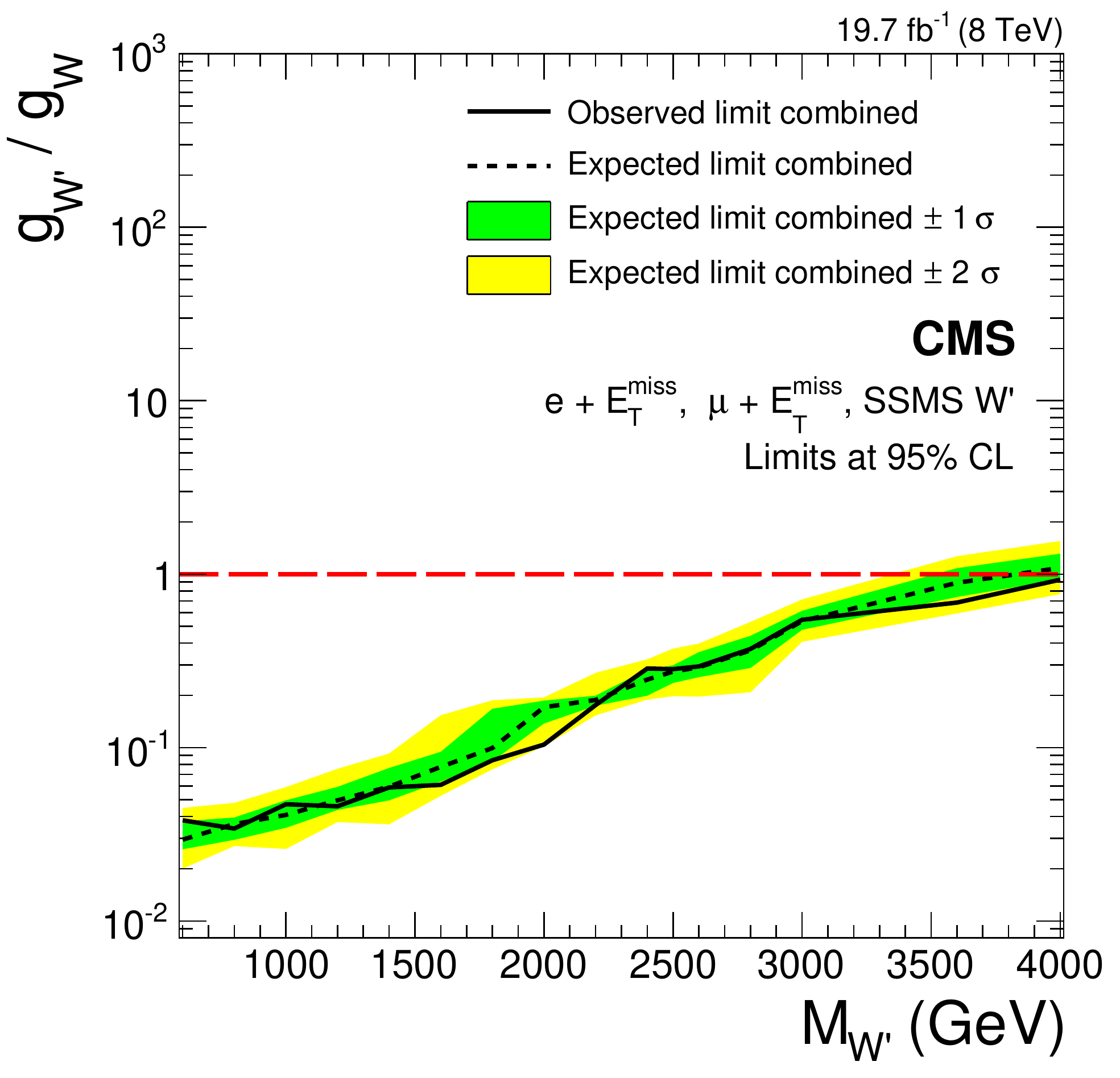}
\includegraphics[width=\cmsFigWidthTwo]{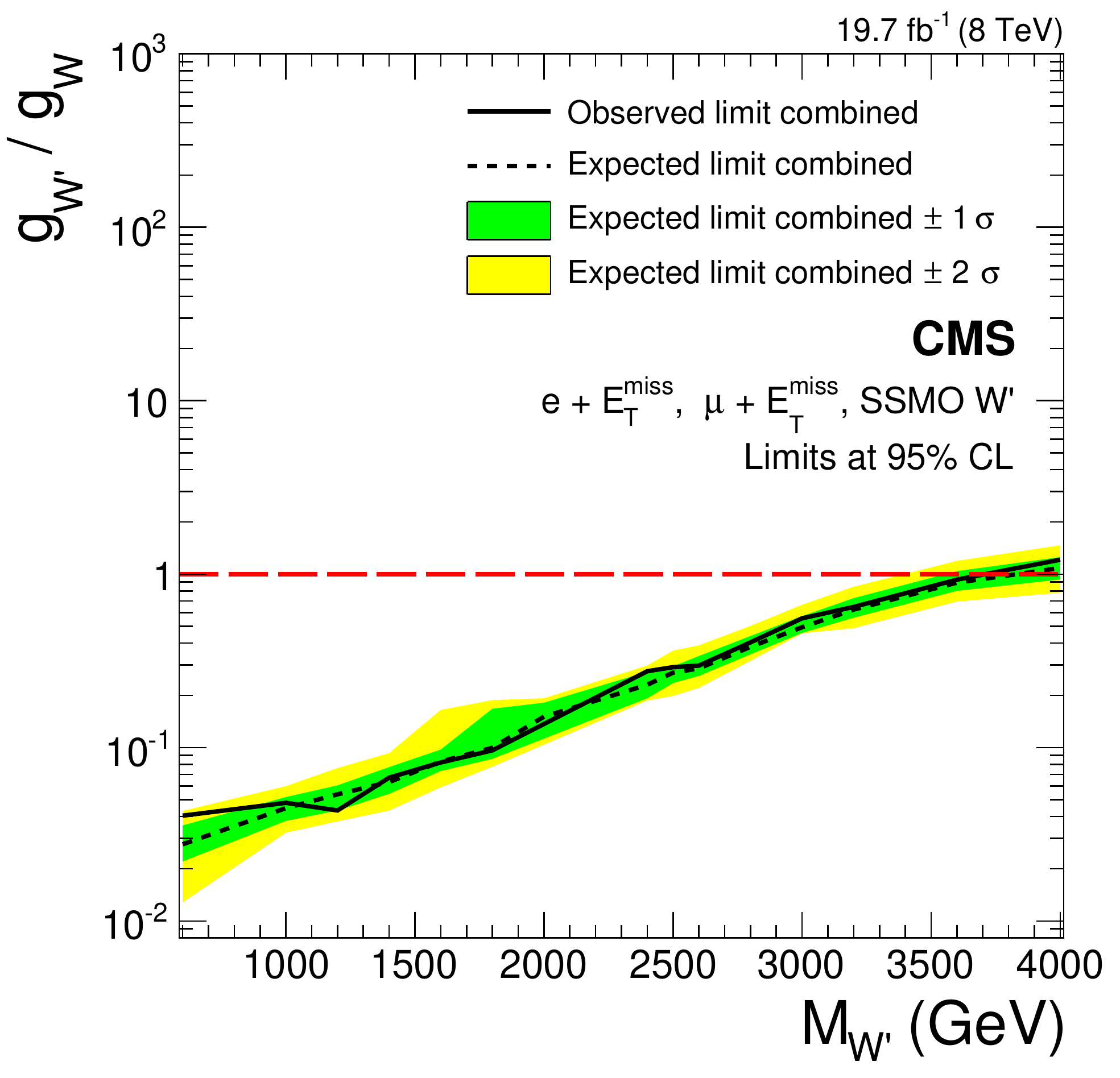}
\caption{Limits on the coupling $g_{\Wprime}$ in terms of the SM coupling $g_{\PW}$ in the electron (top row) and muon (middle row) channels, and their combination (bottom row). Limits for the SSMS model are displayed in the left column, those for the SSMO model in the right column.}
\label{fig:limitinterference}
\end{figure*}

The limit on the coupling strength $g_{\Wprime}$ as a function of the \Wprime-boson mass is shown in Fig.~\ref{fig:limitinterference}. The mass limits in the case where the \Wprime-boson coupling is equal to the \tPW-boson coupling are summarized in Table~\ref{tab:limitinterference}.

\begin{table}[htpb]
\centering
\topcaption{Summary of all SSMS and SSMO exclusion limits on the \Wprime-boson mass in the electron and muon channels, and for their combination assuming $g_{\PW} = g_{\Wprime}$.}
\label{tab:limitinterference}
\begin{scotch}{lccc}
Model & Channel & Observed lower limit & Expected lower limit \\
&&(\TeVns{})&(\TeVns{})\\
\hline
SSMS & $\Pe$       & 3.41 & 3.52  \\
SSMS & $\mu$   & 3.97 & 3.43  \\
SSMS & combined& 4.00 & 3.83  \\
SSMO & $\Pe$       & 3.54 & 3.57  \\
SSMO & $\mu$   & 3.22 & 3.38  \\
SSMO & combined& 3.71 & 3.83  \\
\end{scotch}
\end{table}

\subsection{Interpretation in the split-UED model}

\begin{figure}[hbtp]
\centering
\includegraphics[width=\cmsFigWidth]{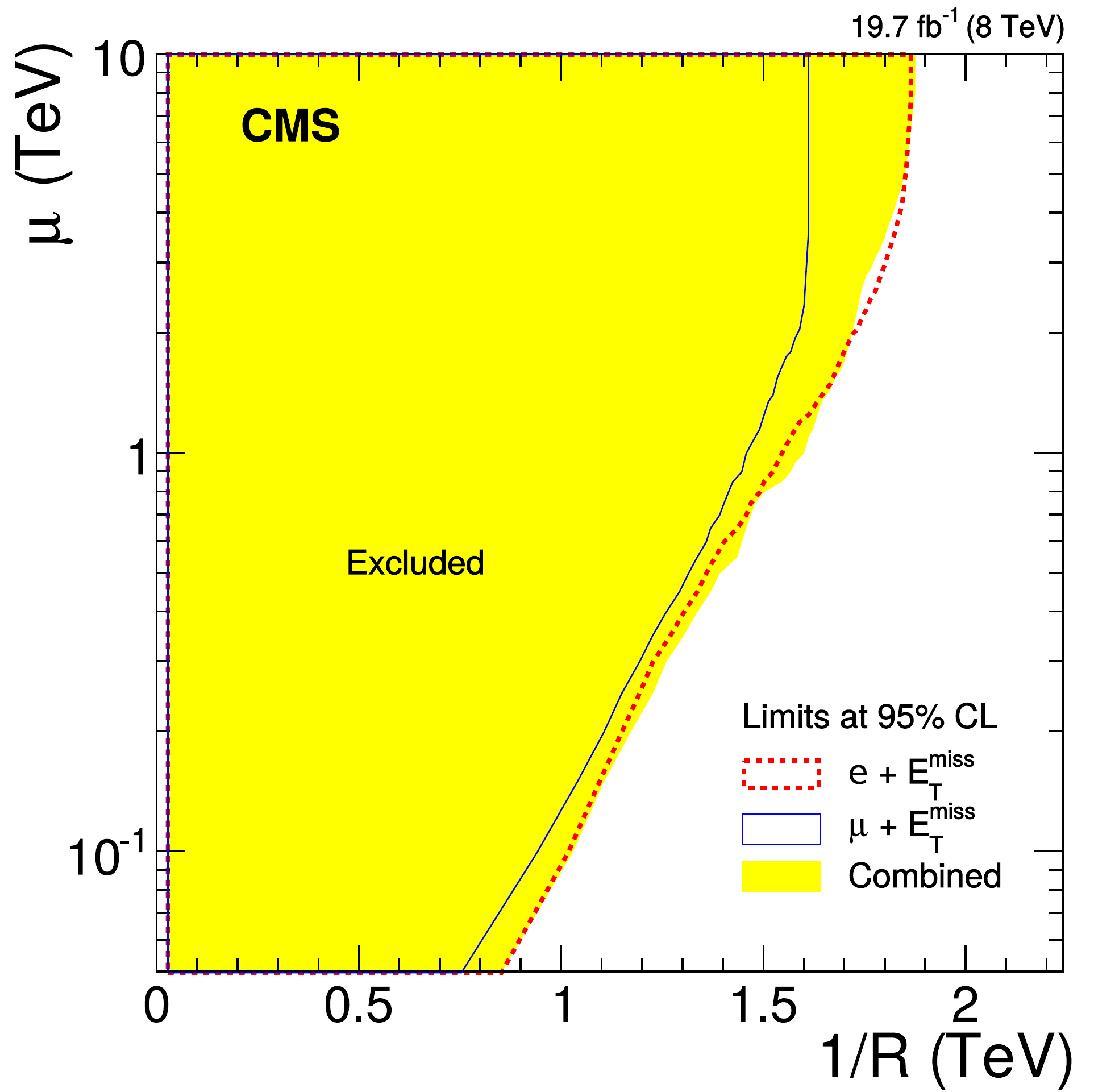}
\caption{Limits on the split-UED parameters $\mu$ and $1/R$, derived from the \Wprime-boson mass limits, taking into account the corresponding width of the \WprimeKKtwo boson.}
\label{fig:UED}
\end{figure}

The observed limits on the SSM \Wprime boson (see Section~\ref{sec:SSMlimit}) can be reinterpreted as limits on the \WprimeKKtwo mass in the framework of split-UED, with the second KK excitation being the only accessible state at present LHC energies with non-zero couplings to SM particles. Figure~\ref{fig:CombLimit2012} shows two examples of \WprimeKKtwo-boson mass limits for values of the bulk mass parameter $\mu$ = 0.05\TeV and $\mu = 10\TeV$. For these two examples, the lower mass limit is 1.74\TeV for $\mu = 0.05\TeV$ and 3.71\TeV for $\mu = 10\TeV$, when combining both channels. The lower limits on the mass can be directly translated into bounds on the split-UED parameter space ($1/R, \mu$) as shown in Fig.~\ref{fig:UED}.

\subsection{Interpretation in the \texorpdfstring{\TeVmOne}{TeV-1} model}

Based on the model-independent cross section limit and the recipe described in
Section~\ref{sec:modelindependentlimit}, a single-bin limit on the compactification scale $M_\mathrm{C}$ of the \TeVmOne model is derived
using the kinematic selection efficiency for this particular model.
The lower bound on $M_\mathrm{C}$ is established as 3.4\TeV. The existing indirect limit on $M_\mathrm{C}$ is 6.8\TeV, which was obtained by fitting results from LEP2, Tevatron, and HERA experiments~\cite{Greg}. The lower limit set by LEP2 experiments is 6.6\TeV, which is the dominant contribution. The sensitivity of this analysis is therefore still less stringent than that based on LEP2 data.

\section{Summary}
\label{sec:summary}

A search for physics beyond the standard model, based on events with a final state containing a charged lepton (electron or muon) and significant missing transverse energy,
has been performed, using proton-proton collision data at $\sqrt{s} = 8\TeV$, corresponding to an integrated luminosity of 19.7\fbinv. No significant deviation from the
standard model expectation has been observed in the transverse mass distribution.

A model-independent upper limit at 95\% \CL on the cross section times branching fraction of additional contributions has been established, ranging from 100 to $0.1\fb$ over
an \MT range of 300\GeV to 2.5\TeV, respectively. The results have been interpreted in the context of various models, as summarized below.

An SSM \Wprime boson that does not interfere with the \tPW boson has been excluded at 95\% \CL for \Wprime-boson masses up to 3.22\,(2.99)\TeV for the electron (muon) channel, where the expected limit is 3.18\,(3.06)\TeV.
When combining both channels, the limit improves to 3.28\TeV.
Lower mass limits in either channel are implicit due to trigger thresholds.

An interpretation in terms of a four-fermion contact interaction yields a limit for the compositeness energy scale $\Lambda$ of 11.3\TeV in the electron channel,
10.9\TeV in the muon channel, and 12.4\TeV for their combination.

Assuming the production of a pair of dark matter particles along with a recoiling \tPW boson that subsequently decays leptonically, the results have been reinterpreted in
terms of an effective dark matter theory.
The effective scale is excluded below 0.3 to 1\TeV, depending on model parameters. This is particularly interesting for low masses of dark matter particles, where the
sensitivity of direct searches is poor.

Building upon earlier versions of this analysis~\cite{2012ichep-limit}, the expected impact of $\tPW$-\Wprime interference on the shape of the $\tPW$ boson \MT distribution is fully taken into account.
Along with the shape, the expected cross section varies, making possible the setting of limits for models with both destructive (SSMS) and constructive (SSMO) interference.

The lower limit on the \Wprime-boson mass is 3.41 (3.97)\TeV in the electron (muon) channels for the SSMS and 3.54 (3.22)\TeV for the SSMO.
For the first time, limits in terms of generalized lepton couplings are given.

An interpretation of the search results has been made in a specific framework of universal extra dimensions
where bulk fermions propagate in the one additional dimension.
The second Kaluza--Klein excitation
\WprimeKKtwo has been excluded for masses below 1.74\TeV, assuming a bulk mass parameter $\mu$ of 0.05\TeV, or for masses below 3.71\TeV, for $\mu = 10\TeV$.
In an alternative model in which only SM gauge bosons propagate in a compactified extra dimension (\TeVmOne model),
a lower bound on the size of the compactified dimension, $M_C$, has been set at 3.40\TeV.

\begin{figure*}[hbtp]
\centering
\includegraphics[width=0.95\textwidth]{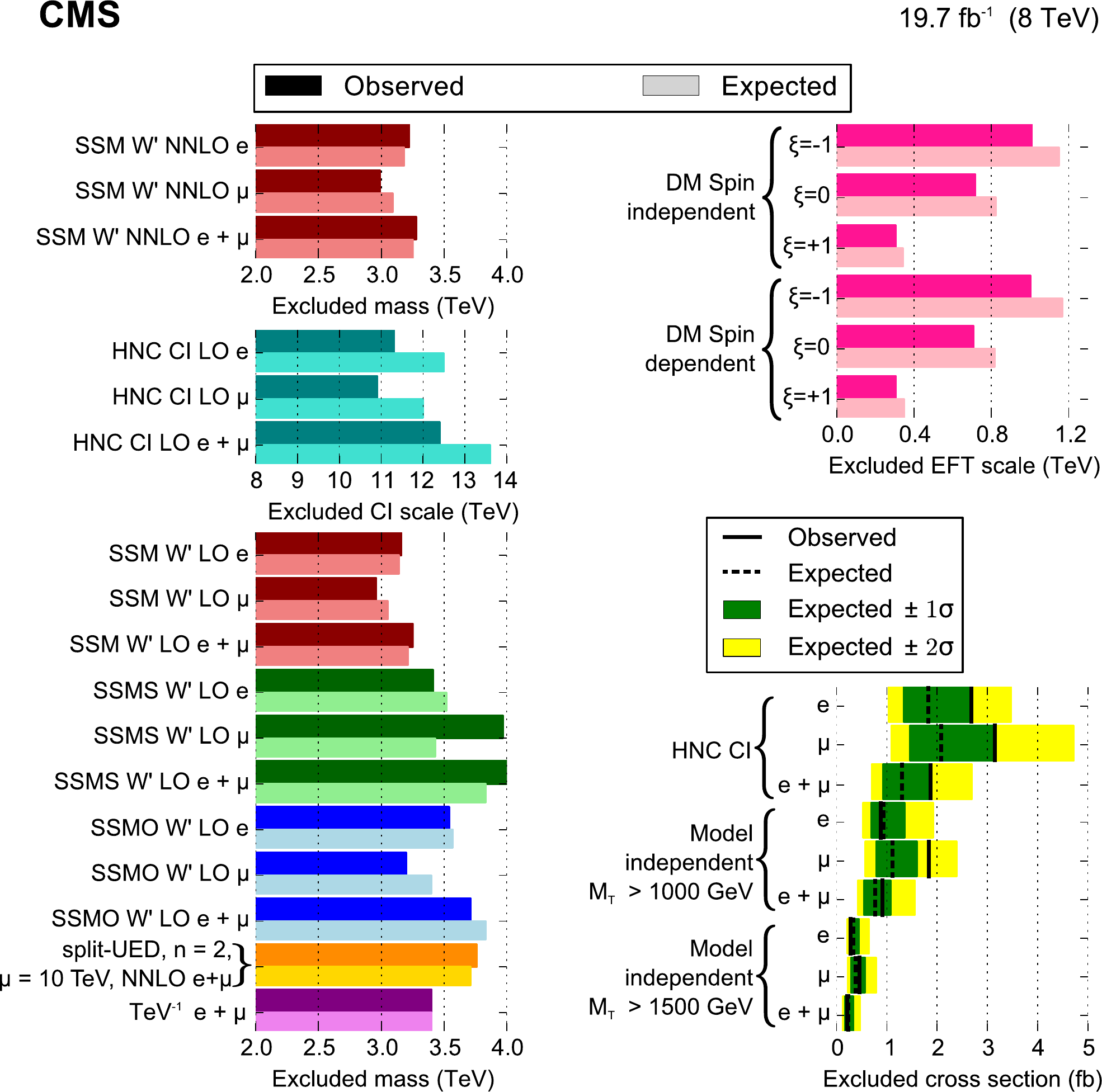}
\caption{Summary of all exclusion limits in the electron and muon channels, and for their combinations. No interference of \Wprime
and $\PW$ bosons is considered in the interpretation labeled as SSM while it is taken into account in the SSMS and SSMO models.
For the HNC contact interaction, the compositeness scale $\Lambda$ is probed.
In the upper rows of the right column, the EFT limits are shown
for the DM interpretations in term of $\Lambda$ for small dark matter masses $M_\chi<100\GeV$.
The reinterpretation in terms of additional extra dimensions is provided in the context of
split-UED, given for a bulk mass parameter $\mu = 10\TeV$, and in the \TeVmOne model.
}
\label{fig:AllLimits}
\end{figure*}

Study of the mono-lepton channel provides a powerful tool to probe for beyond the standard model physics.
All the results of this search are summarized in Fig.~\ref{fig:AllLimits}, including the expected and observed limits.
Fig.~\ref{fig:AllLimits} is structured by theories and the related model parameters (particle mass, compositeness scale $\Lambda$ or dark matter
effective field scale $\Lambda$) as given in Table~\ref{tab:models}.
The three representative signal examples (SSM \Wprime, contact interaction and dark matter) are shown in the upper figure section.
Limits for specific models along with the model independent cross section limit are given in the lower part of the figure.

\begin{acknowledgments}
\hyphenation{Bundes-ministerium Forschungs-gemeinschaft Forschungs-zentren} We congratulate our colleagues in the CERN accelerator departments for the excellent performance
of the LHC and thank the technical and administrative staffs at CERN and at other CMS institutes for their contributions to the success of the CMS effort. In addition, we
gratefully acknowledge the computing centers and personnel of the Worldwide LHC Computing Grid for delivering so effectively the computing infrastructure essential to our
analyses. Finally, we acknowledge the enduring support for the construction and operation of the LHC and the CMS detector provided by the following funding agencies: the
Austrian Federal Ministry of Science, Research and Economy and the Austrian Science Fund; the Belgian Fonds de la Recherche Scientifique, and Fonds voor Wetenschappelijk
Onderzoek; the Brazilian Funding Agencies (CNPq, CAPES, FAPERJ, and FAPESP); the Bulgarian Ministry of Education and Science; CERN; the Chinese Academy of Sciences,
Ministry of Science and Technology, and National Natural Science Foundation of China; the Colombian Funding Agency (COLCIENCIAS); the Croatian Ministry of Science,
Education and Sport, and the Croatian Science Foundation; the Research Promotion Foundation, Cyprus; the Ministry of Education and Research, Estonian Research Council via
IUT23-4 and IUT23-6 and European Regional Development Fund, Estonia; the Academy of Finland, Finnish Ministry of Education and Culture, and Helsinki Institute of Physics;
the Institut National de Physique Nucl\'eaire et de Physique des Particules~/~CNRS, and Commissariat \`a l'\'Energie Atomique et aux \'Energies Alternatives~/~CEA, France;
the Bundesministerium f\"ur Bildung und Forschung, Deutsche Forschungsgemeinschaft, and Helmholtz-Gemeinschaft Deutscher Forschungszentren, Germany; the General Secretariat
for Research and Technology, Greece; the National Scientific Research Foundation, and National Innovation Office, Hungary; the Department of Atomic Energy and the
Department of Science and Technology, India; the Institute for Studies in Theoretical Physics and Mathematics, Iran; the Science Foundation, Ireland; the Istituto Nazionale
di Fisica Nucleare, Italy; the Korean Ministry of Education, Science and Technology and the World Class University program of NRF, Republic of Korea; the Lithuanian Academy
of Sciences; the Ministry of Education, and University of Malaya (Malaysia); the Mexican Funding Agencies (CINVESTAV, CONACYT, SEP, and UASLP-FAI); the Ministry of
Business, Innovation and Employment, New Zealand; the Pakistan Atomic Energy Commission; the Ministry of Science and Higher Education and the National Science Centre,
Poland; the Funda\c{c}\~ao para a Ci\^encia e a Tecnologia, Portugal; JINR, Dubna; the Ministry of Education and Science of the Russian Federation, the Federal Agency of
Atomic Energy of the Russian Federation, Russian Academy of Sciences, and the Russian Foundation for Basic Research; the Ministry of Education, Science and Technological
Development of Serbia; the Secretar\'{\i}a de Estado de Investigaci\'on, Desarrollo e Innovaci\'on and Programa Consolider-Ingenio 2010, Spain; the Swiss Funding Agencies
(ETH Board, ETH Zurich, PSI, SNF, UniZH, Canton Zurich, and SER); the Ministry of Science and Technology, Taipei; the Thailand Center of Excellence in Physics, the
Institute for the Promotion of Teaching Science and Technology of Thailand, Special Task Force for Activating Research and the National Science and Technology Development
Agency of Thailand; the Scientific and Technical Research Council of Turkey, and Turkish Atomic Energy Authority; the National Academy of Sciences of Ukraine, and State
Fund for Fundamental Researches, Ukraine; the Science and Technology Facilities Council, UK; the US Department of Energy, and the US National Science Foundation.

Individuals have received support from the Marie-Curie program and the European Research Council and EPLANET (European Union); the Leventis Foundation; the A. P. Sloan
Foundation; the Alexander von Humboldt Foundation; the Belgian Federal Science Policy Office; the Fonds pour la Formation \`a la Recherche dans l'Industrie et dans
l'Agriculture (FRIA-Belgium); the Agentschap voor Innovatie door Wetenschap en Technologie (IWT-Belgium); the Ministry of Education, Youth and Sports (MEYS) of the Czech
Republic; the Council of Science and Industrial Research, India; the HOMING PLUS program of Foundation for Polish Science, cofinanced from European Union, Regional
Development Fund; the Compagnia di San Paolo (Torino); the Consorzio per la Fisica (Trieste); MIUR project 20108T4XTM (Italy); the Thalis and Aristeia programs cofinanced
by EU-ESF and the Greek NSRF; and the National Priorities Research Program by Qatar National Research Fund.
\end{acknowledgments}

\bibliography{auto_generated}

\providecommand{\href}[2]{#2}\begingroup\raggedright\begin{thebibliography}{10}%
\makeatletter
\providecommand{\hrefCMSnoop }[0]{\@secondoftwo}%
\makeatother
\providecommand{\doi}{\texttt{doi:}\begingroup \urlstyle{tt}\Url}

\bibitem{reference-model}
\hrefCMSnoop {}{G.~Altarelli, B.~Mele, and M.~Ruiz-Altaba, ``Searching for new
  heavy vector bosons in {$\Pp\Pap$} colliders'',} \textit{ Z. Phys. C}
  \textbf{ 45} (1989) 109,
  \href{http://dx.doi.org/10.1007/BF01556677}{\doi{10.1007/BF01556677}}.

\bibitem{HNCM-CI}
\hrefCMSnoop {}{K.~D. Lane, F.~E. Paige, T.~Skwarnicki, and W.~J. Womersley,
  ``Simulations of supercollider physics'',} \textit{ Phys. Rept.} \textbf{
  278} (1997) 291,
  \href{http://dx.doi.org/10.1016/S0370-1573(96)00018-X}{\doi{10.1016/S0370-1573(96)00018-X}},
\href{http://www.arXiv.org/abs/hep-ph/9412280}{\texttt{arXiv:hep-ph/9412280}}.

\bibitem{DMModelSummery}
J.~Goodman\hrefCMSnoop {}{ {et~al.}, ``Constraints on light {Majorana} dark
  matter from colliders'',} \textit{ Phys. Lett. B} \textbf{ 695} (2011) 185,
  \href{http://dx.doi.org/10.1016/j.physletb.2010.11.009}{\doi{10.1016/j.physletb.2010.11.009}},
\href{http://www.arXiv.org/abs/1005.1286}{\texttt{arXiv:1005.1286}}.

\bibitem{DMModelSummery2}
J.~Goodman\hrefCMSnoop {}{ {et~al.}, ``Constraints on dark matter from
  colliders'',} \textit{ Phys. Rev. D} \textbf{ 82} (2010) 116010,
  \href{http://dx.doi.org/10.1103/PhysRevD.82.116010}{\doi{10.1103/PhysRevD.82.116010}},
\href{http://www.arXiv.org/abs/1008.1783}{\texttt{arXiv:1008.1783}}.

\bibitem{DMSummery}
\hrefCMSnoop {}{M.~Beltran, D.~Hooper, E.~W. Kolb, and Z.~A.~C. Krusberg,
  ``{Deducing the nature of dark matter from direct and indirect detection
  experiments in the absence of collider signatures of new physics}'',}
  \textit{ Phys. Rev. D} \textbf{ 80} (2009) 043509,
  \href{http://dx.doi.org/10.1103/PhysRevD.80.043509}{\doi{10.1103/PhysRevD.80.043509}},
\href{http://www.arXiv.org/abs/0808.3384}{\texttt{arXiv:0808.3384}}.

\bibitem{criticalPaper}
E.~Accomando\hrefCMSnoop {}{ {et~al.}, ``{Interference effects in heavy
  W$'$-boson searches at the LHC}'',} \textit{ Phys. Rev. D} \textbf{ 85}
  (2012) 115017,
  \href{http://dx.doi.org/10.1103/PhysRevD.85.115017}{\doi{10.1103/PhysRevD.85.115017}},
\href{http://www.arXiv.org/abs/1110.0713}{\texttt{arXiv:1110.0713}}.

\bibitem{WprimeHelicity}
\hrefCMSnoop {}{T.~G. Rizzo, ``The determination of the helicity of {\Wprime}
  boson couplings at the {LHC}'',} \textit{ JHEP} \textbf{ 05} (2007) 037,
  \href{http://dx.doi.org/10.1088/1126-6708/2007/05/037}{\doi{10.1088/1126-6708/2007/05/037}}.

\bibitem{dudkov}
\hrefCMSnoop {}{E.~Boos, V.~Bunichev, L.~Dudko, and M.~Perfilov,
  ``{Interference between $W^\prime$ and $W$ in single-top quark production
  processes}'',} \textit{ Phys. Lett. B} \textbf{ 655} (2007) 245,
  \href{http://dx.doi.org/10.1016/j.physletb.2007.03.064}{\doi{10.1016/j.physletb.2007.03.064}},
\href{http://www.arXiv.org/abs/hep-ph/0610080}{\texttt{arXiv:hep-ph/0610080}}.

\bibitem{PhysRevD.79.091702}
C.-R. Chen\hrefCMSnoop {}{ {et~al.}, ``{Dark matter and collider phenomenology
  of split-UED}'',} \textit{ JHEP} \textbf{ 09} (2009) 078,
  \href{http://dx.doi.org/10.1088/1126-6708/2009/09/078}{\doi{10.1088/1126-6708/2009/09/078}},
\href{http://www.arXiv.org/abs/0903.1971}{\texttt{arXiv:0903.1971}}.

\bibitem{JHEP04(2010)081}
\hrefCMSnoop {}{K.~Kong, S.~C. Park, and T.~G. Rizzo, ``{Collider phenomenology
  with Split-UED}'',} \textit{ JHEP} \textbf{ 04} (2010) 081,
  \href{http://dx.doi.org/10.1007/JHEP04(2010)081}{\doi{10.1007/JHEP04(2010)081}},
\href{http://www.arXiv.org/abs/1002.0602}{\texttt{arXiv:1002.0602}}.

\bibitem{TeVref1}
\hrefCMSnoop {}{K.~R. Dienes, E.~Dudas, and T.~Gherghetta, ``{Grand unification
  at intermediate mass scales through extra dimensions}'',} \textit{ Nucl.
  Phys. B} \textbf{ 537} (1999) 47,
  \href{http://dx.doi.org/10.1016/S0550-3213(98)00669-5}{\doi{10.1016/S0550-3213(98)00669-5}},
\href{http://www.arXiv.org/abs/hep-ph/9806292}{\texttt{arXiv:hep-ph/9806292}}.

\bibitem{TeVref2}
\hrefCMSnoop {}{A.~Pomarol and M.~Quiros, ``The standard model from extra
  dimensions'',} \textit{ Phys. Lett. B} \textbf{ 438} (1998) 255,
  \href{http://dx.doi.org/10.1016/S0370-2693(98)00979-4}{\doi{10.1016/S0370-2693(98)00979-4}},
\href{http://www.arXiv.org/abs/hep-ph/9806263}{\texttt{arXiv:hep-ph/9806263}}.

\bibitem{TeVref3}
\hrefCMSnoop {}{M.~Masip and A.~Pomarol, ``{Effects of standard model
  Kaluza-Klein excitations on electroweak observables}'',} \textit{ Phys. Rev.
  D} \textbf{ 60} (1999) 096005,
  \href{http://dx.doi.org/10.1103/PhysRevD.60.096005}{\doi{10.1103/PhysRevD.60.096005}},
\href{http://www.arXiv.org/abs/hep-ph/9902467}{\texttt{arXiv:hep-ph/9902467}}.

\bibitem{TeVref4}
\hrefCMSnoop {}{I.~Antoniadis, K.~Benakli, and M.~Quiros, ``{Direct collider
  signatures of large extra dimensions}'',} \textit{ Phys. Lett. B} \textbf{
  460} (1999) 176,
  \href{http://dx.doi.org/10.1016/S0370-2693(99)00764-9}{\doi{10.1016/S0370-2693(99)00764-9}},
\href{http://www.arXiv.org/abs/hep-ph/9905311}{\texttt{arXiv:hep-ph/9905311}}.

\bibitem{d0-limit}
\hrefCMSnoop {}{{D0} Collaboration, ``{Search for $W^\prime$ bosons decaying to
  an electron and a neutrino with the D0 detector}'',} \textit{ Phys. Rev.
  Lett.} \textbf{ 100} (2008) 031804,
  \href{http://dx.doi.org/10.1103/PhysRevLett.100.031804}{\doi{10.1103/PhysRevLett.100.031804}},
  \href{http://www.arXiv.org/abs/0710.2966}{\texttt{arXiv:0710.2966}}.

\bibitem{cdf-limit}
\hrefCMSnoop {}{{CDF} Collaboration, ``{Search for a new heavy gauge boson
  $W^\prime$ with event signature electron + missing transverse energy in
  $\Pp\Pap$ collisions at $\sqrt{s} = 1.96$ TeV}'',} \textit{ Phys. Rev. D}
  \textbf{ 83} (2011) 031102,
  \href{http://dx.doi.org/10.1103/PhysRevD.83.031102}{\doi{10.1103/PhysRevD.83.031102}},
\href{http://www.arXiv.org/abs/1012.5145}{\texttt{arXiv:1012.5145}}.

\bibitem{atlas-2010}
\hrefCMSnoop {}{{ATLAS} Collaboration, ``{Search for high-mass states with one
  lepton plus missing transverse momentum in proton-proton collisions at
  $\sqrt{s} = 7$ TeV with the ATLAS detector}'',} \textit{ Phys. Lett. B}
  \textbf{ 701} (2011) 50,
  \href{http://dx.doi.org/10.1016/j.physletb.2011.05.043}{\doi{10.1016/j.physletb.2011.05.043}},
\href{http://www.arXiv.org/abs/1103.1391}{\texttt{arXiv:1103.1391}}.

\bibitem{atlas-1fb-limit}
\hrefCMSnoop {}{{ATLAS} Collaboration, ``{Search for a heavy gauge boson
  decaying to a charged lepton and a neutrino in 1 \fbinv of pp collisions at
  $\sqrt{s} = 7$ TeV using the ATLAS detector}'',} \textit{ Phys. Lett. B}
  \textbf{ 705} (2011) 28,
  \href{http://dx.doi.org/10.1016/j.physletb.2011.09.093}{\doi{10.1016/j.physletb.2011.09.093}},
\href{http://www.arXiv.org/abs/1108.1316}{\texttt{arXiv:1108.1316}}.

\bibitem{atlas-2011limit}
\hrefCMSnoop {}{{ATLAS} Collaboration, ``{ATLAS search for a heavy gauge boson
  decaying to a charged lepton and a neutrino in $\text{pp}$ collisions at
  $\sqrt{s}=7$ TeV}'',} \textit{ Eur. Phys. J. C} \textbf{ 72} (2012) 2241,
  \href{http://dx.doi.org/10.1140/epjc/s10052-012-2241-5}{\doi{10.1140/epjc/s10052-012-2241-5}},
\href{http://www.arXiv.org/abs/1209.4446}{\texttt{arXiv:1209.4446}}.

\bibitem{ATLAS:2014wra}
\hrefCMSnoop {}{{ATLAS} Collaboration, ``{Search for new particles in events
  with one lepton and missing transverse momentum in $pp$ collisions at
  $\sqrt{s}$ = 8 TeV with the ATLAS detector}'',} \textit{ JHEP} \textbf{ 09}
  (2014) 037,
  \href{http://dx.doi.org/10.1007/JHEP09(2014)037}{\doi{10.1007/JHEP09(2014)037}},
\href{http://www.arXiv.org/abs/1407.7494}{\texttt{arXiv:1407.7494}}.

\bibitem{2010cms-electron-limit}
\hrefCMSnoop {}{{CMS} Collaboration, ``{Search for a heavy gauge boson W$'$ in
  the final state with an electron and large missing transverse energy in pp
  collisions at $\sqrt{s} = 7$ TeV}'',} \textit{ Phys. Lett. B} \textbf{ 698}
  (2011) 21,
  \href{http://dx.doi.org/10.1016/j.physletb.2011.02.048}{\doi{10.1016/j.physletb.2011.02.048}},
\href{http://www.arXiv.org/abs/1012.5945}{\texttt{arXiv:1012.5945}}.

\bibitem{2010cms-muon-limit}
\hrefCMSnoop {}{{CMS} Collaboration, ``{Search for a $W^\prime$ boson decaying
  to a muon and a neutrino in $\text{pp}$ collisions at $\sqrt{s} = 7$ TeV}'',}
  \textit{ Phys. Lett. B} \textbf{ 701} (2011) 160,
  \href{http://dx.doi.org/10.1016/j.physletb.2011.05.048}{\doi{10.1016/j.physletb.2011.05.048}},
\href{http://www.arXiv.org/abs/1103.0030}{\texttt{arXiv:1103.0030}}.

\bibitem{2011limit}
\hrefCMSnoop {}{{CMS} Collaboration, ``{Search for leptonic decays of W$'$
  bosons in pp collisions at $\sqrt{s}=7$ TeV}'',} \textit{ JHEP} \textbf{ 08}
  (2012) 023,
  \href{http://dx.doi.org/10.1007/JHEP08(2012)023}{\doi{10.1007/JHEP08(2012)023}},
\href{http://www.arXiv.org/abs/1204.4764}{\texttt{arXiv:1204.4764}}.

\bibitem{2012ichep-limit}
\hrefCMSnoop {}{{CMS} Collaboration, ``{Search for new physics in final states
  with a lepton and missing transverse energy in pp collisions at the LHC}'',}
  \textit{ Phys. Rev. D} \textbf{ 87} (2013) 072005,
  \href{http://dx.doi.org/10.1103/PhysRevD.87.072005}{\doi{10.1103/PhysRevD.87.072005}},
\href{http://www.arXiv.org/abs/1302.2812}{\texttt{arXiv:1302.2812}}.

\bibitem{LUM-13-001}
\href {http://cdsweb.cern.ch/record/1598864}{{CMS} Collaboration, ``CMS
  Luminosity Based on Pixel Cluster Counting - Summer 2013 Update'',} CMS
  Physics Analysis Summary CMS-PAS-LUM-13-001, 2013.

\bibitem{Chatrchyan:2008zzk}
\hrefCMSnoop {}{{CMS} Collaboration, ``The {CMS} experiment at the {CERN}
  {LHC}'',} \textit{ JINST} \textbf{ 3} (2008) S08004,
\href{http://dx.doi.org/10.1088/1748-0221/3/08/S08004}{\doi{10.1088/1748-0221/3/08/S08004}}.

\bibitem{Chatrchyan:2013dga}
\hrefCMSnoop {}{{CMS} Collaboration, ``{Energy calibration and resolution of
  the CMS electromagnetic calorimeter in pp collisions at $\sqrt{s} = 7$
  TeV}'',} \textit{ JINST} \textbf{ 8} (2013) P09009,
  \href{http://dx.doi.org/10.1088/1748-0221/8/09/P09009}{\doi{10.1088/1748-0221/8/09/P09009}},
\href{http://www.arXiv.org/abs/1306.2016}{\texttt{arXiv:1306.2016}}.

\bibitem{EXO-12-061}
\hrefCMSnoop {}{{CMS} Collaboration, ``{Search for heavy narrow dilepton
  resonances in pp collisions at $\sqrt{s}=7$ TeV and $\sqrt{s}=8$ TeV}'',}
  \textit{ Phys. Lett. B} \textbf{ 720} (2013) 63,
  \href{http://dx.doi.org/10.1016/j.physletb.2013.02.003}{\doi{10.1016/j.physletb.2013.02.003}},
\href{http://www.arXiv.org/abs/1212.6175}{\texttt{arXiv:1212.6175}}.

\bibitem{Chatrchyan:2012xi}
\hrefCMSnoop {}{{CMS} Collaboration, ``Performance of {CMS} muon reconstruction
  in pp collision events at {$\sqrt{s} = 7$\TeV}'',} \textit{ JINST} \textbf{
  7} (2012) P10002,
  \href{http://dx.doi.org/10.1088/1748-0221/7/10/P10002}{\doi{10.1088/1748-0221/7/10/P10002}},
\href{http://www.arXiv.org/abs/1206.4071}{\texttt{arXiv:1206.4071}}.

\bibitem{Agostinelli:2002hh}
\hrefCMSnoop {}{{GEANT4} Collaboration, ``{GEANT4---a simulation toolkit}'',}
  \textit{ Nucl. Instrum. Meth. A} \textbf{ 506} (2003) 250,
\href{http://dx.doi.org/10.1016/S0168-9002(03)01368-8}{\doi{10.1016/S0168-9002(03)01368-8}}.

\bibitem{Allison:2006ve}
\hrefCMSnoop {}{J.~Allison {et~al.}, ``{Geant4 developments and
  applications}'',} \textit{ IEEE Trans. Nucl. Sci.} \textbf{ 53} (2006) 270,
\href{http://dx.doi.org/10.1109/TNS.2006.869826}{\doi{10.1109/TNS.2006.869826}}.

\bibitem{tbanalysis-d0}
\hrefCMSnoop {}{{D0} Collaboration, ``{Search for $W^\prime$ Boson Resonances
  Decaying to a Top Quark and a Bottom Quark}'',} \textit{ Phys. Rev. Lett.}
  \textbf{ 100} (2008) 211803,
  \href{http://dx.doi.org/10.1103/PhysRevLett.100.211803}{\doi{10.1103/PhysRevLett.100.211803}},
\href{http://www.arXiv.org/abs/0803.3256}{\texttt{arXiv:0803.3256}}.

\bibitem{WprimeTBCMS}
\hrefCMSnoop {}{{CMS} Collaboration, ``{Search for W$'$ $\to$ tb decays in the
  lepton + jets final state in pp collisions at $\sqrt{s}$ = 8 TeV}'',}
  \textit{ JHEP} \textbf{ 05} (2014) 108,
  \href{http://dx.doi.org/10.1007/JHEP05(2014)108}{\doi{10.1007/JHEP05(2014)108}},
\href{http://www.arXiv.org/abs/1402.2176}{\texttt{arXiv:1402.2176}}.

\bibitem{WprimeTBAtlas}
\hrefCMSnoop {}{{ATLAS} Collaboration, ``{Search for tb resonances in
  proton-proton collisions at $\sqrt{s} = 7$ TeV with the ATLAS detector}'',}
  \textit{ Phys. Rev. Lett.} \textbf{ 109} (2012) 081801,
  \href{http://dx.doi.org/10.1103/PhysRevLett.109.081801}{\doi{10.1103/PhysRevLett.109.081801}},
\href{http://www.arXiv.org/abs/1205.1016}{\texttt{arXiv:1205.1016}}.

\bibitem{WprimeTBCMSCharge}
\hrefCMSnoop {}{{CMS} Collaboration, ``{Search for charge-asymmetric production
  of W$'$ bosons in top pair + jet events from pp collisions at $\sqrt{s} = 7$
  TeV}'',} \textit{ Phys. Lett. B} \textbf{ 717} (2012) 351,
  \href{http://dx.doi.org/10.1016/j.physletb.2012.09.048}{\doi{10.1016/j.physletb.2012.09.048}},
\href{http://www.arXiv.org/abs/1206.3921}{\texttt{arXiv:1206.3921}}.

\bibitem{WprimeTBCMS7TeV}
\hrefCMSnoop {}{{CMS} Collaboration, ``{Search for a W$'$ boson decaying to a
  bottom quark and a top quark in pp collisions at $\sqrt{s} = 7$ TeV}'',}
  \textit{ Phys. Lett. B} \textbf{ 718} (2013) 1229,
  \href{http://dx.doi.org/10.1016/j.physletb.2012.12.008}{\doi{10.1016/j.physletb.2012.12.008}},
\href{http://www.arXiv.org/abs/1208.0956}{\texttt{arXiv:1208.0956}}.

\bibitem{Aad:2014xra}
\hrefCMSnoop {}{{ATLAS} Collaboration, ``{Search for $W' \rightarrow tb
  \rightarrow qqbb$ Decays in pp Collisions at $\sqrt{s}$ = 8 TeV with the
  ATLAS Detector}'',} (2014).
  \href{http://www.arXiv.org/abs/1408.0886}{\texttt{arXiv:1408.0886}}.
Submitted to Eur. Phys. J. C.

\bibitem{WprimeWZCMS}
\hrefCMSnoop {}{{CMS} Collaboration, ``{Search for a W$'$ or Techni-$\rho$
  Decaying into WZ in pp Collisions at $\sqrt{s} = 7$ TeV}'',} \textit{ Phys.
  Rev. Lett.} \textbf{ 109} (2012) 141801,
  \href{http://dx.doi.org/10.1103/PhysRevLett.109.141801}{\doi{10.1103/PhysRevLett.109.141801}},
\href{http://www.arXiv.org/abs/1206.0433}{\texttt{arXiv:1206.0433}}.

\bibitem{Khachatryan:2014xja}
\hrefCMSnoop {}{{CMS} Collaboration, ``{Search for new resonances decaying via
  WZ to leptons in proton-proton collisions at $\sqrt{s} = 8$\TeV}'',} \textit{
  Phys. Lett. B} \textbf{ 740} (2014) 83,
  \href{http://dx.doi.org/10.1016/j.physletb.2014.11.026}{\doi{10.1016/j.physletb.2014.11.026}},
\href{http://www.arXiv.org/abs/1407.3476}{\texttt{arXiv:1407.3476}}.

\bibitem{FourthColor}
\hrefCMSnoop {}{J.~C. Pati and A.~Salam, ``Lepton number as the fourth
  `color''',} \textit{ Phys. Rev. D} \textbf{ 10} (1974) 275,
  \href{http://dx.doi.org/10.1103/PhysRevD.10.275}{\doi{10.1103/PhysRevD.10.275}}.
  [Erratum: \DOI{10.1103/PhysRevD.11.703.2}].

\bibitem{LeftRightNatur}
\hrefCMSnoop {}{R.~N. Mohapatra and J.~C. Pati, ````{Natural}'' left-right
  symmetry'',} \textit{ Phys. Rev. D} \textbf{ 11} (1975) 2558,
  \href{http://dx.doi.org/10.1103/PhysRevD.11.2558}{\doi{10.1103/PhysRevD.11.2558}}.

\bibitem{LeftRight}
\hrefCMSnoop {}{G.~Senjanovi{\'c} and R.~N. Mohapatra, ``Exact left-right
  symmetry and spontaneous violation of parity'',} \textit{ Phys. Rev. D}
  \textbf{ 12} (1975) 1502,
  \href{http://dx.doi.org/10.1103/PhysRevD.12.1502}{\doi{10.1103/PhysRevD.12.1502}}.

\bibitem{Senjanovic:1978ev}
\hrefCMSnoop {}{G.~Senjanovi{\'c}, ``Spontaneous breakdown of parity in a class
  of gauge theories'',} \textit{ Nucl. Phys. B} \textbf{ 153} (1979) 334,
\href{http://dx.doi.org/10.1016/0550-3213(79)90604-7}{\doi{10.1016/0550-3213(79)90604-7}}.

\bibitem{Min}
\hrefCMSnoop {}{P.~Minkowski, ``{$\mu \rightarrow \text{e} \gamma$ at a rate of
  one out of $10^9$ muon decays?}'',} \textit{ Phys. Lett. B} \textbf{ 67}
  (1977) 421,
  \href{http://dx.doi.org/10.1016/0370-2693(77)90435-X}{\doi{10.1016/0370-2693(77)90435-X}}.

\bibitem{MS}
\hrefCMSnoop {}{R.~N. Mohapatra and G.~Senjanovi\'c, ``Neutrino Mass and
  Spontaneous Parity Nonconservation'',} \textit{ Phys. Rev. Lett.} \textbf{
  44} (1980) 912,
  \href{http://dx.doi.org/10.1103/PhysRevLett.44.912}{\doi{10.1103/PhysRevLett.44.912}}.

\bibitem{Mohapatra:1980yp}
\hrefCMSnoop {}{R.~N. Mohapatra and G.~Senjanovi\'c, ``Neutrino masses and
  mixings in gauge models with spontaneous parity violation'',} \textit{ Phys.
  Rev. D} \textbf{ 23} (1981) 165,
  \href{http://dx.doi.org/10.1103/PhysRevD.23.165}{\doi{10.1103/PhysRevD.23.165}}.

\bibitem{Sjostrand:2006za}
\hrefCMSnoop {}{T.~Sj{\"o}strand, S.~Mrenna, and P.~Z. Skands, ``{PYTHIA} 6.4
  physics and manual'',} \textit{ JHEP} \textbf{ 05} (2006) 026,
  \href{http://dx.doi.org/10.1088/1126-6708/2006/05/026}{\doi{10.1088/1126-6708/2006/05/026}},
\href{http://www.arXiv.org/abs/hep-ph/0603175}{\texttt{arXiv:hep-ph/0603175}}.

\bibitem{CTEQ6L1}
J.~Pumplin\hrefCMSnoop {}{ {et~al.}, ``{New generation of parton distributions
  with uncertainties from global QCD analysis}'',} \textit{ JHEP} \textbf{ 07}
  (2002) 012,
  \href{http://dx.doi.org/10.1088/1126-6708/2002/07/012}{\doi{10.1088/1126-6708/2002/07/012}},
\href{http://www.arXiv.org/abs/hep-ph/0201195}{\texttt{arXiv:hep-ph/0201195}}.

\bibitem{fewz}
\hrefCMSnoop {}{R.~Gavin, Y.~Li, F.~Petriello, and S.~Quackenbush, ``{FEWZ 2.0:
  A code for hadronic Z production at next-to-next-to-leading order}'',}
  \textit{ Comput. Phys. Commun.} \textbf{ 182} (2011) 2388,
  \href{http://dx.doi.org/10.1016/j.cpc.2011.06.008}{\doi{10.1016/j.cpc.2011.06.008}},
\href{http://www.arXiv.org/abs/1011.3540}{\texttt{arXiv:1011.3540}}.

\bibitem{Gavin:2012kw}
\hrefCMSnoop {}{R.~Gavin, Y.~Li, F.~Petriello, and S.~Quackenbush, ``{W Physics
  at the LHC with FEWZ 2.1}'',} \textit{ Comput. Phys. Commun.} \textbf{ 184}
  (2013) 208,
  \href{http://dx.doi.org/10.1016/j.cpc.2012.09.005}{\doi{10.1016/j.cpc.2012.09.005}},
\href{http://www.arXiv.org/abs/1201.5896}{\texttt{arXiv:1201.5896}}.

\bibitem{preons}
\hrefCMSnoop {}{H.~Terazawa, M.~Yasue, K.~Akama, and M.~Hayashi, ``{Observable
  effects of the possible substructure of leptons and quarks}'',} \textit{
  Phys. Lett. B} \textbf{ 112} (1982) 387,
\href{http://dx.doi.org/10.1016/0370-2693(82)91075-9}{\doi{10.1016/0370-2693(82)91075-9}}.

\bibitem{cdf-CIlimit}
\hrefCMSnoop {}{{CDF} Collaboration, ``{Search for quark lepton compositeness
  and a heavy $W^\prime$ boson using the $\mathrm{e} \nu$ channel in
  $\mathrm{p\bar{p}}$ collisions at $\sqrt{s} = 1.8$ TeV}'',} \textit{ Phys.
  Rev. Lett.} \textbf{ 87} (2001) 231803,
  \href{http://dx.doi.org/10.1103/PhysRevLett.87.231803}{\doi{10.1103/PhysRevLett.87.231803}},
\href{http://www.arXiv.org/abs/hep-ex/0107008}{\texttt{arXiv:hep-ex/0107008}}.

\bibitem{CMS-PAS-EXO-12-048}
\hrefCMSnoop {}{{CMS Collaboration}, ``Search for dark matter, extra
  dimensions, and unparticles in monojet events in proton-proton collisions at
  $\sqrt{s} = 8${\TeV}'',} (2014).
  \href{http://www.arXiv.org/abs/1408.3583}{\texttt{arXiv:1408.3583}}.
Submitted to {Eur. Phys. J. C}.

\bibitem{CMS-monophoton-2012}
\hrefCMSnoop {}{{CMS Collaboration}, ``Search for new phenomena in monophoton
  final states in proton-proton collisions at $\sqrt{s} = 8${\TeV}'',} (2014).
  \href{http://www.arXiv.org/abs/1410.8812}{\texttt{arXiv:1410.8812}}.
Submitted to {Phys. Lett. B}.

\bibitem{monoLepton}
\hrefCMSnoop {}{Y.~Bai and T.~M.~P. Tait, ``Searches with mono-leptons'',}
  \textit{ Phys. Lett. B} \textbf{ 723} (2013) 384,
  \href{http://dx.doi.org/10.1016/j.physletb.2013.05.057}{\doi{10.1016/j.physletb.2013.05.057}},
\href{http://www.arXiv.org/abs/1208.4361}{\texttt{arXiv:1208.4361}}.

\bibitem{pdg}
\hrefCMSnoop {}{{Particle Data Group}, J.~Beringer {et~al.}, ``{Review of
  Particle Physics}'',} \textit{ Phys. Rev. D} \textbf{ 86} (2012) 010001,
\href{http://dx.doi.org/10.1103/PhysRevD.86.010001}{\doi{10.1103/PhysRevD.86.010001}}.

\bibitem{madgraph5}
J.~Alwall\hrefCMSnoop {}{ {et~al.}, ``{MadGraph 5: going beyond}'',} \textit{
  JHEP} \textbf{ 06} (2011) 128,
  \href{http://dx.doi.org/10.1007/JHEP06(2011)128}{\doi{10.1007/JHEP06(2011)128}},
\href{http://www.arXiv.org/abs/1106.0522}{\texttt{arXiv:1106.0522}}.

\bibitem{madgraph}
J.~Alwall\hrefCMSnoop {}{ {et~al.}, ``{MadGraph/MadEvent v4}: the new web
  generation'',} \textit{ JHEP} \textbf{ 09} (2007) 028,
  \href{http://dx.doi.org/10.1088/1126-6708/2007/09/028}{\doi{10.1088/1126-6708/2007/09/028}}.

\bibitem{Field:2010bc}
\hrefCMSnoop {}{R.~Field, ``{Early LHC Underlying Event Data - Findings and
  Surprises}'',} in \textit{ 22nd Hadron Collider Physics Symposium (HCP
  2010)}, W.~Trischuk, ed.
\newblock Toronto, 2010.
\newblock
\href{http://www.arXiv.org/abs/1010.3558}{\texttt{arXiv:1010.3558}}.
\newblock

\bibitem{CMS-PAS-PFT-09-001}
\href {http://cdsweb.cern.ch/record/1194487}{{CMS} Collaboration,
  ``Particle--Flow Event Reconstruction in {CMS} and Performance for Jets,
  Taus, and {\MET}'',} CMS Physics Analysis Summary CMS-PAS-PFT-09-001, 2009.

\bibitem{CMS-PAS-PFT-10-001}
\href {http://cdsweb.cern.ch/record/1247373}{{CMS} Collaboration,
  ``Commissioning of the Particle-flow Event Reconstruction with the first
  {LHC} collisions recorded in the {CMS} detector'',} CMS Physics Analysis
  Summary CMS-PAS-PFT-10-001, 2010.

\bibitem{CMS-PAS-PFT-10-002}
\href {http://cdsweb.cern.ch/record/1279341}{{CMS} Collaboration,
  ``Commissioning of the Particle-Flow Reconstruction in Minimum-Bias and Jet
  Events from {\Pp\Pp} Collisions at 7 {TeV}'',} CMS Physics Analysis Summary
  CMS-PAS-PFT-10-002, 2010.

\bibitem{CMS-JME-10-011}
\hrefCMSnoop {}{{CMS} Collaboration, ``{Determination of jet energy calibration
  and transverse momentum resolution in CMS}'',} \textit{ JINST} \textbf{ 6}
  (2011) P11002,
  \href{http://dx.doi.org/10.1088/1748-0221/6/11/P11002}{\doi{10.1088/1748-0221/6/11/P11002}},
\href{http://www.arXiv.org/abs/1107.4277}{\texttt{arXiv:1107.4277}}.

\bibitem{Chatrchyan:2011tn}
\hrefCMSnoop {}{{CMS} Collaboration, ``{Missing transverse energy performance
  of the CMS detector}'',} \textit{ JINST} \textbf{ 6} (2011) P09001,
  \href{http://dx.doi.org/10.1088/1748-0221/6/09/P09001}{\doi{10.1088/1748-0221/6/09/P09001}},
\href{http://www.arXiv.org/abs/1106.5048}{\texttt{arXiv:1106.5048}}.

\bibitem{WZcrosssectionPaper}
\hrefCMSnoop {}{{CMS} Collaboration, ``{Measurements of Inclusive W and Z Cross
  Sections in pp Collisions at $\sqrt{s} = 7$ TeV}'',} \textit{ JHEP} \textbf{
  01} (2011) 080,
  \href{http://dx.doi.org/10.1007/JHEP01(2011)080}{\doi{10.1007/JHEP01(2011)080}},
\href{http://www.arXiv.org/abs/1012.2466}{\texttt{arXiv:1012.2466}}.

\bibitem{CMS-DP-2013-009}
\href {http://cds.cern.ch/record/1536406}{{CMS} Collaboration, ``{Single Muon
  efficiencies in 2012 Data}'',} CMS Detector Performance Summary
  CMS-DP-2013-009, 2013.

\bibitem{Garwood1936}
\hrefCMSnoop {}{F.~Garwood, ``Fiducial Limits for the Poisson Distribution'',}
  \textit{ Biometrika} \textbf{ 28} (1936), no.~3-4, 437,
  \href{http://dx.doi.org/10.1093/biomet/28.3-4.437}{\doi{10.1093/biomet/28.3-4.437}}.

\bibitem{mcatnlo}
\hrefCMSnoop {}{S.~Frixione and B.~R. Webber, ``Matching {NLO QCD} computations
  and parton shower simulations'',} \textit{ JHEP} \textbf{ 06} (2002) 029,
  \href{http://dx.doi.org/10.1088/1126-6708/2002/06/029}{\doi{10.1088/1126-6708/2002/06/029}},
  \href{http://www.arXiv.org/abs/hep-ph/0204244}{\texttt{arXiv:hep-ph/0204244}}.

\bibitem{mcatnloheavyflavour}
\hrefCMSnoop {}{S.~Frixione, P.~Nason, and B.~R. Webber, ``{Matching NLO QCD
  and parton showers in heavy flavor production}'',} \textit{ JHEP} \textbf{
  08} (2003) 007,
  \href{http://dx.doi.org/10.1088/1126-6708/2003/08/007}{\doi{10.1088/1126-6708/2003/08/007}},
\href{http://www.arXiv.org/abs/hep-ph/0305252}{\texttt{arXiv:hep-ph/0305252}}.

\bibitem{herwig}
G.~Corcella\hrefCMSnoop {}{ {et~al.}, ``{HERWIG} 6: an event generator for
  hadron emission reactions with interfering gluons (including supersymmetric
  processes)'',} \textit{ JHEP} \textbf{ 01} (2001) 010,
  \href{http://dx.doi.org/10.1088/1126-6708/2001/01/010}{\doi{10.1088/1126-6708/2001/01/010}},
  \href{http://www.arXiv.org/abs/hep-ph/0011363}{\texttt{arXiv:hep-ph/0011363}}.

\bibitem{powheg_1}
\hrefCMSnoop {}{P.~Nason, ``{A new method for combining NLO QCD with shower
  Monte Carlo algorithms}'',} \textit{ JHEP} \textbf{ 11} (2004) 040,
  \href{http://dx.doi.org/10.1088/1126-6708/2004/11/040}{\doi{10.1088/1126-6708/2004/11/040}},
\href{http://www.arXiv.org/abs/hep-ph/0409146}{\texttt{arXiv:hep-ph/0409146}}.

\bibitem{powheg_2}
\hrefCMSnoop {}{S.~Frixione, P.~Nason, and C.~Oleari, ``{Matching NLO QCD
  computations with Parton Shower simulations: the POWHEG method}'',} \textit{
  JHEP} \textbf{ 11} (2007) 070,
  \href{http://dx.doi.org/10.1088/1126-6708/2007/11/070}{\doi{10.1088/1126-6708/2007/11/070}},
\href{http://www.arXiv.org/abs/0709.2092}{\texttt{arXiv:0709.2092}}.

\bibitem{powheg_3}
\hrefCMSnoop {}{S.~Alioli, P.~Nason, C.~Oleari, and E.~Re, ``{A general
  framework for implementing NLO calculations in shower Monte Carlo programs:
  the POWHEG BOX}'',} \textit{ JHEP} \textbf{ 06} (2010) 043,
  \href{http://dx.doi.org/10.1007/JHEP06(2010)043}{\doi{10.1007/JHEP06(2010)043}},
\href{http://www.arXiv.org/abs/1002.2581}{\texttt{arXiv:1002.2581}}.

\bibitem{Re:2010bp}
\hrefCMSnoop {}{E.~Re, ``{Single-top Wt-channel production matched with parton
  showers using the POWHEG method}'',} \textit{ Eur. Phys. J. C} \textbf{ 71}
  (2011) 1547,
  \href{http://dx.doi.org/10.1140/epjc/s10052-011-1547-z}{\doi{10.1140/epjc/s10052-011-1547-z}},
\href{http://www.arXiv.org/abs/1009.2450}{\texttt{arXiv:1009.2450}}.

\bibitem{Czakon:2013goa}
\hrefCMSnoop {}{M.~Czakon, P.~Fiedler, and A.~Mitov, ``{Total Top-Quark
  Pair-Production Cross Section at Hadron Colliders Through
  $O(\alpha^{4}_{S})$}'',} \textit{ Phys. Rev. Lett.} \textbf{ 110} (2013)
  252004,
  \href{http://dx.doi.org/10.1103/PhysRevLett.110.252004}{\doi{10.1103/PhysRevLett.110.252004}},
\href{http://www.arXiv.org/abs/1303.6254}{\texttt{arXiv:1303.6254}}.

\bibitem{powheg_DY}
\hrefCMSnoop {}{S.~Alioli, P.~Nason, C.~Oleari, and E.~Re, ``{Vector boson plus
  one jet production in POWHEG}'',} \textit{ JHEP} \textbf{ 01} (2011) 095,
  \href{http://dx.doi.org/10.1007/JHEP01(2011)095}{\doi{10.1007/JHEP01(2011)095}},
\href{http://www.arXiv.org/abs/1009.5594}{\texttt{arXiv:1009.5594}}.

\bibitem{horace}
\hrefCMSnoop {}{C.~M.~C. Calame, G.~Montagna, O.~Nicrosini, and A.~Vicini,
  ``Precision electroweak calculation of the production of a high
  transverse-momentum lepton pair at hadron colliders'',} \textit{ JHEP}
  \textbf{ 10} (2007) 109,
  \href{http://dx.doi.org/10.1088/1126-6708/2007/10/109}{\doi{10.1088/1126-6708/2007/10/109}},
  \href{http://www.arXiv.org/abs/0710.1722}{\texttt{arXiv:0710.1722}}.

\bibitem{CT10}
H.-L. Lai\hrefCMSnoop {}{ {et~al.}, ``New parton distributions for collider
  physics'',} \textit{ Phys. Rev. D} \textbf{ 82} (2010) 074024,
  \href{http://dx.doi.org/10.1103/PhysRevD.82.074024}{\doi{10.1103/PhysRevD.82.074024}},
\href{http://www.arXiv.org/abs/1007.2241}{\texttt{arXiv:1007.2241}}.

\bibitem{Wkfactors}
G.~Balossini\hrefCMSnoop {}{ {et~al.}, ``Combination of electroweak and {QCD}
  corrections to single {W} production at the {Fermilab Tevatron and the CERN
  LHC}'',} \textit{ JHEP} \textbf{ 01} (2010) 013,
  \href{http://dx.doi.org/10.1007/JHEP01(2010)013}{\doi{10.1007/JHEP01(2010)013}},
  \href{http://www.arXiv.org/abs/0907.0276}{\texttt{arXiv:0907.0276}}.

\bibitem{Alekhin:2011sk}
\hrefCMSnoop {}{S.~Alekhin {et~al.}, ``{The PDF4LHC Working Group Interim
  Report}'',} (2011).
\href{http://www.arXiv.org/abs/1101.0536}{\texttt{arXiv:1101.0536}}.

\bibitem{Botje:2011sn}
M.~Botje\hrefCMSnoop {}{ {et~al.}, ``{The PDF4LHC Working Group Interim
  Recommendations}'',} (2011).
\href{http://www.arXiv.org/abs/1101.0538}{\texttt{arXiv:1101.0538}}.

\bibitem{NNPDF2.3}
\hrefCMSnoop {}{{NNPDF} Collaboration, ``Impact of heavy quark masses on parton
  distributions and {LHC} Phenomenology'',} \textit{ Nucl. Phys. B} \textbf{
  849} (2011) 296,
  \href{http://dx.doi.org/10.1016/j.nuclphysb.2011.03.021}{\doi{10.1016/j.nuclphysb.2011.03.021}},
\href{http://www.arXiv.org/abs/1101.1300}{\texttt{arXiv:1101.1300}}.

\bibitem{MSTW2008}
\hrefCMSnoop {}{A.~D. Martin, W.~J. Stirling, R.~S. Thorne, and G.~Watt,
  ``Parton distributions for the {LHC}'',} \textit{ Eur. Phys. J. C} \textbf{
  63} (2009) 189,
  \href{http://dx.doi.org/10.1140/epjc/s10052-009-1072-5}{\doi{10.1140/epjc/s10052-009-1072-5}},
\href{http://www.arXiv.org/abs/0901.0002}{\texttt{arXiv:0901.0002}}.

\bibitem{CLSRead}
\hrefCMSnoop {}{A.~L. Read, ``Presentation of search results: the {$CL_s$}
  technique'',} \textit{ J. Phys. G} \textbf{ 28} (2002) 2693,
  \href{http://dx.doi.org/10.1088/0954-3899/28/10/313}{\doi{10.1088/0954-3899/28/10/313}}.

\bibitem{Junk:1999kv}
\hrefCMSnoop {}{T.~Junk, ``{Confidence level computation for combining searches
  with small statistics}'',} \textit{ Nucl. Instrum. Meth. A} \textbf{ 434}
  (1999) 435,
  \href{http://dx.doi.org/10.1016/S0168-9002(99)00498-2}{\doi{10.1016/S0168-9002(99)00498-2}},
\href{http://www.arXiv.org/abs/hep-ex/9902006}{\texttt{arXiv:hep-ex/9902006}}.

\bibitem{TevatronDMFrontier}
\hrefCMSnoop {}{Y.~Bai, P.~J. Fox, and R.~Harnik, ``{The Tevatron at the
  frontier of dark matter direct detection}'',} \textit{ JHEP} \textbf{ 12}
  (2010) 048,
  \href{http://dx.doi.org/10.1007/JHEP12(2010)048}{\doi{10.1007/JHEP12(2010)048}},
\href{http://www.arXiv.org/abs/1005.3797}{\texttt{arXiv:1005.3797}}.

\bibitem{BuchmuellerDarkMatterAssumptions}
\hrefCMSnoop {}{O.~Buchmueller, M.~J. Dolan, and C.~McCabe, ``{Beyond effective
  field theory for dark matter searches at the LHC}'',} \textit{ JHEP} \textbf{
  01} (2014) 025,
  \href{http://dx.doi.org/10.1007/JHEP01(2014)025}{\doi{10.1007/JHEP01(2014)025}},
\href{http://www.arXiv.org/abs/1308.6799}{\texttt{arXiv:1308.6799}}.

\bibitem{Greg}
\hrefCMSnoop {}{K.-M. Cheung and G.~L. Landsberg, ``{Kaluza-Klein states of the
  standard model gauge bosons: constraints from high energy experiments}'',}
  \textit{ Phys. Rev. D} \textbf{ 65} (2002) 076003,
  \href{http://dx.doi.org/10.1103/PhysRevD.65.076003}{\doi{10.1103/PhysRevD.65.076003}},
\href{http://www.arXiv.org/abs/hep-ph/0110346}{\texttt{arXiv:hep-ph/0110346}}.

\end{thebibliography}\endgroup

\cleardoublepage \appendix\section{The CMS Collaboration \label{app:collab}}\begin{sloppypar}\hyphenpenalty=5000\widowpenalty=500\clubpenalty=5000\textbf{Yerevan Physics Institute,  Yerevan,  Armenia}\\*[0pt]
V.~Khachatryan, A.M.~Sirunyan, A.~Tumasyan
\vskip\cmsinstskip
\textbf{Institut f\"{u}r Hochenergiephysik der OeAW,  Wien,  Austria}\\*[0pt]
W.~Adam, T.~Bergauer, M.~Dragicevic, J.~Er\"{o}, C.~Fabjan\cmsAuthorMark{1}, M.~Friedl, R.~Fr\"{u}hwirth\cmsAuthorMark{1}, V.M.~Ghete, C.~Hartl, N.~H\"{o}rmann, J.~Hrubec, M.~Jeitler\cmsAuthorMark{1}, W.~Kiesenhofer, V.~Kn\"{u}nz, M.~Krammer\cmsAuthorMark{1}, I.~Kr\"{a}tschmer, D.~Liko, I.~Mikulec, D.~Rabady\cmsAuthorMark{2}, B.~Rahbaran, H.~Rohringer, R.~Sch\"{o}fbeck, J.~Strauss, A.~Taurok, W.~Treberer-Treberspurg, W.~Waltenberger, C.-E.~Wulz\cmsAuthorMark{1}
% ...
\vskip\cmsinstskip
% ...
1:~~Also at Vienna University of Technology, Vienna, Austria\\
2:~~Also at CERN, European Organization for Nuclear Research, Geneva, Switzerland\\
% ...

\end{sloppypar}
\end{document}